\documentclass{llncs}

\usepackage{mathptmx}
\usepackage{helvet}
\usepackage{courier}
\usepackage{multicol}
\usepackage[bottom]{footmisc}
\usepackage{amssymb}
\usepackage{rotating}

\usepackage{epsfig}
\usepackage{latexsym}
\usepackage{graphicx}
\usepackage{color}
\usepackage{url}
\usepackage{amsfonts}
\usepackage{pgf,pgfarrows,pgfnodes,pgfautomata,pgfheaps,pgfshade}
\usepackage{tikz}

\usetikzlibrary{arrows,decorations.pathmorphing,backgrounds,positioning,fit,petri}
\usepackage{etex}

\usepackage{stmaryrd}
\usepackage{amsmath}

\usepackage{bussproofs}

\newcommand{\post}[1]{\mbox{$#1^{\bullet}$}}
\newcommand{\pre}[1]{\mbox{$^{\bullet}#1$}}

\newcommand{\fine}{{\mbox{ }\nolinebreak\hfill{$\Box$}}}

\newcommand{\deriv}[1]{{\mbox{${\:\stackrel{#1}{\longrightarrow}\:}$}}}

\newcommand{\Deriv}[1]{{\mbox{${\:\stackrel{#1}{\Longrightarrow}\:}$}}}

\newcommand{\nderiv}[1]{\nrightarrow}

\newcommand{\bigfrac}[2]{
\renewcommand{\arraystretch}{1.5}
\begin{array}{c}#1\\
\hline
#2
\end{array}}
\newcommand{\restr}[1]{\mbox{$({\bf\nu} #1)$}}
\newcommand{\para}{\mbox{$\,|\,$}}

\renewcommand{\mid}{\;\;\big|\;\;}

\newcommand{\nat}{{\mathbb N}}

\begin{document}

 \pagestyle{headings}

\title{Branching Place Bisimilarity}
\author{Roberto Gorrieri\\
\institute{Dipartimento di Informatica --- Scienza e Ingegneria\\
Universit\`a di Bologna, \\Mura A. Zamboni 7,
40127 Bologna, Italy}
\email{{\small roberto.gorrieri@unibo.it}}
}

\maketitle

\begin{abstract}
Place bisimilarity $\sim_p$ is a behavioral equivalence for finite Petri nets, proposed in \cite{ABS91}
and proved decidable in \cite{Gor21}. In this paper we propose an extension to finite Petri nets with silent moves
of the place bisimulation idea, yielding {\em branching} place bisimilarity $\approx_p$, following the intuition 
of branching bisimilarity \cite{vGW96} on labeled transition systems. We also propose a slightly coarser variant,
called branching {\em d-place} bisimilarity $\approx_d$, following the intuition of d-place bisimilarity in \cite{Gor21}.
We prove that $\approx_p$ and $\approx_d$ are decidable equivalence relations. Moreover, we prove that $\approx_d$ is strictly finer than 
branching fully-concurrent bisimilarity \cite{Pin93,Gor20c}, essentially because $\approx_d$ does not consider 
as unobservable those $\tau$-labeled net transitions
with pre-set size larger than one, i.e., those resulting from (multi-party) interaction.
\end{abstract}

%
\section{Introduction}
%

Place bisimilarity,
originating from an idea by Olderog \cite{Old} (under the name of strong bisimilarity) and then refined by Autant, 
Belmesk and Schnoebelen \cite{ABS91},  
is a behavioral equivalence over finite Place/Transition Petri nets (P/T nets, for short), 
based on relations over the {\em finite set of net places}, rather than over the 
(possibly infinite) set of net markings. 
This equivalence does respect the expected causal behavior of Petri nets; in fact, van Glabbeek 
proved in \cite{G15}
that place bisimilarity is slightly finer than 
{\em structure preserving bisimilarity} \cite{G15}, in turn
slightly finer than {\em fully-concurrent bisimilarity} \cite{BDKP91}.
Place bisimilarity was proved decidable in \cite{Gor21} and it is the first {\em sensible} (i.e., fully respecting causality 
and the branching structure) 
behavioral equivalence
which was proved decidable over finite (possibly unbounded) Petri nets (with the exception of net isomorphism).
In \cite{Gor21}, a sligthly coarser variant is proposed, called {\em d-}place bisimilarity, which allows to relate not only places 
to places, but also
places to the empty marking. D-place bisimilarity was proved to be finer than fully-concurrent bisimilarity and, to date, it is 
the coarsest sensible behavioral relation to be decidable on finite Petri nets (when all the transition labels are considered as observable).

This paper aims at extending the place bisimulation idea to Petri nets {\em with silent transitions}, i.e., transitions with 
unobservable label, usually denoted by $\tau$.
To this aim, we take inspiration from {\em branching} bisimilarity, proposed in \cite{vGW96} over 
labeled transition systems \cite{Kel76,GV15} (LTSs, for short),
a behavioral relation more appropriate than weak bisimilarity \cite{Mil89}, as it better respects the timing of choices. 
In fact, this crucial property is enjoyed by branching bisimilarity (but not by weak bisimilarity):
when in the branching bisimulation game a transition $q_1 \deriv{\mu} q_1'$ is matched by a computation, 
say, $q_2 \Deriv{\epsilon} q_2' \deriv{\mu} q_2''$, 
all the states traversed by the silent computation from $q_2$ to $q_2'$ are 
branching bisimilar, so that they all belong to the same equivalence class. We call this property by {\em weak stuttering property}.

The main problem we had to face, in order to transpose this idea in the realm of Petri nets, 
was to properly understand if and when a silent net transition can 
be really considered as potentially
unobservable. In fact, while in the theory of sequential, nondeterministic systems, modeled by means of LTSs, 
all the $\tau$-labeled transitions can, to some extent, be abstracted away,
in the theory of Petri nets (and of distributed systems, in general), it is rather questionable whether this is the case. 
For sure a silent net transition with pre-set and post-set of size 1 may be abstracted away, as it represents some internal 
computation, local to a single sequential component of the distributed system. However, a $\tau$-labeled net transition
with pre-set of size 2 or more, which models a (possibly, multi-party) interaction, is really observable: as to 
establish the synchronization
it is necessary to use some communication infrastructure, for sure one observer can see that such a synchronization takes place.
This is, indeed, what happens over the Internet: a communication via IP is an observable event, 
even if the actual content of the message may be unobservable (in case it is encrypted).
For this reason, our definition of {\em branching place bisimulation} considers as potentially unobservable only the so-called
{\em $\tau$-sequential} transitions, i.e., those silent transitions whose pre-set and post-set have size 1.

We define branching place bisimulation in such a way that it enjoys the weak stuttering property mentioned above, so that
it really respects the timing of choices. We prove that  the induced
branching place bisimilarity $\approx_p$ is an equivalence relation, where the crucial step in this proof
is to prove that the relational composition of two branching place bisimulations is a branching place bisimulation.
We also define a slightly coarser variant, called branching d-place bisimilarity $\approx_d$, that allows to relate
a place not only to another place, but also to the empty marking.
Of course, $\approx_d$ is rather discriminating if compared to other behavioral semantics; in particular, we prove
that it is strictly finer than {\em branching fully-concurrent bisimilarity} \cite{Pin93,Gor20c}, essentially because the latter
may also abstract w.r.t. silent transitions that are not $\tau$-sequential (and also may relate markings of different size).

The main contribution of this paper is to show that $\approx_p$  is decidable for finite P/T nets
(and, in a similar manner, that also $\approx_d$ is so).
The proof idea is as follows. As a place relation $R \subseteq S \times S$ is finite if the set $S$ 
of places is finite,
there are finitely many place relations for a finite net. We can list all these relations, say $R_1, R_2, \ldots R_n$. 
It is decidable whether a place relation $R_i$ is a branching
place bisimulation by checking two {\em finite} conditions over 
a {\em finite} number of marking pairs: this is a non-obvious observation, as a branching place bisimulation requires
that the place bisimulation game holds for the infinitely many pairs $m_1$ and $m_2$
which are {\em bijectively} related via $R_i$ (denoted by $(m_1, m_2) \in R_i^\oplus$). 
Hence, to decide whether 
$m_1 \approx_p m_2$, it is enough to check, for $i = $ $1, \ldots n$, whether $R_i$  is a branching place 
bisimulation and, in such a case, whether $(m_1, m_2) \in R_i^\oplus$.

The paper is organized as follows. Section \ref{def-sec} recalls the basic definitions about Petri nets, their sequential semantics and also their
causal semantics. A particular care is devoted to the definition of branching interleaving bisimilarity, showing that it really 
enjoys the weak stuttering property, so that it respects the timing of choices; we also recall 
branching fully-concurrent bisimilarity from \cite{Pin93,Gor20c},
but we were not able to prove (or disprove) that it really enjoys the weak stuttering property.
Section \ref{place-sec} recalls the main definitions and results about place bisimilarity and d-place bisimilarity from \cite{Gor21}; in particular, it shows that
place bisimulation is not defined coinductively, as the union of place bisimulations may be not a place bisimulation.
Section \ref{br-place-sec} introduces the concept of 
branching place bisimulation, proves that the induced place bisimilarity $\approx_p$ is an equivalence relation, and 
shows that it really enjoys the weak stuttering property.
Section \ref{decid-br-place-sec} shows that $\approx_p$ is decidable.
Section \ref{case-sec} presents a small case study about a producer-consumer system in order to show the real applicability of the approach.
Section \ref{br-d-place-sec} introduces branching d-place bisimilarity $\approx_d$, hints that it is a coarser, decidable equivalence relation
and proves that it is strictly finer than branching fully-concurrent bisimilarity.
Finally, in Section \ref{conc-sec} we discuss the pros and cons of branching (d-)place bisimilarity,  
and describe related literature and some future research.

\begin{quote}
{\em This paper is the extended and revised version of \cite{Gor-forte21}.}
\end{quote}

%
\section{Basic Definitions} \label{def-sec}
%

\begin{definition}\label{multiset}{\bf (Multiset)}\index{Multiset}
Let $\nat$ be the set of natural numbers. 
Given a finite set $S$, a {\em multiset} over $S$ is a function $m: S \rightarrow\nat$. 
The {\em support} set $dom(m)$ of $m$ is $\{ s \in S \mid m(s) \neq 0\}$. 
The set of all multisets 
over $S$,  denoted by ${\mathcal M}(S)$, is ranged over by $m$. 
We write $s \in m$ if $m(s)>0$.
The {\em multiplicity} of $s$ in $m$ is given by the number $m(s)$. The {\em size} of $m$, denoted by $|m|$,
is the number $\sum_{s\in S} m(s)$, i.e., the total number of its elements.
A multiset $m$ such 
that $dom(m) = \emptyset$ is called {\em empty} and is denoted by $\theta$.
We write $m \subseteq m'$ if $m(s) \leq m'(s)$ for all $s \in S$. 

{\em Multiset union} $\_ \oplus \_$ is defined as follows: $(m \oplus m')(s)$ $ = m(s) + m'(s)$; 
it is commutative, associative and has $\theta$ as neutral element. 
{\em Multiset difference} $\_ \ominus \_$ is defined as follows: 
$(m_1 \ominus m_2)(s) = max\{m_1(s) - m_2(s), 0\}$.
The {\em scalar product} of a number $j$ with $m$ is the multiset $j \cdot m$ defined as
$(j \cdot m)(s) = j \cdot (m(s))$. By $s_i$ we also denote the multiset with $s_i$ as its only element.
Hence, a multiset $m$ over $S = \{s_1, \ldots, s_n\}$
can be represented as $k_1\cdot s_{1} \oplus k_2 \cdot s_{2} \oplus \ldots \oplus k_n \cdot s_{n}$,
where $k_j = m(s_{j}) \geq 0$ for $j= 1, \ldots, n$.
\fine
\end{definition}

\begin{definition}\label{pt-net-def}{\bf (Place/Transition net)}
A labeled {\em Place/Transition} Petri net (P/T net for short) is a tuple $N = (S, A, T)$, where
\begin{itemize}
\item 
$S$ is the finite set of {\em places}, ranged over by $s$ (possibly indexed),
\item 
$A$ is the finite set of {\em labels}, ranged over by $\ell$ (possibly indexed), and
\item 
$T \subseteq {(\mathcal M}(S) \setminus \{\theta\}) \times A \times {\mathcal M}(S)$ 
is the finite set of {\em transitions}, 
ranged over by $t$ (possibly indexed).
\end{itemize}

Given a transition $t = (m, \ell, m')$,
we use the notation:
\begin{itemize}
\item  $\pre t$ to denote its {\em pre-set} $m$ (which cannot be  empty) of tokens to be consumed;
\item $l(t)$ for its {\em label} $\ell$, and
\item $\post t$ to denote its {\em post-set} $m'$ of tokens to be produced.
\end{itemize} 
Hence, transition $t$ can be also represented as $\pre t \deriv{l(t)} \post t$.
We also define the {\em flow function}
{\mbox  flow}$: (S \times T) \cup (T \times S) \rightarrow \nat$ as follows:
for all $s \in S$, for all $t \in T$,
{\mbox  flow}$(s,t) = \pre{t}(s)$ and {\mbox  flow}$(t,s) = \post{t}(s)$.
We will use $F$ to denote the {\em flow relation} 
$\{(x,y) \mid x,y \in S \cup T \, \wedge \, ${\mbox  flow}$(x,y) > 0\}$.
Finally, we define pre-sets and post-sets also for places as: $\pre s = \{t \in T \mid s \in \post t\}$
and $\post s = \{t \in T \mid s \in \pre t\}$. Note that while the pre-set (post-set) of a transition is, in general, 
a multiset, the pre-set (post-set) of a place is a set.
 \fine
\end{definition}

Graphically, a place is represented by a little circle and a transition by a little box. These are 
connected by directed arcs, which
may be labeled by a positive integer, called the {\em weight}, to denote the number of tokens 
consumed (when the arc goes from a place to the transition) or produced (when the arc goes form the transition to a
place) by the execution of the transition;
if the number is omitted, then the weight default value is $1$.

\begin{definition}\label{net-system}{\bf (Marking, P/T net system)}
A multiset over $S$  is called a {\em marking}. Given a marking $m$ and a place $s$, 
we say that the place $s$ contains $m(s)$ {\em tokens}, graphically represented by $m(s)$ bullets
inside place $s$.
A {\em P/T net system} $N(m_0)$ is a tuple $(S, A, T, m_{0})$, where $(S,A, T)$ is a P/T net and $m_{0}$ is  
a marking over $S$, called
the {\em initial marking}. We also say that $N(m_0)$ is a {\em marked} net.
\fine
\end{definition}

%
\subsection{Sequential Semantics}\label{seql-sem-sec}
%

\begin{definition}\label{firing-system}{\bf (Enabling, firing sequence, reachable marking, safe net)}
Given a P/T net $N = (S, A, T)$, a transition $t $ is {\em enabled} at $m$, 
denoted by $m[t\rangle$, if $\pre t \subseteq m$. 
The execution (or {\em firing}) of  $t$ enabled at $m$ produces the marking $m' = (m \ominus  \pre t) \oplus \post t$. 
This is written $m[t\rangle m'$. 
A {\em firing sequence} starting at $m$ is defined inductively as follows:
\begin{itemize}
\item $m[\epsilon\rangle m$ is a firing sequence (where $\epsilon$ denotes an empty sequence of transitions) and
\item if $m[\sigma\rangle m'$ is a firing sequence and $m' [t\rangle m''$, then
$m [\sigma t\rangle m''$ is a firing sequence. 
\end{itemize}
If $\sigma = t_1 \ldots t_n$ (for $n \geq 0$) and $m[\sigma\rangle m'$ is a firing sequence, then there exist $m_1,  \ldots, m_{n+1}$ such that
$m = m_1[t_1\rangle m_2 [t_2\rangle \ldots  m_n [t_n\rangle m_{n+1} = m'$, and 
$\sigma = t_1 \ldots t_n$ is called a {\em transition sequence} starting at $m$ and ending 
at $m'$.
The definition of pre-set and post-set can be extended to transition sequences as follows:
$\pre{\epsilon} = \theta$, $\pre{(t \sigma)} = \pre{t} \oplus (\pre{\sigma} \ominus \post{t})$, $\post{\epsilon} = \theta$,
$\post{(t \sigma)} = \post{\sigma} \oplus (\post{t} \ominus \pre{\sigma})$.

The set of {\em reachable markings} from $m$ is 
$[m\rangle = \{m' \mid  \exists \sigma.$ $
m[\sigma\rangle m'\}$. 
The P/T net system $N = $ $(S, A, T, m_0)$ is  {\em safe} if  for each 
$m \in [m_0\rangle$ and for all $s \in S$, we have that $m(s) \leq 1$.
\fine
\end{definition}

Note that the reachable markings of a P/T net can be countably infinitely many when the net is not bounded, i.e.,
when the number of tokens on some places can grow unboundedly.

Now we recall a simple behavioral equivalence on P/T nets, defined directly over the markings of the net, which 
compares two markings with respect to their sequential behavior.

\begin{definition}\label{def-int-bis}{\bf (Interleaving Bisimulation)}
Let $N = (S, A, T)$ be a P/T net. 
An {\em interleaving bisimulation} is a relation
$R\subseteq {\mathcal M}(S) \times {\mathcal M}(S)$ such that if $(m_1, m_2) \in R$
then
\begin{itemize}
\item $\forall t_1$ such that  $m_1[t_1\rangle m'_1$, $\exists t_2$ such that $m_2[t_2\rangle m'_2$ 
with $l(t_1) = l(t_2)$ and $(m'_1, m'_2) \in R$,
\item $\forall t_2$ such that  $m_2[t_2\rangle m'_2$, $\exists t_1$ such that $m_1[t_1\rangle m'_1$ 
with $l(t_1) = l(t_2)$ and $(m'_1, m'_2) \in R$.
\end{itemize}

Two markings $m_1$ and $m_2$ are {\em interleaving bisimilar}, 
denoted by $m_1 \sim_{int} m_2$, if there exists an interleaving bisimulation $R$ such that $(m_1, m_2) \in R$.
\fine
\end{definition}

Interleaving bisimilarity was proved undecidable in \cite{Jan95} for P/T nets having at least two unbounded places, 
with a proof based on the comparison of two {\em sequential} P/T nets, 
where a P/T net is sequential if it does not offer any concurrent behavior. Hence, interleaving bisimulation equivalence is 
undecidable even for the subclass of sequential finite P/T nets. Esparza observed in \cite{Esp98} that all the non-interleaving 
bisimulation-based equivalences (in the spectrum ranging from interleaving bisimilarity to fully-concurrent bisimilarity \cite{BDKP91})
collapse to interleaving bisimilarity over sequential P/T nets. Hence, the proof in \cite{Jan95} applies to all these
non-interleaving bisimulation equivalences as well.

\begin{definition}\label{pt-silent-def}{\bf (P/T net with silent moves)}
A P/T net $N = (S, A, T)$ such that $\tau \in A$, where $\tau$ 
is the only invisible action that can be used to label transitions, is called a P/T net {\em with silent moves}.
\fine
\end{definition}

We now extend the behavioral equivalence above to P/T nets with silent transitions, following the intuition
of {\em branching} bisimulation \cite{vGW96} on LTSs.

\begin{definition}\label{br-int-bis}{\bf (Branching interleaving bisimulation)}
Let $N = (S, A, T)$ be a P/T net with silent moves.
A {\em branching} interleaving bisimulation is a relation
$R\subseteq {\mathcal M}(S) \times {\mathcal M}(S)$ such that if $(m_1, m_2) \in R$
then
\begin{itemize}
\item $\forall t_1$ such that  $m_1[t_1\rangle m_1'$,
\begin{itemize}
    \item[--]  either $l(t_1) = \tau$ and $\exists \sigma_2$ such that $o(\sigma_2) = \epsilon$, 
    $m_2[\sigma_2\rangle m_2'$ with $(m_1, m_2') \in R$ and $(m_1', m_2') \in R$,
     \item[--] or $\exists \sigma,  t_2$ such that $o(\sigma) = \epsilon$, $l(t_1) = l(t_2)$, 
     $m_2[\sigma\rangle m [t_2\rangle m_2'$ with $(m_1, m) \in R$ and $(m_1', m_2') \in R$,
\end{itemize}
\item and, symmetrically, $\forall t_2$ such that  $m_2[t_2 \rangle m_2'$.
\end{itemize}

Two markings $m_1$ and $m_2$ are {\em branching interleaving bisimilar}, 
denoted $m_1 \approx_{bri} m_2$, if there exists a branching interleaving bisimulation $R$
that relates them.
\fine
\end{definition}

This definition is not a rephrasing on nets of the original definition on LTSs in \cite{vGW96}, rather of a slight variant 
called {\em semi-branching bisimulation} \cite{vGW96,Bas96}, which gives rise to the same equivalence
as the original definition but has better mathematical properties; in particular it ensures \cite{Bas96} that the relational composition of 
branching bisimulations is a branching bisimulation.
Note that a silent transition performed by one of the two markings may be matched by the other one also by idling: 
this is due to the {\em either} case when $\sigma_2 = \epsilon$
(or $\sigma_1 = \epsilon$ for the symmetric case). 
Branching interleaving bisimilarity $\approx_{bri}$, which is defined as the union of all the branching interleaving bisimulations, 
is the largest branching interleaving bisimulation and also an equivalence relation.
Of course, also branching interleaving bisimilarity is undecidable for finite P/T nets.

\begin{theorem}\label{br-int-bis-fix}{\bf (Branching interleaving bisimilarity is a fixpoint)}
Let $N = (S, A, T)$ be a P/T net with silent moves.
Branching interleaving bisimilarity $\approx_{bri}$ is a relation
such that $m_1 \approx_{bri} m_2$
{\em if and only if}
\begin{itemize}
\item $\forall t_1$ such that  $m_1[t_1\rangle m_1'$,
\begin{itemize}
    \item[--]  either $l(t_1) = \tau$ and $\exists \sigma_2$ such that $o(\sigma_2) = \epsilon$, 
    $m_2[\sigma_2\rangle m_2'$ with $m_1 \approx_{bri} m_2'$ and $m_1' \approx_{bri} m_2'$,
     \item[--] or $\exists \sigma,  t_2$ such that $o(\sigma) = \epsilon$, $l(t_1) = l(t_2)$, 
     $m_2[\sigma\rangle m [t_2\rangle m_2'$ with $m_1 \approx_{bri} m$ and $m_1' \approx_{bri} m_2'$,
\end{itemize}
\item and, symmetrically, $\forall t_2$ such that  $m_2[t_2 \rangle m_2'$.
\fine
\end{itemize}
\proof
Note that in Definition \ref{br-int-bis}, we have {\em ``implies"} instead of {\em  ``if and only if"}.
Hence, the implication from left to right is due to the fact that $\approx_{bri}$ is itself a branching interleaving bisimulation. 

For the implication from right to left, first, define a new relation $\approx'$ in terms of $\approx_{bri}$ as follows:
$m_1 \approx' m_2$ {\em  if and only if} 
\begin{itemize}
\item $\forall t_1$ such that  $m_1[t_1\rangle m_1'$,
\begin{itemize}
    \item[--]  either $l(t_1) = \tau$ and $\exists \sigma_2$ such that $o(\sigma_2) = \epsilon$, 
    $m_2[\sigma_2\rangle m_2'$ with $m_1 \approx_{bri} m_2'$ and $m_1' \approx_{bri} m_2'$,
     \item[--] or $\exists \sigma,  t_2$ such that $o(\sigma) = \epsilon$, $l(t_1) = l(t_2)$, 
     $m_2[\sigma\rangle m [t_2\rangle m_2'$ with $m_1 \approx_{bri} m$ and $m_1' \approx_{bri} m_2'$,
\end{itemize}
\item and, symmetrically, $\forall t_2$ such that  $m_2[t_2 \rangle m_2'$.
\end{itemize}

Now we want to prove that $\approx_{bri} = \approx'$, hence proving the property stated above. 
First, if $m_1 \approx_{bri} m_2$, then (as $\approx_{bri}$ is a branching interleaving bisimulation)
\begin{itemize}
\item $\forall t_1$ such that  $m_1[t_1\rangle m_1'$,
\begin{itemize}
    \item[--]  either $l(t_1) = \tau$ and $\exists \sigma_2$ such that $o(\sigma_2) = \epsilon$, 
    $m_2[\sigma_2\rangle m_2'$ with $m_1 \approx_{bri} m_2'$ and $m_1' \approx_{bri} m_2'$,
     \item[--] or $\exists \sigma,  t_2$ such that $o(\sigma) = \epsilon$, $l(t_1) = l(t_2)$, 
     $m_2[\sigma\rangle m [t_2\rangle m_2'$ with $m_1 \approx_{bri} m$ and $m_1' \approx_{bri} m_2'$,
\end{itemize}
\item and, symmetrically, $\forall t_2$ such that  $m_2[t_2 \rangle m_2'$,
\end{itemize}
and so (by using the implication from right to left in the definition of $\approx'$) we have that $m_1 \approx' m_2$.
It remains to prove the reverse implication, i.e., that $m_1 \approx' m_2$ implies $m_1 \approx_{bri} m_2$. To obtain this, we prove that $\approx'$
is a branching interleaving bisimulation. 

Assume that $m_1 \approx' m_2$ and $m_1[t_1\rangle m_1'$ (the symmetric case when $m_2$ moves first is analogous, 
hence omitted). By definition of $\approx'$, we have that 
\begin{itemize}
    \item[--]  either $l(t_1) = \tau$ and $\exists \sigma_2$ such that $o(\sigma_2) = \epsilon$, 
    $m_2[\sigma_2\rangle m_2'$ with $m_1 \approx_{bri} m_2'$ and $m_1' \approx_{bri} m_2'$;
    but, by what we just proved, we have also that $m_1 \approx' m_2'$ and $m_1' \approx' m_2'$, and we are done;
     \item[--] or $\exists \sigma,  t_2$ such that $o(\sigma) = \epsilon$, $l(t_1) = l(t_2)$, 
     $m_2[\sigma\rangle m [t_2\rangle m_2'$ with $m_1 \approx_{bri} m$ and $m_1' \approx_{bri} m_2'$;
     but, by what we just proved, we have also that $m_1 \approx' m$ and $m_1' \approx' m_2'$, and we are done.
\end{itemize}
Hence, $\approx'$ is a branching interleaving bisimulation, indeed. And this completes the proof.
\fine
\end{theorem}

\begin{remark}\label{stutt1-rem-int}{\bf (Strong stuttering property)}
By means of Theorem \ref{br-int-bis-fix}, it is not difficult to prove that, given a silent firing sequence 
$m_1 [t_1\rangle m_2 [t_2\rangle m_3  \ldots m_n [t_n\rangle m_{n+1}$, with $l(t_i) = \tau$ for $i = 1, \ldots, n$, 
if $m_1 \approx_{bri} m_{n+1}$,
then $m_i \approx_{bri} m_j$ for $i, j = 1, \ldots, n+1$.
This is sometimes called the {\em strong stuttering property}. 

For the sake of the argument, let $n = 2$, so that $m_1 [t_1\rangle m_2 [t_2\rangle m_3$ and that $m_1 \approx_{bri} m_3$.
We want to prove that $m_1 \approx_{bri} m_2$ (and, symmetrically, we can prove that $m_2 \approx_{bri} m_3$).
Assume $m_1[t_1'\rangle m_1'$. Then, as $m_1 \approx_{bri} m_3$, we have that:
\begin{itemize}
    \item[--]  either $l(t_1') = \tau$ and $\exists \sigma_3$ such that $o(\sigma_3) = \epsilon$, 
    $m_3[\sigma_3\rangle m_3'$ with $m_1 \approx_{bri} m_3'$ and also $m_1' \approx_{bri} m_3'$;
     \item[--] or $\exists \sigma,  t_3$ such that $o(\sigma) = \epsilon$, $l(t_1') = l(t_3)$, 
     $m_3[\sigma\rangle m [t_3\rangle m_3'$ with $m_1 \approx_{bri} m$ and $m_1' \approx_{bri} m_3'$.
\end{itemize}
Hence, in the either-case, $m_2$ can reply with $m_2[t_2\rangle m_3[\sigma_3\rangle m_3'$ with $m_1 \approx_{bri} m_3'$ and $m_1' \approx_{bri} m_3'$;
while in the or-case, $m_2$ can reply with $m_2[t_2\rangle m_3[\sigma\rangle m [t_3\rangle m_3'$ with $m_1 \approx_{bri} m$ and $m_1' \approx_{bri} m_3'$.
Now, assume $m_2[t_2'\rangle m_2'$. Then, $m_1$ can reply with $m_1[t_1\rangle m_2[t_2'\rangle m_2'$, with $m_2 \approx_{bri} m_2$ and $m_2' \approx_{bri} m_2'$. In all the cases, we have checked that the branching interleaving bisimulation game holds for $m_1$ and $m_2$, so that, by 
using the implication from right to left of Theorem \ref{br-int-bis-fix}, we get the thesis $m_1 \approx_{bri} m_2$.
\fine
\end{remark}

\begin{remark}\label{stutt2-rem-int}{\bf (Weak stuttering property)}
By using the strong stuttering property above, another, quite interesting property can be proved for $\approx_{bri}$, we 
call {\em weak stuttering property}.
Consider the {\em either} case: since $(m_1, m_2) \in \approx_{bri}$ 
by hypothesis, and $m_2[\sigma_2\rangle m_2'$ with  $(m_1, m_2') \in \approx_{bri}$, 
it follows that $(m_2, m_2') \in \approx_{bri}$ because $\approx_{bri}$ is an equivalence relation. This implies that
all the markings in the silent path from $m_2$ to $m_2'$ are branching interleaving bisimilar (by the 
{\em strong stuttering property}).
Similarly for the  {\em or} case: if $m_1 [t_1\rangle m'_1$ (with $l(t_1)$ that can be $\tau$) and $m_2$ responds by performing 
$m_2 [\sigma\rangle m [t_2\rangle m'_2$  with $m_1 \approx_{bri} m$, then, by transitivity, $m_2 \approx_{bri} m$;
hence, by the strong stuttering property, 
$m_1$ is branching interleaving bisimilar to each marking in the path from $m_2$ to $m$. Summing up, this means that in the 
branching interleaving bisimilarity game, 
while matching a transition with a computation, all the intermediate states in the computation are equivalent, so that $\approx_{bri}$ strictly respects
the timing of choices.
\fine
\end{remark}

%
\subsection{Causality-based Semantics}\label{causal-sem-sec}
%

We outline some definitions, adapted from the literature
(cf., e.g., \cite{GR83,BD87,Old,G15,Gor22}).

\begin{definition}\label{acyc-def}{\bf (Acyclic net)}
A P/T net $N = (S, A, T)$ is
 {\em acyclic} if its flow relation $F$ is acyclic (i.e., $\not \exists x$ such that $x F^+ x$, 
 where $F^+$ is the transitive closure of $F$).
 \fine
\end{definition}

The causal semantics of a marked P/T net is defined by a class of particular acyclic safe nets, 
where places are not branched (hence they represent a single run) and all arcs have weight 1. 
This kind of net is called {\em causal net}. 
We use the name $C$ (possibly indexed) to denote a causal net, the set $B$ to denote its 
places (called {\em conditions}), the set $E$ to denote its transitions 
(called {\em events}), and
$L$ to denote its labels.

\begin{definition}\label{causalnet-def}{\bf (Causal net)}
A causal net is a finite marked net $C(\mathsf{m}_0) = (B,L, 
E,  \mathsf{m}_0)$ satisfying
the following conditions:
\begin{enumerate}
\item $C$ is acyclic;
\item $\forall b \in B \; \; | \pre{b} | \leq 1\, \wedge \, | \post{b} | \leq 1$ (i.e., the places are not branched);
\item  $ \forall b \in B \; \; \mathsf{m}_0(b)   =  \begin{cases}
 1 & \mbox{if $\; \pre{b} = \emptyset$}\\ 
  0  & \mbox{otherwise;}   
   \end{cases}$\\
\item $\forall e \in E \; \; \pre{e}(b) \leq 1 \, \wedge \, \post{e}(b) \leq 1$ for all $b \in B$ (i.e., all the arcs have weight $1$).
\end{enumerate}
We denote by $Min(C)$ the set $\mathsf{m}_0$, and by $Max(C)$ the set
$\{b \in B \mid \post{b} = \emptyset\}$.
\fine
\end{definition}

Note that any reachable marking of a causal net is a set, i.e., 
this net is {\em safe}; in fact, the initial marking is a set and, 
assuming by induction that a reachable marking $\mathsf{m}$ is a set and enables $e$, i.e., 
$\mathsf{m}[e\rangle \mathsf{m}'$,
then also
$\mathsf{m}' =  (\mathsf{m} \ominus \pre{e}) \oplus \post{e}$ is a set, 
as the net is acyclic and because
of the condition on the shape of the post-set of $e$ (weights can only be $1$).

 As the initial marking of a causal net is fixed by its shape (according to item $3$ of 
Definition \ref{causalnet-def}), in the following, in order to make the 
 notation lighter, we often omit the indication of the initial marking, 
 so that the causal 
 net $C(\mathsf{m}_0)$ is denoted by $C$.

\begin{definition}\label{trans-causal}{\bf (Moves of a causal net)}
Given two causal nets $C = (B, L, E,  \mathsf{m}_0)$
and $C' = (B', L, E',  \mathsf{m}_0)$, we say that $C$
moves in one step to $C'$ through $e$, denoted by
$C [e\rangle C'$, if $\; \pre{e} \subseteq Max(C)$, $E' = E \cup \{e\}$
and $B' = B \cup \post{e}$. 
\fine
\end{definition}

\begin{definition}\label{folding-def}{\bf (Folding and Process)}
A {\em folding} from a causal net $C = (B, L, E, \mathsf{m}_0)$ into a net system
$N(m_0) = (S, A, T, m_0)$ is a function $\rho: B \cup E \to S \cup T$, which is type-preserving, i.e., 
such that $\rho(B) \subseteq S$ and $\rho(E) \subseteq T$, satisfying the following:
\begin{itemize}
\item $L = A$ and $\mathsf{l}(e) = l(\rho(e))$ for all $e \in E$;
\item $\rho(\mathsf{m}_0) = m_0$, i.e., $m_0(s) = | \rho^{-1}(s) \cap \mathsf{m}_0 |$;
\item $\forall e \in E, \rho(\pre{e}) = \pre{\rho(e)}$, i.e., $\rho(\pre{e})(s) = | \rho^{-1}(s) \cap \pre{e} |$
for all $s \in S$;
\item $\forall e \in E, \, \rho(\post{e}) = \post{\rho(e)}$,  i.e., $\rho(\post{e})(s) = | \rho^{-1}(s) \cap \post{e} |$
for all $s \in S$.
\end{itemize}
A pair $(C, \rho)$, where $C$ is a causal net and $\rho$ a folding from  
$C$ to a net system $N(m_0)$, is a {\em process} of $N(m_0)$. 
\fine
\end{definition}

\begin{definition}\label{trans-process}{\bf (Moves of a process)}
Let $N(m_0) = (S, A, T, m_0)$ be a net system 
and let $(C_i, \rho_i)$, for $i = 1, 2$, be two processes of $N(m_0)$.
We say that $(C_1, \rho_1)$
moves in one step to $(C_2, \rho_2)$ through $e$, denoted by
$(C_1, \rho_1) \deriv{e} (C_2, \rho_2)$, if $C_1 [e\rangle C_2$
and $\rho_1 \subseteq \rho_2$.
\noindent
If $\pi_1 = (C_1, \rho_1)$ and $\pi_2 = (C_2, \rho_2)$, we denote
the move as $\pi_1 \deriv{e} \pi_2$.
We can extend the definition of move to transition sequences as follows:
\begin{itemize}
\item $\pi \Deriv{\epsilon} \pi$, where $\epsilon$ is the empty transition sequence, is a move sequence 
and
\item if $\pi \deriv{e} \pi'$ and 
$\pi' \Deriv{\sigma} \pi''$, then
$\pi \Deriv{e \sigma} \pi''$ is a move sequence. \\[-1.1cm]
\end{itemize}
\fine
\end{definition}

\begin{definition}\label{po-process-def}{\bf (Partial orders of events from a process)}
From a causal net $C = (B, L, E, \mathsf{m}_0)$,  we can 
extract the {\em partial order of its events}
$\mathsf{E}_{\mathsf{C}} = (E, \preceq)$,
where $e_1 \preceq e_2$ if there is a path in the net from $e_1$ to $e_2$, i.e., if $e_1 \mathsf{F}^* e_2$, where
$\mathsf{F}^*$ is the reflexive and transitive closure of
$\mathsf{F}$, which is the flow relation for $C$.
Given a process $\pi = (C, \rho)$, we denote $\preceq$ as $\leq_\pi$, 
i.e. given $e_1, e_2 \in E$, $e_1 \leq_\pi e_2$ if and only if $e_1 \preceq e_2$.

We can also extract the
{\em abstract} partial order of its {\em observable} events $\mathsf{O}_{C} = (E', \preceq')$,
where $E' = \{e \in E \mid \mathsf{l}(e) \neq \tau\}$ and 
$\preceq' = \preceq \upharpoonright E'$.

Two partial orders $(E_1, \preceq_1)$ and $(E_2, \preceq_2)$ are isomorphic 
if there is a label-preserving, order-preserving  bijection $g: E_1 \to E_2$, i.e., a bijection such that
$\mathsf{l}_1(e) = \mathsf{l}_2(g(e))$ and $e \preceq_1 e'$ if and only if $g(e) \preceq_2 g(e')$.

We also say that $g$ is an {\em abstract} (or {\em concrete}) 
{\em event isomorphism} between 
$C_1$ and 
$C_2$ if it is an isomorphism between their associated abstract (or concrete)
partial orders of events $\mathsf{O}_{C_1}$
and $\mathsf{O}_{C_2}$ (or $\mathsf{E}_{C_1}$
and $\mathsf{E}_{C_2}$).
\fine
\end{definition}

In case of P/T nets without silent transitions, the coarsest behavioral equivalence fully respecting causality and the branching time is
the the largest
{\em fully-concurrent bisimulation} (fc-bisimulation, for short) \cite{BDKP91}, whose definition
was inspired by previous notions of equivalence on other models of concurrency:
{\em history-preserving bisimulation}, originally defined in \cite{RT88} under the name of {\em behavior-structure bisimulation}, and 
then elaborated on in \cite{vGG89} (who called it by this name) and also independently defined in \cite{DDM89} 
(who called it by {\em mixed ordering bisimulation}). If two markings are fully-concurrent bisimilar, then they generate
processes with isomorphic concrete partial orders. Its definition follows.

\begin{definition}\label{sfc-bis-def}{\bf (Fully-concurrent bisimulation)}
Given a P/T net $N = (S, A, T)$, a {\em fully-concurrent bisimulation}
is a relation $R$, composed of 
triples of the form $(\pi_1, g, \pi_2) $, where, for $i = 1,2$, 
$\pi_i = (C_i, \rho_i)$ is a process of $N(m_{0i})$ for some $m_{0i}$ and
$g$ is a concrete event isomorphism between $C_1$ and $C_2$, such that  
if $(\pi_1, g, \pi_2) \in R$ then

\begin{itemize}
\item[$i)$] 
$\forall t_1, \pi_1'$ such that $\pi_1 \deriv{e_1} \pi_1'$ with $\rho_1'(e_1) = t_1$, $\exists t_2, \pi_2', g'$ such that

\begin{enumerate}
\item $\pi_2 \deriv{e_2} \pi_2'$ with $\rho_2'(e_2) = t_2$;
 \item $g' = g \cup \{(e_1, e_2)\}$, and finally,
\item $(\pi_1', g', \pi_2') \in R$;
\end{enumerate}

\item[$ii)$] and symmetrically, if $\pi_2$ moves first.
\end{itemize}

Two markings $m_1, m_2$ are fc-bisimilar,
denoted by $m_1 \sim_{fc} m_2$, if a fully-concurrent bisimulation R exists,
containing a triple $(\pi^0_1, \emptyset, \pi^0_2)$ where 
$\pi^0_i = (C^0_i, \rho^0_i)$ such that 
$C^0_i$ contains no events and 
$\rho^0_i(Min(C^0_i))  = \rho^0_i(Max(C^0_i))$ $ = m_i\;$ for $i = 1, 2$.
\fine
\end{definition}

Fully-concurrent bisimilarity $\sim_{fc}$ is an equivalence relation, that is strictly finer than interleaving bisimilarity $\sim_{int}$
and also undecidable for finite P/T nets.
An extension to P/T nets with silent transitions can be the following branching fully-concurrent bisimilarity \cite{Pin93,Gor20c}.

\begin{definition}\label{brfc-bis-def}{\bf (Branching fc-bisimulation)}
Given a net $N = (S, A, T)$, a {\em branching fully-concurrent bisimulation}
is a relation $R$, composed of 
triples of the form $(\pi_1, g, \pi_2) $, where, for $i = 1,2$, 
$\pi_i = (C_i, \rho_i)$ is a process of $N(m_{0i})$ for some $m_{0i}$, 
and $g$ is an abstract event isomorphism between $C_1$ and $C_2$, such that 
if $(\pi_1, g, \pi_2) \in R$ then

\begin{itemize}
\item[$i)$] 
$\forall t_1, \pi_1'$ such that $\pi_1 \deriv{e_1} \pi_1'$ with $\rho_1'(e_1) = t_1$, 

     \begin{itemize}
     \item {\em either} $l(e_1) = \tau$ and there exist $\sigma_2$ (with $o(\sigma_2) = \epsilon$) and $\pi_2'$ 
     such that $\pi_2 \Deriv{\sigma_2} \pi_2'$, $(\pi_1, g, \pi_2') \in R$ and $(\pi_1', g, \pi_2') \in R$;
   
 \item {\em or}  $\exists \sigma$ (with $o(\sigma) = \epsilon$), $e_2, \pi_2', \pi_2'', g'$ 
     such that 
        
      \begin{enumerate}
           \item $\pi_2 \Deriv{\sigma} \pi_2' \deriv{e_2} \pi_2''$;
       \item if $l(e_1) = \tau$, then $l(e_2) = \tau$ and $g' = g$;
                otherwise, $l(e_1) = l(e_2)$ and  
                 $g' = g \cup \{(e_1, e_2)\}$; 
       \item and finally, $(\pi_1, g, \pi_2') \in R$ and  
              $(\pi_1', g', \pi_2'') \in R$;
    \end{enumerate}
  \end{itemize}

\item[$ii)$] symmetrically, if $\pi_2$ moves first.
\end{itemize}

Two markings $m_{1}$ and $m_2$ of $N$ are bfc-bisimilar, 
denoted by $m_{1} \approx_{bfc} m_{2}$, 
if there exists a branching fully-concurrent bisimulation $R$ with a triple 
$((C^0_1, \rho_1), g_0, (C^0_2, \rho_2))$, where 
$C^0_i$ contains no transitions, $g_0$ is empty and 
$\rho_i(Min( C^0_i)) = \rho_i(Max( C^0_i)) = m_i\;$ for $i = 1, 2$.
\fine
\end{definition}

Branching fully-concurrent bisimilarity $\approx_{bfc}$ is an equivalence relation \cite{Gor20c}, that is strictly finer than
branching interleaving bisimilarity $\approx_{bri}$ and also undecidable for finite P/T nets.
Even if its definition is in {\em branching-style} (cf. Definition \ref{br-int-bis}), it is an open problem to see whether it fully respects the timing of choices,
i.e., whether it enjoys the weak stuttering property.

%
\section{Place Bisimilarity} \label{place-sec}
%

We now present place bisimulation, introduced in \cite{ABS91} as an 
improvement of {\em strong bisimulation}, a behavioral relation proposed by Olderog in \cite{Old} on safe 
nets which fails to induce an equivalence relation.
Our definition is formulated in a slightly different way, but it is coherent with the original one. 
First, an auxiliary definition.
 
\begin{definition}\label{add-eq}{\bf (Additive closure)}
Given a P/T net $N = (S, A, T)$ and a {\em place relation} $R \subseteq S \times S$, we define a {\em marking relation}
$R^\oplus \, \subseteq \, {\mathcal M}(S) \times {\mathcal M}(S)$, called 
the {\em additive closure} of $R$,
as the least relation induced by the following axiom and rule.

$\begin{array}{lllllllllll}
 \bigfrac{}{(\theta, \theta) \in  R^\oplus} & \; \; \; & \; \; \; 
 \bigfrac{(s_1, s_2) \in R \; \; \; (m_1, m_2) \in R^\oplus }{(s_1 \oplus m_1, s_2 \oplus m_2) \in  R^\oplus }  \\
\end{array}$
\\[-.2cm]
\fine
\end{definition}

Note that, by definition, two markings are related by $R^\oplus$ only if they have the same size; 
in fact, the axiom states that
the empty marking is related to itself, while the rule, assuming by induction 
that $m_1$ and $m_2$ have the same size, ensures that $s_1 \oplus m_1$ and
$s_2 \oplus m_2$ have the same size.

\begin{proposition}\label{fin-k-add}
For each relation $R \subseteq S \times S$,  if $(m_1, m_2) \in R^\oplus$, 
then $|m_1| = |m_2|$.
\fine
\end{proposition}

Note also that there may be several proofs of $(m_1, m_2) \in R^\oplus$, 
depending on the chosen order of the elements
of the two markings and on the definition of $R$. For instance, if 
$R = \{(s_1, s_3), (s_1, s_4), (s_2, s_3), (s_2, s_4)\}$,
then $(s_1 \oplus s_2, s_3 \oplus s_4) \in R^\oplus$ can be proved by 
means of the pairs $(s_1, s_3)$ and $(s_2, s_4)$,
as well as by means of $(s_1, s_4), (s_2, s_3)$.
An alternative way to define that two markings $m_1$ and $m_2$
are related by $R^\oplus$ is to state that $m_1$ can be represented 
as $s_1 \oplus s_2 \oplus \ldots \oplus s_k$, 
$m_2$ can be represented as $s_1' \oplus s_2' \oplus \ldots \oplus s_k'$ 
and $(s_i, s_i') \in R$ for $i = 1, \ldots, k$. 
In fact, a naive algorithm for checking 
whether $(m_1, m_2) \in R^\oplus$ would simply consider 
$m_1$ represented as $s_1 \oplus s_2 \oplus \ldots \oplus s_k$ and then scan all the possible permutations of 
$m_2$, each represented as $s'_1 \oplus s'_2 \oplus \ldots \oplus s'_k$, 
to check that $(s_i, s_i') \in R$ for $i = 1, \ldots, k$. Of course, this naive algorithm is in $O(k!)$.

\begin{example}\label{nsubtractive}
Consider $R = \{(s_1, s_3),$  $(s_1, s_4), (s_2, s_4)\}$, which is not an equivalence relation.
Suppose we want to check that $(s_1 \oplus s_2, s_4 \oplus s_3) \in R^\oplus$.
If we start by matching $(s_1, s_4) \in R$, then we fail because the residual $(s_2, s_3)$ is not in $R$.
However, if we permute the second marking to $s_3 \oplus s_4$, then we succeed because the required pairs
$(s_1, s_3)$ and $(s_2, s_4)$ are both in $R$.
\fine
\end{example}

Nonetheless, the problem of checking whether $(m_1, m_2) \in R^\oplus$ has polynomial time complexity
because it can be considered as an instance of
the problem of finding a perfect matching in a bipartite graph, 
where the nodes of the two partitions are the tokens in the 
two markings, and the edges
are defined by the relation $R$. 
In fact, 
the definition of the bipartite graph takes $O(k^2)$ time (where $k = |m_1| = |m_2|$) and, then, 
the Hopcroft-Karp-Karzanov algorithm  \cite{HK73,Kar73} for computing the maximum matching has
worst-case time complexity $O(h\sqrt{k})$, where $h$ is the number of the edges in the bipartire graph ($h \leq k^2$) and
to check whether the maximum matching is perfect can be done simply by checking that the size of the matching equals the number of nodes in each partition, i.e., $k$.
Hence, in evaluating the complexity of the algorithm in Section \ref{decid-br-place-sec}, we assume that the complexity of 
checking whether $(m_1, m_2) \in R^\oplus$ is in $O(k^2 \sqrt{k})$.

A related problem is that of computing, given a marking $m_1$ of size $k$, the set of all the markings $m_2$ such that 
$(m_1, m_2) \in R^\oplus$. This problem can be solved with a worst-case time complexity of $O(n^k)$ because each of the $k$
tokens in $m_1$ can be related via $R$ to $n$ places at most.

\begin{proposition}\label{add-prop1}\cite{Gor17b}
For each place relation $R \subseteq S \times S$, the following hold:
\begin{enumerate}
\item If $R$ is an equivalence relation, then $R^\oplus$ is an equivalence relation.
\item If $R_1 \subseteq R_2$, then $R_1^\oplus \subseteq R_2^\oplus$, i.e., the additive closure is monotone.
\item If $(m_1, m_2) \in R^\oplus$ and $(m_1', m_2') \in R^\oplus$,
then $(m_1 \oplus m_1', m_2 \oplus m_2') \in R^\oplus$, i.e., the additive closure is additive.\\[-1.1cm]
\end{enumerate}
\fine
\end{proposition}

Now we list some useful, and less obvious, properties of additively closed place relations (proof in \cite{Gor17b}).

\begin{proposition}\label{add-prop2}
For each family of place relations $R_i \subseteq S \times S$, the following hold:
\begin{enumerate}
\item $\emptyset^\oplus = \{(\theta, \theta)\}$, i.e., the additive closure of the empty place relation
is a singleton marking relation, relating the empty marking to itself.
\item $(\mathcal{I}_S)^\oplus = \mathcal{I}_M$, i.e., the additive closure of the
identity relation on places $\mathcal{I}_S = \{(s, s) \mid s \in S\}$ is the identity relation on markings
$\mathcal{I}_M = \{(m, m) \mid m \in  {\mathcal M}(S)\}$.
\item $(R^\oplus)^{-1} = (R^{-1})^\oplus$, i.e., the inverse of an additively closed relation $R$ is the additive closure
of its inverse $R^{-1}$.
\item $(R_1 \circ R_2)^\oplus = (R_1^\oplus) \circ (R_2^\oplus)$, i.e., the additive closure of the composition of two 
place relations is the compositions of their additive closures.\\[-1.1cm]
\end{enumerate}
\fine
\end{proposition}

\begin{definition}\label{def-place-bis}{\bf (Place Bisimulation)}
Let $N = (S, A, T)$ be a P/T net. 
A {\em place bisimulation} is a relation
$R\subseteq S \times S$ such that if $(m_1, m_2) \in R^\oplus$
then
\begin{itemize}
\item $\forall t_1$ such that  $m_1[t_1\rangle m'_1$, $\exists t_2$ such that $m_2[t_2\rangle m'_2$ 
with $(\pre{t_1}, \pre{t_2}) \in R^\oplus$, $l(t_1) = l(t_2)$,  $(\post{t_1}, \post{t_2}) \in R^\oplus$ and $(m'_1, m'_2) \in R^\oplus$,
\item $\forall t_2$ such that  $m_2[t_2\rangle m'_2$, $\exists t_1$ such that $m_1[t_1\rangle m'_1$ 
with $(\pre{t_1}, \pre{t_2}) \in R^\oplus$, $l(t_1) = l(t_2)$,  $(\post{t_1}, \post{t_2}) \in R^\oplus$ and $(m'_1, m'_2) \in R^\oplus$.
\end{itemize}
Two markings $m_1$ and $m_2$ are  {\em place bisimilar}, denoted by
$m_1 \sim_p m_2$, if there exists a place bisimulation $R$ such that $(m_1, m_2) \in R^\oplus$.
\fine
\end{definition}

\begin{proposition}\label{place-bis-eq}
For each P/T net $N = (S, A, T)$, relation $\sim_p \; \subseteq  \mathcal{M}(S) \times  \mathcal{M}(S)$ is an equivalence relation.
\proof
It follows directly from the followings facts: $(i)$ the identity place relation $\mathcal{I}_S = \{(s, s) \mid s \in S\}$ is a place bisimulation,
$(ii)$ the inverse $R^{-1}$ of a place bisimulation $R$ is a place bisimulation and $(iii)$ the relational composition
$R_1 \circ R_2$ of two place bisimulations $R_1$ and $R_2$, is a place bisimulation. Details in \cite{Gor21}.
\fine
\end{proposition}

By Definition \ref{def-place-bis}, place bisimilarity can be defined as follows:

$\sim_p = \bigcup \{ R^\oplus \mid R \mbox{ is a place bisimulation}\}.$

\noindent
By monotonicity of the additive closure (Proposition \ref{add-prop1}(2)), if $R_1 \subseteq R_2$, then
$R_1^\oplus \subseteq R_2^\oplus$. Hence, we can restrict our attention to maximal place bisimulations only:

$\sim_p = \bigcup \{ R^\oplus \mid R \mbox{ is a {\em maximal} place bisimulation}\}.$

\noindent
However, it is not true that 

$\sim_p = (\bigcup \{ R \mid R \mbox{ is a {\em maximal} place bisimulation}\})^\oplus$

\noindent 
because the union of place bisimulations may  not be a place bisimulation. 
We illustrate this fact by means of the following tiny example.

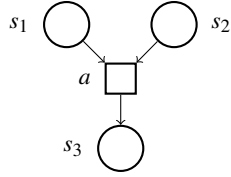
\begin{figure}[t]
\centering
\begin{tikzpicture}[
every place/.style={draw,thick,inner sep=0pt,minimum size=6mm},
every transition/.style={draw,thick,inner sep=0pt,minimum size=4mm},
bend angle=30,
pre/.style={<-,shorten <=1pt,>=stealth,semithick},
post/.style={->,shorten >=1pt,>=stealth,semithick}
]
\def\eofigdist{4cm}
\def\eodist{0.4cm}
\def\eodisty{0.8cm}

\node (p1) [place]  [label=left:$s_1\;$] {};
\node (t1) [transition] [below right=\eodist of p1,label=left:$a\;$] {};
\node (p2) [place] [right=\eodisty of p1,label=right:$\;s_2$] {};
\node (p3) [place] [below=\eodist of t1,label=left:$s_3\;$] {};

\draw  [->] (p1) to (t1);
\draw  [->] (p2) to (t1);
\draw  [->] (t1) to (p3);
\end{tikzpicture}
\caption{A simple net}
\label{net-tau1}
\end{figure}

\begin{example}\label{primo-tau-ex}
Consider the simple P/T net in Figure \ref{net-tau1}, with $S = \{s_1, s_2, s_3\}$. It is rather easy to realize the following two are  maximal place bisimulations:

$R_1 = \mathcal{I}_S = \{(s_1, s_1), (s_2, s_2), (s_3, s_3)\}$ and

$R_2 = (R_1 \setminus  \mathcal{I}_{\{s_1, s_2\}}) \cup \{(s_1, s_2), (s_2, s_1)\} =  \{(s_1, s_2), (s_2, s_1), (s_3, s_3)\}$,

\noindent However, note that 
their union $R = R_1 \cup R_2$ is not a place bisimulation. In fact, on the one hand $(s_1 \oplus s_1, s_1 \oplus s_2) \in R^\oplus$,
but, on the other hand, these two markings do not satisfy the place bisimulation game, because 
$s_1 \oplus s_1$ is stuck, while $s_1 \oplus s_2$ can fire 
the $a$-labeled transition, reaching $s_3$.
\fine
\end{example}

Since the union of place bisimulations may  not be a place bisimulation, its definition is not coinductive, 
so that we cannot adapt the well-known algorithms for computing
the largest bisimulation (which is an equivalence) \cite{PT87,KS83}, as there is not one largest place bisimulation.
Nonetheless, place bisimilarity $\sim_p$ is decidable \cite{Gor21} and also sensible, 
i.e., it fully respects causality and the branching structure,
because it is slightly finer than {\em causal-net bisimilarity} \cite{G15,Gor22} (or, equivalently, {\em structure-preserving} bisimilarity \cite{G15}), in turn slightly finer than fully-concurrent bisimilarity \cite{BDKP91}.

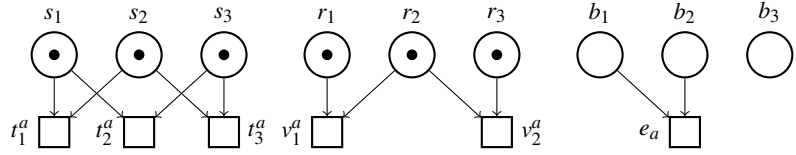
\begin{figure}[t]
\centering

\begin{tikzpicture}[
every place/.style={draw,thick,inner sep=0pt,minimum size=6mm},
every transition/.style={draw,thick,inner sep=0pt,minimum size=4mm},
bend angle=42,
pre/.style={<-,shorten <=1pt,>=stealth,semithick},
post/.style={->,shorten >=1pt,>=stealth,semithick}
]
\def\eofigdist{3cm}
\def\eodist{0.5cm}
\def\eodisty{1.5cm}

\node (p1) [place,tokens=1]  [label=above:$s_1$] {};
\node (p2) [place,tokens=1]  [right=\eodist of p1,label=above:$s_2$] {};
\node (p3) [place,tokens=1] [right=\eodist of p2,label=above:$s_3$] {};
\node (t1) [transition] [below=\eodist of p1,label=left:$t_1^a$] {};
\node (t2) [transition] [below=\eodist of p2,label=left:$t_2^a$] {};
\node (t3) [transition] [below=\eodist of p3,label=right:$t_3^a$] {};

\draw  [->] (p1) to (t1);
\draw  [->] (p2) to (t1);
\draw  [->] (p1) to (t2);
\draw  [->] (p3) to (t2);
\draw  [->] (p2) to (t3);
\draw  [->] (p3) to (t3);

\node (q1) [place,tokens=1]  [right=\eofigdist of p1,label=above:$r_1$] {};
\node (q2) [place,tokens=1]  [right=\eodist of q1,label=above:$r_2$] {};
\node (q3) [place,tokens=1]  [right=\eodist of q2,label=above:$r_3$] {};
\node (s1) [transition] [below=\eodist of q1,label=left:$v_1^a$] {};
\node (s2) [transition] [below=\eodist of q3,label=right:$v_2^a$] {};

\draw  [->] (q1) to (s1);
\draw  [->] (q2) to (s1);
\draw  [->] (q2) to (s2);
\draw  [->] (q3) to (s2);

\node (b1) [place]  [right=\eofigdist of q1,label=above:$b_1$] {};
\node (b2) [place]  [right=\eodist of b1,label=above:$b_2$] {};
\node (b3) [place]  [right=\eodist of b2,label=above:$b_3$] {};
\node (u1) [transition] [below=\eodist of b2,label=left:$e_a$] {};

\draw  [->] (b1) to (u1);
\draw  [->] (b2) to (u1);

\end{tikzpicture}
\caption{Two non-place-bisimilar markings, but with the same causal nets}
\label{cn-vs-icn-fig}
\end{figure}

\begin{remark}\label{rem-no-fix}{\bf (Place bisimilarity is not a fixpoint)}
Even if place bisimilarity $\sim_p$ is not a place bisimulation, it supports the place bisimulation game, i.e.,
if $m_1 \sim_p m_2$ then
\begin{itemize}
\item $\forall t_1$ such that  $m_1[t_1\rangle m'_1$, $\exists t_2$ such that $m_2[t_2\rangle m'_2$ 
with $\pre{t_1} \sim_p \pre{t_2}$, $l(t_1) = l(t_2)$,  $\post{t_1} \sim_p \post{t_2}$ and $m'_1 \sim_p m'_2$,
\item $\forall t_2$ such that  $m_2[t_2\rangle m'_2$, $\exists t_1$ such that $m_1[t_1\rangle m'_1$ 
with $\pre{t_1} \sim_p \pre{t_2}$, $l(t_1) = l(t_2)$,  $\post{t_1} \sim_p \post{t_2}$ and $m'_1 \sim_p m'_2$.
\end{itemize}
However, the reverse implication does not hold, i.e., even if two markings satisfy the place bisimulation game for $\sim_p$, 
they may be not place bisimilar. 
Consider the nets in Figure \ref{cn-vs-icn-fig} (where all the transitions are labeled by $a$) and the markings $m_1 = s_1 \oplus s_2 \oplus s_3$
and $m_2 = r_1 \oplus r_2 \oplus r_3$. For each transition executable by $m_1$ (e.g., $t_2^a$), $m_2$ can reply with a transition (say, 
$v_1^a$) such that the place bisimulation game is satisfied (i.e., $s_1 \oplus s_3 \sim_p r_1 \oplus r_2$ because $R_1 = \{(s_1, r_1), (s_3, r_2)\}$
is a place bisimulation; $\theta \sim_p \theta$ trivially; $s_2 \sim_p r_3$ because $R_2 = \{(s_2, r_3)\}$ is a place bisimulation). And, symmetrically,
for each transition executable by $m_2$, $m_1$ can reply with a suitable transition respecting the place bisimulation game.
However, there is no place bisimulation $R$ such that $(m_1, m_2) \in R^\oplus$. For instance, consider the trivial place relation
$R' = \{(s_1, r_1), (s_2, r_2), (s_3, r_3)\}$; this is not a place bisimulation because if $m_1$ performs $t_2^a$, whose preset is $s_1 \oplus s_3$,
$m_2$ cannot reply with a transition with preset $r_1 \oplus r_3$. And the same problem arises whatever is the place
 relation $\overline{R}$ we consider
such that $(m_1, m_2) \in \overline{R}^\oplus$. Hence, $m_1 \nsim_p m_2$, even if these two markings generate the same causal nets, that are
isomorphic to that on the right of Figure \ref{cn-vs-icn-fig}. Summing up,
we have showed that, contrary to branching interleaving bisimilarity (cf. Theorem \ref{br-int-bis-fix}), place bisimilarity is not a fixpoint.
\fine
\end{remark}

%
\section{Branching Place Bisimilarity} \label{br-place-sec}
%

Now we define a variant of place bisimulation, which is insensitive, to some extent, to $\tau$-sequential transitions,
i.e., $\tau$-labeled transitions whose pre-set and post-set have size one. In order to properly define this relation,
called {\em branching} place bisimulation, we need some auxiliary definitions.

\begin{definition}\label{pt-silent-def}{\bf ($\tau$-sequential)}
Give a P/T net $N = (S, A, T)$ with silent moves,
a transition $t \in T$ is {\em $\tau$-sequential} if $l(t) = \tau$ and $|\post{t}| = 1 =  |\pre{t}|$.
A P/T net $N$ with silent moves is {\em $\tau$-sequential} if 
 $\forall t \in T$ if $l(t) = \tau$, then $t$ is $\tau$-sequential.
\fine
\end{definition}

\begin{definition}\label{tr-seq-silent-def}{\bf (Idling transitions, $\tau$-sequential (acyclic) transition sequence)}
Given a P/T net $N = (S, A, T)$ with silent moves, the set  of  {\em idling transitions} is $I(S) = \{i(s) \mid $ $ s \in S, i(s) = (s, \tau, s)\}$.
In defining {\em silent transition sequences}, we take the liberty of using also the fictitious idling transitions, so that, e.g.,
if $\sigma = i(s_1) i(s_2)$, then $s_1 \oplus s_2 [\sigma \rangle s_1 \oplus s_2$.
For simplicity sake, we sometimes denote by $i(m)$ the sequence $i(s_1) i(s_2) \ldots i(s_n)$, 
where $m = s_1 \oplus s_2 \oplus \ldots \oplus s_n$ (with $i(\theta) = \epsilon$).
Given a transition sequence $\sigma$, its {\em observable label} $o(\sigma)$ is computed inductively as:

 $\begin{array}{lllllllll}
 & o(\epsilon) & = & \epsilon \\ 
& o(t \sigma) & = & \begin{cases}
  l(t) o(\sigma) & \mbox{if $l(t) \neq \tau$}\\ 
  o(\sigma)  & \mbox{otherwise.}   
   \end{cases}
\end{array}$

A transition sequence $\sigma = t_1 t_2 \ldots t_n$ (where $n \geq 1$ and some of the $t_i$ can be idling transitions) 
is {\em $\tau$-1-sequential} if $l(t_i) = \tau$, $|\post{t_i}| = 1 = |\pre{t_i}|$  for $i = 1, \ldots, n$, 
and  $\post{t_i} = \pre{t_{i+1}} $   for $i = 1, \ldots, n-1$, so that 
$o(\sigma) = \epsilon$ and $|\post{\sigma}| = 1 =  |\pre{\sigma}|$. 

A transition sequence 
$\sigma = \sigma_1 \sigma_2 \ldots \sigma_k$ is {\em $\tau$-k-sequential} if $\sigma_i$ is $\tau$-1-sequential for $i = 1, \ldots, k$,   
$\pre{\sigma} = \pre{\sigma_1} \oplus \pre{\sigma_2} \oplus \ldots \oplus \pre{\sigma_k}$ and
$\post{\sigma} = \post{\sigma_1} \oplus \post{\sigma_2} \oplus \ldots \oplus \post{\sigma_k}$, so that 
$o(\sigma) = \epsilon$ and $|\post{\sigma}| = k =  |\pre{\sigma}|$. 
We say that $\sigma$ is $\tau$-sequential if it is $\tau$-k-sequential for some $k \geq 1$.

A $\tau$-1-sequential 
$\sigma = t_1 t_2 \ldots t_n$ is {\em acyclic} if $\pre{\sigma} =$ 
$ m_0 [t_1\rangle m_1 [t_2\rangle m_2$ $ \ldots m_{n-1}[t_n\rangle 
m_{n} = \post{\sigma}$ and $m_i \neq m_j$ for all $i \neq j$, with $i,j \in \{1, 2, \ldots, n\}$
and $m_0 \neq m_i$ for $i = 1, \ldots, n-1$.

A $\tau$-k-sequential  
$\sigma = \sigma_1 \sigma_2 \ldots \sigma_k$ is {\em acyclic} 
if $\sigma_i$ is $\tau$-1-sequential and acyclic for $i = 1, \ldots, k$.
We say that $\sigma$ is an acyclic $\tau$-sequential transition sequence if it is  $\tau$-k-sequential and acyclic
for some $k \geq 1$.
\fine
\end{definition}

\begin{remark}\label{tr-seq-silent-rem}{\bf (Acyclic $\tau$-sequential transition sequence)}
The definition of acyclic $\tau$-1-sequential transition sequence is a bit non-standard as it may allow for a cycle 
when the initial marking $m_0$ and the final one $m_n$ are the same. 
For instance, $\sigma = i(s) i(s)$ is cyclic, while the apparently cyclic subsequence $\sigma' = i(s)$ is actually acyclic, according to our definition.
Note that, given a $\tau$-1-sequential transition sequence $\sigma$, it is always possible to find an acyclic
$\tau$-1-sequential transition sequence $\sigma'$ such that $\pre{\sigma} = \pre{\sigma'}$ and $\post{\sigma} = \post{\sigma'}$.
For instance, if $\pre{\sigma} = m_0 [t_1\rangle m_1 [t_2\rangle m_2 \ldots m_{n-1}[t_n\rangle 
m_{n} = \post{\sigma}$ and the only cycle is given by $m_i [t_{i+1}\rangle m_{i+1} \ldots m_{j-1}[t_j\rangle m_j$ with $m_i = m_j$ and $i \geq 1$, then
$\sigma' = t_1 t_2 \ldots t_i t_{j+1} \ldots t_n$ is acyclic and $\pre{\sigma} = \pre{\sigma'}$ and $\post{\sigma} = \post{\sigma'}$.

Note that the definition of acyclic $\tau$-k-sequential transition sequence does not ensure the absence of cycles even if all the 
$\tau$-1-sequential transition sequences composing it are acyclic. For instance, consider 
$\sigma = \sigma_1 \sigma_2$, where $\sigma_1 = i(s_1)$
and $\sigma_2 = i(s_2)$. According to our definition, $\sigma$ is $\tau$-2-sequential and
acyclic because both $\sigma_1$ and $\sigma_2$ are $\tau$-1-sequential and acyclic 
(according to our definition); however,
the execution of the two idling transitions generates a cycle.

Note also that, given a $\tau$-k-sequential transition sequence $\sigma = \sigma_1 \sigma_2 \ldots \sigma_k$, it is always 
possible to find an acyclic
$\tau$-k-sequential transition sequence $\sigma' = \sigma'_1 \sigma'_2 \ldots \sigma'_k$,
where $\sigma_i'$ is the acyclic $\tau$-1-sequential transition sequence corresponding to $\sigma_i$ for $i = 1, 2, \ldots, k$,
in such a way that $\pre{\sigma} = \pre{\sigma'}$ and $\post{\sigma} = \post{\sigma'}$.

Finally, we remark that, given two markings $m_1$ and $m_2$ of equal size $k$, it is decidable whether there exists an acyclic $\tau$-k-sequential transition
$\sigma$ such that $\pre{\sigma} = m_1$ and $\post{\sigma} = m_2$, essentially because this is similar to the reachability problem
(limited by using only $\tau$-sequential transitions), which is decidable \cite{Mayr84}.
\fine
\end{remark}

Now we want to introduce a definition of branching place bisimilarity that 
satisfies the {\em weak stuttering property} (cf. Remark \ref{stutt2-rem-int}), as this ensures that the timing of choices is fully respected.
For sure, the original definition of branching place bisimulation in the preliminary version of this paper \cite{Gor-forte21}
enjoys the weak stuttering property if the {\em strong stuttering property} holds (cf. Remark \ref{stutt1-rem-int}), following 
an argument similar to that in Remark \ref{stutt2-rem-int}.
Unfortunately, the observation in Remark \ref{rem-no-fix} explains that, whatever is the actual definition of branching 
place bisimilarity $\approx_p$, since it has to coincide with place 
bisimilarity $\sim_p$ on nets without silent transitions, $\approx_p$ cannot be a fixpoint. Therefore, 
we cannot prove the strong stuttering property for $\approx_p$ with the same proof technique used in
Remark \ref{stutt1-rem-int} for $\approx_{bri}$. Actually, we were not able to prove (nor to disprove)
the strong stuttering property for the original proposal in \cite{Gor-forte21}, so that, in order to achieve our goal (i.e., defining a suitable variant of
branching place bisimilarity enjoying the weak stuttering property), 
here we strengthen 
slightly the definition in \cite{Gor-forte21}, by adding an extra condition expressed by the following predicates $\Psi$ and $\Phi$.
Given a $\tau$-sequential transition sequence $\sigma = t_1, \, t_2, \ldots t_n$  (i.e., sequences composed of $\tau$-sequential
transitions in $T \cup I(S)$) such that 

$\pre{\sigma} = m_0 [t_1 \rangle m_1 [t_2\rangle \ldots m_{n-1} [t_n\rangle m_n = \post{\sigma}$, 

\noindent we say that predicate $\Psi(m, \sigma, R^\oplus)$
holds if $(m, m_i) \in R^\oplus$ for $i = 0, 1, \ldots, n-1$ and  that predicate $\Phi(\sigma, m, R^\oplus)$
holds if $(m_i, m) \in R^\oplus$ for $i = 0, 1, \ldots, n-1$. Note that 
$\Psi(m, \sigma, R^\oplus)$ holds if and only if $\Phi(\sigma, m, (R^\oplus)^{-1})$, hence, by Proposition \ref{add-prop2}(3),
iff $\Phi(\sigma, m, (R^{-1})^\oplus)$ holds.


\begin{definition}\label{bpb-bis-def}{\bf (Branching place bisimulation)}
Given a P/T net $N = (S, A, T)$, a {\em branching place bisimulation} is a relation
$R\subseteq S \times S$ such that if $(m_1, m_2) \in R^\oplus$
\begin{enumerate}
\item $\forall t_1$ such that $m_1[t_1\rangle m_1'$
    \begin{itemize}
    \item[$(i)$] either $t_1$ is $\tau$-sequential and
                      $\exists \sigma, m_2'$ such that $\sigma$ is $\tau$-sequential, $m_2 [\sigma\rangle m_2'$,
                      $\Psi(\pre{t_1}, \sigma, R^\oplus)$, $(\pre{t_1}, \post{\sigma}) \in R^\oplus$,
                      $(\post{t_1}, \post{\sigma}) \in R^\oplus$ and 
                      $(m_1 \ominus \pre{t_1}, m_2 \ominus \pre{\sigma}) \in R^\oplus$;
     \item[$(ii)$] or $\exists \sigma, t_2, m, m_2'$ such that 
                     $\sigma$ is $\tau$-sequential, $m_2 [\sigma\rangle m [t_2\rangle m_2'$,
                     $\post{\sigma} = \pre{t_2}$, 
                     $l(t_1) = l(t_2)$, 
                     $\Psi(\pre{t_1}, \sigma, R^\oplus)$, $(\pre{t_1}, \post{\sigma}) \in R^\oplus$,
                     $(\post{t_1}, \post{t_2}) \in R^\oplus$, and 
                     $(m_1 \ominus \pre{t_1}, m_2 \ominus \pre{\sigma}) \in R^\oplus$; 
   \end{itemize}
\item and, symmetrically, $\forall t_2$ such that $m_2[t_2\rangle m_2'$
   \begin{itemize}
    \item[$(i)$] either $t_2$ is $\tau$-sequential and
                      $\exists \sigma, m_1'$ such that $\sigma$ is $\tau$-sequential, $m_1 [\sigma\rangle m_1'$,
                      $\Phi(\sigma, \pre{t_2}, R^\oplus)$, $(\post{\sigma}, \pre{t_2}) \in R^\oplus$,
                      $(\post{\sigma}, \post{t_2}) \in R^\oplus$ and 
                      $(m_1 \ominus  \pre{\sigma}, m_2 \ominus  \pre{t_2}) \in R^\oplus$;
   \item[$(ii)$] or $\exists \sigma, t_1, m, m_1'$ such that 
                     $\sigma$ is $\tau$-sequential, $m_1 [\sigma\rangle m [t_1\rangle m_1'$,
                     $\post{\sigma} = \pre{t_1}$,         
                     $l(t_1) = l(t_2)$,
                     $\Phi(\sigma, \pre{t_2}, R^\oplus)$, $(\post{\sigma}, \pre{t_2}) \in R^\oplus$,
                     $(\post{t_1}, \post{t_2}) \in R^\oplus$, and
                     $(m_1 \ominus \pre{\sigma}, m_2 \ominus \pre{t_2}) \in R^\oplus$.
\end{itemize}
\end{enumerate}

Two markings $m_1$ and $m_2$ are branching place bisimulation equivalent, 
denoted by $m_1 \approx_{p} m_2$,
if there exists a branching place bisimulation $R$ such that $(m_1, m_2) \in R^\oplus$.
\fine
\end{definition}

We can derive some expected relations: in the either case of item 1, by additivity of $R^\oplus$ 
(cf. Proposition \ref{add-prop1}(3)), from 
$(m_1 \ominus \pre{t_1}, m_2 \ominus \pre{\sigma}) \in R^\oplus$ and 
$(\pre{t_1}, \post{\sigma})  \in R^\oplus$, we get $(m_1, m_2') \in R^\oplus$, 
as well as, from
$(\post{t_1}, \post{\sigma}) \in R$ we get  $(m_1', m_2') \in R^\oplus$. 
Similarly, for the or case of item 1,
from $(m_1 \ominus \pre{t_1}, m_2 \ominus \pre{\sigma}) \in R^\oplus$, $\post{\sigma} = \pre{t_2}$ and 
$(\pre{t_1}, \pre{t_2}) \in R^\oplus$, we get $(m_1, m) \in R^\oplus$, as well as, from $(\post{t_1}, \post{t_2}) \in R^\oplus$,
we get  $(m_1', m_2') \in R^\oplus$. Symmetrically for item 2.

Note also that a $\tau$-sequential  transition performed by one of the two markings may be matched by the other one also by idling: 
this is due to the {\em either} case when $\sigma = i(s_2)$ for a suitable token $s_2$ such that the required properties are satisfied
(i.e., such that $(\pre{t_1}, \pre{\sigma})  \in R^\oplus$, 
$(\pre{t_1}, \post{\sigma})  \in R$, $(\post{t_1}, \post{\sigma}) \in R^\oplus$ and 
$(m_1 \ominus \pre{t_1}, m_2 \ominus \pre{\sigma}) \in R^\oplus$, where $\pre{\sigma} = \post{\sigma} = s_2$).

\begin{proposition}\label{prop-bpb-bis1}
For each P/T net $N = (S, A, T)$, the following hold:
\begin{itemize}
\item[$(i)$] The identity relation ${\mathcal I}_S = \{(s, s) \mid s \in S\}$ is a branching place bisimulation.
\item[$(ii)$] The inverse relation $R^{-1}$ of a branching place bisimulation $R$ is a  branching place bisimulation.
\end{itemize}

\proof
Case $(i)$ is obvious: If $(m_1, m_2) \in {\mathcal I}_S^\oplus$, then $m_1 = m_2$, so that the branching
place bisimulation game can be mimicked trivially: given $(m, m) \in {\mathcal I}_S^\oplus$, 
for all $t$ such that $m[t\rangle m'$, the other instance of $m$ in the pair replies with $m[t\rangle m'$
(case 1($ii$), with $\sigma = i(\pre{t})$) and all the required conditions are trivially satisfied.

For case $(ii)$, assume $(m_2, m_1) \in  (R^{-1})^\oplus$ and $m_2 [t_2\rangle m_2'$.
By Proposition \ref{add-prop2}(3), we have that
$(m_2, m_1) \in (R^\oplus)^{-1}$ and so $(m_1, m_2) \in R^\oplus$.
Since $R$ is a branching place bisimulation, we have that

   \begin{itemize}
    \item[$(i)$] either $t_2$ is $\tau$-sequential and
                      there exist $\sigma, m_1'$ such that $\sigma$ is $\tau$-sequential, $m_1 [\sigma\rangle m_1'$,
                      $\Phi(\sigma, \pre{t_2}, R^\oplus)$, $(\post{\sigma}, \pre{t_2}) \in R^\oplus$
                      $(\post{\sigma}, \post{t_2}) \in R^\oplus$ and 
                      $(m_1 \ominus  \pre{\sigma}, m_2 \ominus  \pre{t_2}) \in R^\oplus$;
   \item[$(ii)$] or $\exists \sigma, t_1, m, m_1'$ such that 
                     $\sigma$ is $\tau$-sequential, $m_1 [\sigma\rangle m [t_1\rangle m_1'$,
                     $\post{\sigma} = \pre{t_1}$,         
                     $l(t_1) = l(t_2)$,
                     $\Phi(\sigma, \pre{t_2}, R^\oplus)$, $(\post{\sigma}, \pre{t_2}) \in R^\oplus$
                     $(\post{t_1}, \post{t_2}) \in R^\oplus$, and
                     $(m_1 \ominus \pre{\sigma}, m_2 \ominus \pre{t_2}) \in R^\oplus$.
\end{itemize}
Summing up, if $(m_2, m_1) \in  (R^{-1})^\oplus$ and $m_2 [t_2\rangle m_2'$ (the case when $m_1$ 
moves first is symmetric, and so omitted), then 
\begin{itemize}
\item[$(i)$] either $t_2$ is $\tau$-sequential and
                      there exist $\sigma, m_1'$ such that $\sigma$ is $\tau$-sequential, 
                      $m_1[\sigma\rangle m_1'$,  
                      $\Psi(\pre{t_2}, \sigma, (R^{-1})^\oplus)$, $(\pre{t_2}, \post{\sigma}) \in (R^{-1})^\oplus$,
                      $(\post{t_2}, \post{\sigma}) \in (R^{-1})^\oplus$ and, moreover,
                      $(m_2 \ominus  \pre{t_2}, m_1 \ominus  \pre{\sigma}) \in (R^{-1})^\oplus$;
\item[$(ii)$] or there exist $\sigma, t_1, m, m_1'$ such that 
                     $\sigma$ is $\tau$-sequential, $m_1[\sigma\rangle m [t_1\rangle m_1'$, 
                     $\post{\sigma} = \pre{t_1}$, $l(t_1) = l(t_2)$, 
                     $\Psi(\pre{t_2}, \sigma, (R^{-1})^\oplus)$, $(\pre{t_2}, \post{\sigma}) \in (R^{-1})^\oplus$,
                     $(\post{t_2}, \post{t_1}) \in (R^{-1})^\oplus$ and, moreover, 
                     $(m_2 \ominus \pre{t_2}, m_1 \ominus \pre{\sigma}) \in (R^{-1})^\oplus$
\end{itemize}
so that $R^{-1}$ is a branching place bisimulation, indeed. 
\fine
\end{proposition}

More challenging is to prove that the relational composition of two branching place bisimulations is a 
branching place bisimulation. We need an auxiliary notation and a technical lemma.
Given a $\tau$-sequential transition sequence $\overline{\sigma}_1 = t_1, \, t_2, \ldots t_n$ such that 

$\pre{\overline{\sigma}_1} = m_0 [t_1 \rangle m_1 [t_2\rangle \ldots m_{n-1} [t_n\rangle m_n = \post{\overline{\sigma}_1}$, 

\noindent
and a $\tau$-sequential  transition sequence 
         $\overline{\sigma}_2 = \sigma_1 \sigma_2 \ldots \sigma_n$,
such that 

$\pre{\overline{\sigma}_2} = \overline{m}_0 [\sigma_1 \rangle \overline{m}_1 [\sigma_2\rangle \ldots \overline{m}_{n-1} 
[\sigma_n\rangle \overline{m}_n = \post{\overline{\sigma}_2}, \quad$ with  $\pre{\sigma}_i = \overline{m}_{i-1}$ for $i = 1, \ldots, n$,

\noindent
we say that predicate $\overline{\Psi}(\overline{\sigma}_1, \overline{\sigma}_2, R^\oplus)$ holds
 iff $\Psi(m_{i-1}, \sigma_{i}, R^\oplus)$ holds for $i = 1, \ldots, n$; similarly, we say that  
 $\overline{\Phi}(\overline{\sigma}_2, \overline{\sigma}_1, R^\oplus)$ holds
 iff $\Phi(\sigma_{i}, m_{i-1}, R^\oplus)$ holds for $i = 1, \ldots, n$.

\begin{lemma}\label{tau-lemma}
Let $N = (S, A, T)$ be a P/T net, and $R$ be a place bisimulation.
\begin{enumerate}
\item For each $\tau$-sequential transition sequence $\overline{\sigma}_1 = t_1, \, t_2, \ldots t_n$ such that 

$\pre{\overline{\sigma}_1} = m_0 [t_1 \rangle m_1 [t_2\rangle \ldots m_{n-1} [t_n\rangle m_n = \post{\overline{\sigma}_1}$, 

         for all $m$ such that $(\pre{\overline{\sigma}_1}, m) \in R^\oplus$, a $\tau$-sequential  transition sequence 
         $\overline{\sigma}_2 = \sigma_1 \sigma_2 \ldots \sigma_n$ exists
         such that $m = \pre{\overline{\sigma}_2}$, $\overline{\Psi}(\overline{\sigma}_1, \overline{\sigma}_2, R^\oplus)$  and
                                   $(\post{\overline{\sigma}_1}, \post{\overline{\sigma}_2})  \in R^\oplus$;
\item and symmetrically, for each $\tau$-sequential transition sequence $\overline{\sigma}_2 = t_1 t_2 \ldots t_n$, such that 
$\pre{\overline{\sigma}_2} = m_0 [t_1 \rangle m_1 [t_2\rangle \ldots m_{n-1} [t_n\rangle m_n = \post{\overline{\sigma}_2}$,

         for all $m$ such that $(m, \pre{\overline{\sigma}_2}) \in R^\oplus$,  a $\tau$-sequential  transition sequence 
        $\overline{\sigma}_1 = \sigma_1 \sigma_2 \ldots \sigma_n$ exists
         such that $m = \pre{\overline{\sigma}_1}$,  $\overline{\Phi}(\overline{\sigma}_1, \overline{\sigma}_2, R^\oplus)$ and
                                   $(\post{\overline{\sigma}_1}, \post{\overline{\sigma}_2})  \in R^\oplus$.
\end{enumerate}
\proof By symmetry, we prove only case $1$, by induction on the length of $\overline{\sigma}_1$.

{\em Base case}: $\overline{\sigma}_1 = \epsilon$. In this trivial case, $\pre{\overline{\sigma}_1} = \theta$ and so the only possible
$m$ is $\theta$ as well. We just take $\overline{\sigma}_2 = \epsilon$ and all the required conditions are trivially satisfied; in particular,
$\overline{\Psi}(\epsilon, \epsilon, R^\oplus)$ vacuously holds (as it requires $\Psi(\theta, \epsilon, R^\oplus)$ for $n = 0$).

{\em Inductive case}: $\overline{\sigma}_1 = \delta_1 t_1$, where $t_1 \in T \cup I(S)$. 
Hence, by inductive hypothesis, for each $m$ such that $(\pre{\delta_1}, m) \in R^\oplus$, we know that there
exists a $\tau$-sequential transition sequence 
$\delta_2$ such that $m = \pre{\delta_2}$, $\overline{\Psi}(\delta_1, \delta_2, R^\oplus)$ holds and
                                   $(\post{\delta_1}, \post{\delta_2})  \in R^\oplus$.  

                             
 \noindent
 If $t_1 = i(s_1)$, then we have to consider two subcases:
\begin{itemize}
\item if $s_1 \in \post{\delta_1}$, then $\pre{\delta_1 t_1} = \pre{\delta_1}$ and $\post{\delta_1 t_1} = \post{\delta_1}$. 
Hence, we can take $\overline{\sigma}_2 = \delta_2 i(\post{\delta_2})$ and all the required conditions are trivially satisfied;
in fact, transition $\post{\delta_1}[t_1\rangle \post{\delta_1 t_1} = \post{\delta_1}$ is matched by 
$\post{\delta_2}[i(\post{\delta_2})\rangle \post{\delta_2}$,
so that
the predicate $\overline{\Psi}(\delta_1t_1, \delta_2 i(\post{\delta_2}), R^\oplus)$ holds, and also 
$(\post{\delta_1 t_1}, \post{(\delta_2 i(\post{\delta_2}))})  \in R^\oplus$, as required.

\item if $s_1 \not\in \post{\delta_1}$, then $\pre{\delta_1 t_1} = \pre{\delta_1} \oplus s_1$ and $\post{\delta_1 t_1} = \post{\delta_1} \oplus s_1$.
Then, $\forall s_2$ such that $(s_1, s_2) \in R$, we can take $\overline{\sigma}_2 = \delta_2 i(\post{\delta_2}) i(s_2)$ with 
$\pre{\overline{\sigma}_2} = \pre{\delta_2} \oplus s_2$ and $\post{\overline{\sigma}_2} = \post{\delta_2} \oplus s_2$;
in fact, transition $\post{\delta_1}\oplus s_1[t_1\rangle \post{\delta_1 t_1} = \post{\delta_1} \oplus s_1$ is matched by 
$\post{\delta_2} \oplus s_2[i(\post{\delta_2}) i(s_2))\rangle \post{\delta_2} \oplus s_2$,
so that
the predicate $\overline{\Psi}(\delta_1t_1, \delta_2 i(\post{\delta_2}) i(s_2), R^\oplus)$ holds, and also 
$(\post{\delta_1 t_1}, \post{(\delta_2 i(\post{\delta_2}) i(s_2))})  \in R^\oplus$, as required.
\end{itemize}

\noindent
Also if $t_1 \in T$, we have consider two subcases:
\begin{itemize}
\item If $s_1 = \pre{t_1} \in \post{\delta_1}$, then, since $(\post{\delta_1}, \post{\delta_2})  \in R^\oplus$,
there exists $s_2 \in  \post{\delta_2}$ such that $(s_1, s_2) \in R$ and 
$(\post{\delta_1} \ominus s_1, \post{\delta_2} \ominus s_2) \in R^\oplus$. 
Then, by 
Definition \ref{bpb-bis-def}, it follows that to the move $t_1 = s_1 \deriv{\tau} s_1'$:
    \begin{itemize} 
    \item[$(i)$] Either $\exists \sigma, s_2'$ such that $\sigma$ is $\tau$-sequential, 
                      $s_2[\sigma\rangle s_2'$,  $\Psi(s_1, \sigma, R^\oplus)$, 
                                   $(s_1, s_2')  \in R^\oplus$ and $(s_1', s_2') \in R^\oplus$.
 
                                    In this case, we take $\overline{\sigma}_2 = \delta_2 i(\post{\delta_2}) \sigma$, so that 
                                   $\overline{\Psi}(\delta_1 t_1, \delta_2 i(\post{\delta_2}) \sigma, R^\oplus)$ holds, 
                                   (by additivity, because $(\post{\delta_1} \ominus s_1, \post{\delta_2} \ominus s_2) \in R^\oplus$
                                   and $\Psi(s_1, \sigma, R^\oplus)$) and, moreover,
                                   $(\post{\delta_1 t_1}, \post{(\delta_2 i(\post{\delta_2}) \sigma)})  \in R^\oplus$
                                   (because $\post{\delta_1 t_1} = (\post{\delta_1} \ominus s_1)\oplus s_1'$
                                   and $\post{(\delta_2 i(\post{\delta_2}) \sigma)} = (\post{\delta_2}\ominus s_2)\oplus s_2'$), as required.                                                                       
         \item[$(ii)$] Or there exist $\sigma, t_2, \overline{s}, s_2'$ such that 
                     $\sigma t_2$ is $\tau$-sequential, $\post{\sigma} = \pre{t_2}$,
                     $s_2[\sigma\rangle \overline{s} [t_2\rangle s_2'$, $\Psi(s_1, \sigma, R^\oplus)$,
                     $(s_1, \overline{s}) \in R^\oplus$ and
                     $(s_1', s_2') \in R^\oplus$. 
      
                     In this case, we take $\overline{\sigma}_2 = \delta_2 i(\post{\delta_2}) \sigma t_2$, so that 
                    $\overline{\Psi}(\delta_1 t_1, \delta_2 i(\post{\delta_2}) \sigma t_2, R^\oplus)$ holds (by additivity, because 
                    $(\post{\delta_1} \ominus s_1, \post{\delta_2} \ominus s_2) \in R^\oplus$, $\Psi(s_1, \sigma, R^\oplus)$ and
                    $(s_1, \overline{s}) \in R^\oplus$) 
                    and, moreover,
                    $(\post{\delta_1 t_1}, \post{(\delta_2 i(\post{\delta_2}) \sigma t_2)})  \in R^\oplus$ (because $(s_1', s_2') \in R^\oplus$), as required.
     
        \end{itemize}

\item If $s_1 = \pre{t_1} \not\in \post{\delta_1}$, then, for each $s_2$ such that $(s_1, s_2) \in R$,
we consider the marking $\post{\delta_2} \oplus s_2$. Following the same step as above (by Definition \ref{bpb-bis-def})
we have that to the move $t_1 = s_1 \deriv{\tau} s_1'$:
    \begin{itemize} 
    \item[$(i)$] Either $\exists \sigma, s_2'$ such that $\sigma$ is $\tau$-sequential, 
                      $s_2[\sigma\rangle s_2'$,  $\Psi(s_1, \sigma, R^\oplus)$, 
                                   $(s_1, s_2')  \in R^\oplus$ and $(s_1', s_2') \in R^\oplus$.
                                   
                                   In this case, we take $\overline{\sigma}_2 = \delta_2 i(\post{\delta_2}) \sigma$, so that 
                                    $\overline{\Psi}(\delta_1 t_1, \delta_2 i(\post{\delta_2}) \sigma, R^\oplus)$ holds, 
                                   (by additivity, because $(\post{\delta_1}, \post{\delta_2}) \in R^\oplus$, $(s_1, s_2) \in R^\oplus$
                                   and $\Psi(s_1, \sigma, R^\oplus)$), and
                                   $(\post{\delta_1 t_1}, \post{(\delta_2 i(\post{\delta_2})\sigma)})  \in R^\oplus$
                                   (by additivity, because $\post{\delta_1 t_1} = \post{\delta_1} \oplus s_1'$, 
                                   $\post{(\delta_2 i(\post{\delta_2}) \sigma)} = \post{\delta_2}\oplus s_2'$ and $(s_1', s_2') \in R^\oplus$), as required.  
                                  
  \item[$(ii)$] Or there exist $\sigma, t_2, \overline{s}, s_2'$ such that 
                     $\sigma t_2$ is $\tau$-sequential, $\post{\sigma} = \pre{t_2}$,
                     $s_2[\sigma\rangle \overline{s} [t_2\rangle s_2'$, $\Psi(s_1, \sigma, R^\oplus)$
                     $(s_1, \overline{s}) \in R^\oplus$ and
                     $(s_1', s_2') \in R^\oplus$. 

                    In this case, we take $\overline{\sigma}_2 = \delta_2 i(\post{\delta_2}) \sigma t_2$, so that 
                    $\overline{\Psi}(\delta_1 t_1, \delta_2 i(\post{\delta_2}) \sigma t_2, R^\oplus)$ holds and, moreover,
                    $(\post{\delta_1 t_1}, \post{(\delta_2i(\post{\delta_2}) \sigma t_2)})  \in R^\oplus$, as required.
        \end{itemize}
\end{itemize}

\noindent
       And so the proof is complete.
\fine
\end{lemma}

\begin{proposition}\label{prop-bpb-bis2}
For each P/T net $N = (S, A, T)$, the relational composition $R_1 \circ R_2$ of
two branching place bisimulations $R_1$ and $R_2$ is a branching place bisimulation.

\proof
Assume $(m_1, m_3) \in (R_1 \circ R_2)^\oplus$ and $m_1 [t_1\rangle m'_1$.
By Proposition \ref{add-prop2}(4), we have that 
$(m_1, m_3) \in (R_1)^\oplus \circ (R_2)^\oplus$, and so  $m_2$ exists such that 
$(m_1, m_2) \in R_1^\oplus$ and $(m_2, m_3) \in R_2^\oplus$.

As $(m_1, m_2) \in R_1^\oplus$ and $R_1$ is a branching place bisimulation, 
if $m_1 [t_1\rangle m_1'$, then 
   \begin{itemize}
    \item[$(i)$] either $t_1$ is $\tau$-sequential and
                      $\exists \sigma, m_2'$ such that $\sigma$ is $\tau$-sequential, $m_2 [\sigma\rangle m_2'$,
                      $\Psi(\pre{t_1}, \sigma, R_1^\oplus)$, $(\pre{t_1}, \post{\sigma}) \in R_1^\oplus$,
                      $(\post{t_1}, \post{\sigma}) \in R_1^\oplus$ and 
                      $(m_1 \ominus \pre{t_1}, m_2 \ominus \pre{\sigma}) \in R_1^\oplus$;
     \item[$(ii)$] or $\exists \sigma, t_2, m, m_2'$ such that 
                     $\sigma$ is $\tau$-sequential, $m_2 [\sigma\rangle m [t_2\rangle m_2'$,
                     $\post{\sigma} = \pre{t_2}$, 
                     $l(t_1) = l(t_2)$, 
                     $\Psi(\pre{t_1}, \sigma, R_1^\oplus)$, $(\pre{t_1}, \post{\sigma}) \in R_1^\oplus$,
                     $(\post{t_1}, \post{t_2}) \in R_1^\oplus$, and 
                     $(m_1 \ominus \pre{t_1}, m_2 \ominus \pre{\sigma}) \in R_1^\oplus$; 
   \end{itemize}

\begin{itemize}
\item 
Consider case $(i)$, i.e., assume that to the move $m_1 [t_1\rangle m_1'$, $m_2$ replies with
 $m_2 [\sigma \rangle m_2'$ such that $\sigma$ is $\tau$-sequential, $m_2 [\sigma\rangle m_2'$,
                      $\Psi(\pre{t_1}, \sigma, R_1^\oplus)$, $(\pre{t_1}, \post{\sigma}) \in R_1^\oplus$,
                      $(\post{t_1}, \post{\sigma}) \in R_1^\oplus$ and 
                      $(m_1 \ominus \pre{t_1}, m_2 \ominus \pre{\sigma}) \in R_1^\oplus$.
 Since $(m_2, m_3) \in R_2^\oplus$, there exists a submarking $\overline{m} \subseteq m_3$
  such that $(\pre{\sigma}, \overline{m}) \in R_2^\oplus$ and
  $(m_2 \ominus \pre{\sigma}, m_3 \ominus \overline{m}) \in R_2^\oplus$.
   By Lemma \ref{tau-lemma},  a $\tau$-sequential transition sequence $\sigma'$ exists
         such that $\overline{m} = \pre{\sigma'}$,  $\overline{\Psi}(\sigma, \sigma', R_2^\oplus)$ and
                                   $(\post{\sigma}, \post{\sigma'})  \in R_2^\oplus$. Hence, 
    $m_3 [\sigma'\rangle m_3'$, where $m_3' = (m_3 \ominus \pre{\sigma'}) \oplus \post{\sigma'}$.
                      
Summing up, considering that $R_1^\oplus \circ R_2^\oplus = (R_1 \circ R_2)^\oplus$ by Proposition \ref{add-prop2}(4),
 to the move $m_1 [t_1\rangle m_1'$, $m_3$ can reply with $m_3 [\sigma'\rangle m_3'$, in such a way that the predicate
$\Psi(\pre{t_1}, \sigma', (R_1 \circ R_2)^\oplus)$ holds (because both $\Psi(\pre{t_1}, \sigma, R_1^\oplus)$ and 
$\overline{\Psi}(\sigma, \sigma', R_2^\oplus)$ hold),
 $(\pre{t_1}, \post{\sigma'})  \in (R_1 \circ R_2)^\oplus$, $(\post{t_1}, \post{\sigma'}) \in (R_1 \circ R_2)^\oplus$ and 
 $(m_1 \ominus \pre{t_1}, m_3 \ominus \pre{\sigma'}) \in (R_1 \circ R_2)^\oplus$, as required.
   
\item 
Consider case $(ii)$, i.e., assume that to the move $m_1 [t_1\rangle m_1'$, $m_2$ replies with the move
$m_2[\sigma\rangle m [t_2\rangle m_2'$, where $\sigma$ is $\tau$-sequential, 
 $l(t_1) = l(t_2)$, $\post{\sigma} = \pre{t_2}$,  $\Psi(\pre{t_1}, \sigma, R_1^\oplus)$, $(\pre{t_1}, \post{\sigma}) \in R_1^\oplus$,
                     $(\post{t_1}, \post{t_2}) \in R_1^\oplus$, and 
                     $(m_1 \ominus \pre{t_1}, m_2 \ominus \pre{\sigma}) \in R_1^\oplus$.
  
    Since $(m_2, m_3) \in R_2^\oplus$, there exists a submarking $\overline{m} \subseteq m_3$
  such that $(\pre{\sigma}, \overline{m}) \in R_2^\oplus$ and
  $(m_2 \ominus \pre{\sigma}, m_3 \ominus \overline{m}) \in R_2^\oplus$.
  By Lemma \ref{tau-lemma}, there exists a $\tau$-sequential transition sequence $\sigma'$
   such that $\overline{m} = \pre{\sigma'}$, $\overline{\Psi}(\sigma, \sigma', R_2^\oplus)$ and
                                   $(\post{\sigma}, \post{\sigma'})  \in R_2^\oplus$.            
    Hence, 
    $m_3 [\sigma'\rangle m'$, where $m' = (m_3 \ominus \pre{\sigma'}) \oplus \post{\sigma'}$ and,
    moreover, $(m, m') \in R_2^\oplus$.
 
Since $(m, m') \in R_2^\oplus$, $\post{\sigma} = \pre{t_2}$ and  $(\post{\sigma}, \post{\sigma'})  \in R_2^\oplus$, there exists 
$\underline{m} = \post{\sigma'} \subseteq m'$ such that 
$(\pre{t_2}, \underline{m}) \in R_2^\oplus$ and $(m \ominus \pre{t_2}, m' \ominus \underline{m}) \in R_2^\oplus$. 
Hence, by Definition \ref{bpb-bis-def},
to the move $\pre{t_2} [t_2\rangle \post{t_2}$, 
$\underline{m}$ can reply as follows:
    \begin{itemize}
    \item[$(a)$] Either $t_2$ is $\tau$-sequential and
                      $\exists \overline{\sigma}$ 
                      such that $ \overline{\sigma}$ is $\tau$-sequential, $\underline{m}= \pre{\overline{\sigma}}$,
                      $\underline{m}[ \overline{\sigma}\rangle \post{\overline{\sigma}}$,  
                      and $\Psi(\pre{t_2}, \overline{\sigma}, R_2^\oplus)$, 
                      $(\pre{t_2}, \post{ \overline{\sigma}})  \in R_2^\oplus$, 
                      $(\post{t_2}, \post{ \overline{\sigma}}) \in R_2^\oplus$ and 
                      $(m \ominus \pre{t_2}, m' \ominus \pre{ \overline{\sigma}}) \in R_2^\oplus$.
       
                     In this case,  to the move $m_1 [t_1\rangle m_1'$, $m_3$ can reply with 
                     $m_3 [\sigma' \rangle m'[ \overline{\sigma}\rangle m_3'$, with 
                     $m_3' = (m' \ominus \pre{ \overline{\sigma}})  \oplus \post{ \overline{\sigma}}$,  
                      such that $\Psi(\pre{t_1}, \sigma' \overline{\sigma}, (R_1 \circ R_2)^\oplus)$ holds
                      (because the validity of $\Psi(\pre{t_1}, \sigma, R_1^\oplus)$ and
                      $\overline{\Psi}(\sigma, \sigma', R_2^\oplus)$ imply that   $\Psi(\pre{t_1}, \sigma', (R_1 \circ R_2)^\oplus)$ holds,
                      and moreover, since
                      $(\pre{t_1}, \pre{t_2}) \in R_1^\oplus$ and $\Psi(\pre{t_2}, \overline{\sigma}, R_2^\oplus)$,
                      we get that predicate $\Psi(\pre{t_1}, \overline{\sigma}, (R_1 \circ R_2)^\oplus)$ holds),
                       $(\pre{t_1}, \post{\sigma' \overline{\sigma}}) \in (R_1 \circ R_2)^\oplus$
                      (as $(\pre{t_1}, \pre{t_2}) \in R_1^\oplus$, $\post{\sigma'} = \pre{\overline{\sigma}}$  
                      and $(\pre{t_2}, \post{ \overline{\sigma}})  \in R_2^\oplus$),
                     $(\post{t_1}, \post{\sigma' \overline{\sigma}'}) \in (R_1 \circ R_2)^\oplus$
                     (as $(\post{t_1}, \post{t_2}) \in R_1^\oplus$ and $(\post{t_2}, \post{ \overline{\sigma}}) \in R_2^\oplus$),
                     and, moreover, 
                    $(m_1 \ominus \pre{t_1}, m_3 \ominus \pre{\sigma' \overline{\sigma}}) \in (R_1 \circ R_2)^\oplus$.

    \item[$(b)$] or $\exists  \overline{\sigma}, t_3, \overline{m}$ such that 
                     $ \overline{\sigma}$ is $\tau$-sequential, $\underline{m} = \pre{ \overline{\sigma}}$,
                     $\underline{m}[ \overline{\sigma}\rangle \overline{m} [t_3\rangle \post{t_3}$, 
                     $l(t_2) = l(t_3)$, $\overline{m} = \post{ \overline{\sigma}} = \pre{t_3}$,
                     $\Psi(\pre{t_2}, \overline{\sigma}, R_2^\oplus)$ holds,
                     $(\pre{t_2}, \pre{t_3}) \in R_2^\oplus$,
                     $(\post{t_2}, \post{t_3}) \in R_2^\oplus$ and, moreover,  
                     $(m \ominus \pre{t_2}, m' \ominus \pre{ \overline{\sigma}}) \in R_2^\oplus$.

                 In this case, to the move $m_2 [\sigma \rangle m [t_2\rangle m_2'$, $m_3$ replies with
                 $m_3 [\sigma' \rangle m'[ \overline{\sigma}\rangle m'' [t_3 \rangle m_3'$, with 
                 $m_3' = (m' \ominus \pre{ \overline{\sigma}}) \oplus \post{t_3}$, such that               
                 $\overline{\sigma}$ is $\tau$-sequential,  $\pre{\overline{\sigma}} = \post{\sigma'}$,
                 and therefore $\overline{\Psi}(\sigma t_2, \sigma' \overline{\sigma}t_3, R_2^\oplus)$
                 (because $\overline{\Psi}(\sigma, \sigma', R_2^\oplus)$ and $\Psi(\pre{t_2}, \overline{\sigma}, R_2^\oplus)$)
                 and
               $(\post{\sigma t_2},$ $ \post{\sigma' \overline{\sigma} t_3}) \in R_2^\oplus$ (because
               $\post{\sigma t_2} = \post{t_2}$, $ \post{\sigma' \overline{\sigma} t_3} =  \post{t_3}$
               and $( \post{t_2},  \post{t_3}) \in R_2^\oplus$).
               
                                Summing up, to the move $m_1 [t_1\rangle m_1'$, $m_3$ can reply with 
                 $m_3 [\sigma' \rangle m'[ \overline{\sigma}\rangle m'' [t_3 \rangle m_3'$,
                 such that $\Psi(\pre{t_1}, \sigma' \overline{\sigma}, (R_1 \circ R_2)^\oplus)$
                 (because $\Psi(\pre{t_1}, \sigma, R_1^\oplus)$ and $\overline{\Psi}(\sigma, \sigma', R_2^\oplus)$
                 imply $\Psi(\pre{t_1}, \sigma', (R_1 \circ R_2)^\oplus)$; moreover, $(\pre{t_1}, \pre{t_2}) \in R_1^\oplus$
                 and $\Psi(\pre{t_2}, \overline{\sigma}, R_2^\oplus)$ imply $\Psi(\pre{t_1}, \overline{\sigma}, (R_1 \circ R_2)^\oplus)$),
                $(\pre{t_1}, \pre{t_3}) \in (R_1 \circ R_2)^\oplus$ (because $(\pre{t_1}, \pre{t_2}) \in R_1^\oplus$, 
                 and $(\pre{t_2}, \pre{t_3}) \in R_2^\oplus$), 
                 $(\post{t_1}, \post{t_3}) \in (R_1 \circ R_2)^\oplus$ (because $(\post{t_1}, \post{t_2}) \in R_1^\oplus$, 
                 and $(\post{t_2}, \post{t_3}) \in R_2^\oplus$), and 
                 $(m_1 \ominus \pre{t_1}, m_3 \ominus \pre{\sigma' \overline{\sigma}}) \in (R_1 \circ R_2)^\oplus$
                 (because $(m_1 \ominus \pre{t_1}, m_2 \ominus \pre{\sigma}) \in R_1^\oplus$ and
                 $(m_2 \ominus \pre{\sigma}, m_3 \ominus  \pre{\sigma'}) \in R_2^\oplus$).

   \end{itemize}
 \end{itemize}
  The case when $m_2$ moves first is symmetric, and so omitted. Hence, $R_1 \circ R_2$ is a branching place bisimulation, indeed.
\fine
\end{proposition}

\begin{theorem}\label{ssb-bis-eq}
For each P/T net $N = (S, A, T)$, relation $\approx_{p} \; \subseteq  \mathcal{M}(S) \times  \mathcal{M}(S)$ is an equivalence relation.
\proof
As the identity relation ${\mathcal I}_S$ is a branching place bisimulation by Proposition \ref{prop-bpb-bis1}(i), 
we have that ${\mathcal I}_S^\oplus \subseteq \; \approx_p$, and so $\approx_p$ is reflexive.
Symmetry derives from the following argument.
For any $(m, m') \in \; \approx_p$, there exists a branching place bisimulation $R$ such that $(m, m') \in R^\oplus$; 
by Proposition \ref{prop-bpb-bis1}(ii), relation $R^{-1}$ is a branching place bisimulation, and  by Proposition \ref{add-prop2}(3)
we have that $(m', m) \in (R^{-1})^\oplus$; hence,
$(m', m) \in \; \approx_p$.
Transitivity also holds for $\approx_p$. Let $(m, m') \in \; \approx_p$ and $(m', m'') \in \; \approx_p$; hence, there 
exist two branching place bisimulations $R_1$ and $R_2$ such that $(m, m') \in R_1^\oplus$ and $(m', m'') \in R_2^\oplus$. By 
Proposition \ref{prop-bpb-bis2},  $R_1 \circ R_2$ is a branching place bisimulation such that the 
pair $(m, m'') \in (R_1 \circ R_2)^\oplus$  
by Proposition \ref{add-prop2}(4); hence, $(m, m'') \in \; \approx_p$.
\fine
\end{theorem}

\begin{remark}\label{rem-weak-stutt-place-bis}{\bf (Place bisimilarity enjoys the weak stuttering property)}
If $m_1 \approx_p m_2$, then  a branching place bisimulation $R$ exists such that $(m_1, m_2) \in R^\oplus$.
If $(m_1, m_2) \in R^\oplus$, then by Definition \ref{bpb-bis-def} we have that
if $m_1 [t_1\rangle m_1'$, then 
   \begin{itemize}
    \item[$(i)$] either $t_1$ is $\tau$-sequential and
                      $\exists \sigma, m_2'$ such that $\sigma$ is $\tau$-sequential, $m_2 [\sigma\rangle m_2'$,
                      $\Psi(\pre{t_1}, \sigma, R^\oplus)$, $(\pre{t_1}, \post{\sigma}) \in R^\oplus$,
                      $(\post{t_1}, \post{\sigma}) \in R^\oplus$ and 
                      $(m_1 \ominus \pre{t_1}, m_2 \ominus \pre{\sigma}) \in R^\oplus$;
     \item[$(ii)$] or $\exists \sigma, t_2, m, m_2'$ such that 
                     $\sigma$ is $\tau$-sequential, $m_2 [\sigma\rangle m [t_2\rangle m_2'$,
                     $\post{\sigma} = \pre{t_2}$, 
                     $l(t_1) = l(t_2)$, 
                     $\Psi(\pre{t_1}, \sigma, R^\oplus)$, $(\pre{t_1}, \post{\sigma}) \in R^\oplus$,
                     $(\post{t_1}, \post{t_2}) \in R^\oplus$, and 
                     $(m_1 \ominus \pre{t_1}, m_2 \ominus \pre{\sigma}) \in R^\oplus$; 
   \end{itemize}
Consider the either-case: we have that for all the markings in the silent path from $m_2$ to $m_2'$, say $m_2 = \overline{m}_0, \overline{m}_1, \ldots, \overline{m}_n = m_2'$, we have that $(m_1, \overline{m}_i) \in R^\oplus$ for $i = 0, \ldots, n$, by additivity as  
$(m_1 \ominus \pre{t_1}, m_2 \ominus \pre{\sigma}) \in R^\oplus$, $\Psi(\pre{t_1}, \sigma, R^\oplus)$ and 
$(\pre{t_1}, \post{\sigma}) \in R^\oplus$.
By Proposition \ref{prop-bpb-bis1}(ii), we have that also $R^{-1}$ is a branching place bisimulation, so that
$(\overline{m}_i, m_1) \in (R^{-1})^\oplus$ for $i = 0, \ldots, n$.
By Proposition \ref{prop-bpb-bis2}, we have that $R^{-1} \circ R$ is a branching place bisimulation, so that
$(\overline{m}_i, \overline{m}_j) \in (R^{-1} \circ R)^\oplus$ for $i, j = 0, \ldots, n$.
Hence, we have proved that all the markings in the silent path from $m_2$ to $m_2'$ are branching place bisimilar, i.e., 
$\overline{m}_i \approx_p \overline{m}_j$ for $i, j = 0, \ldots, n$.
In a similar manner, we can prove, for the or-case, that all the markings in the silent path from $m_2$ to $m$ are branching place bisimilar.
Also we can similarly prove the analogous property in the symmetric case when $m_2$ moves first.
Therefore, we can conclude that the weak stuttering property holds for branching place bisimilarity, and so $\approx_p$ 
fully respects the timing of choices.
\fine
\end{remark}

\begin{proposition}{\bf (Branching place bisimilarity is finer than branching interleaving bisimilarity)}
For each P/T net $N = (S, A, T)$,  $m_1 \approx_{p} m_2$ implies $m_1 \approx_{bri} m_2$.
\proof  If $m_1 \approx_{p} m_2$, then  $(m_1, m_2) \in R^\oplus$ for some branching place bisimulation $R$.
Note that $R^\oplus$ is a branching interleaving bisimilarity, so that $m_1 \approx_{bri} m_2$.
\fine
\end{proposition}

Branching place bisimilarity $\approx_p$ is also finer than branching fully-concurrent bisimilarity $\approx_{bfc}$. The proof of this fact is postponed to Section \ref{br-d-place-sec}.

\begin{figure}[t]
\centering
\begin{tikzpicture}[
every place/.style={draw,thick,inner sep=0pt,minimum size=6mm},
every transition/.style={draw,thick,inner sep=0pt,minimum size=4mm},
bend angle=45,
pre/.style={<-,shorten <=1pt,>=stealth,semithick},
post/.style={->,shorten >=1pt,>=stealth,semithick}
]
\def\eofigdist{1.2cm}
\def\eodist{0.35}
\def\eodisty{0.65}

\node (a) [label=left:$a)\quad $]{};

\node (q1) [place] [label={above:$s_1$} ] {};


\node (b) [right={1.5cm} of a,label=left:$b)\;\;$] {};

\node (p1) [place]  [right=\eofigdist of q1,label=above:$s_2$] {};
\node (s1) [transition] [below=\eodist of p1,label=right:$\;\tau$] {};
\node (p2) [place] [below=\eodist of s1,label=below:$s_3$]{};
\draw  [->] (p1) to (s1);
\draw  [->] (s1) to (p2);


\node (c) [right={1.6cm} of b,label=left:$c)\;\;$] {};

\node (v1) [place]  [right=\eofigdist of p1,label=above:$s_4$] {};
\node (t1) [transition] [below=\eodist of v1,label=right:$\;\tau$] {};

\draw  [->, bend left] (v1) to (t1);
\draw  [->, bend left] (t1) to (v1);


\node (d) [right={1.6cm} of c,label=left:$d)\;\;$] {};

\node (p3) [place]  [right=\eofigdist of v1,label=above:$s_5$] {};
\node (s2) [transition] [below=\eodist of p3,label=right:$\;\tau$] {};
\draw  [->] (p3) to (s2);


\node (e) [right={1.7cm} of d,label=left:$e)\;\;$] {};

\node (v2) [place]  [right=\eofigdist of p3,label=above:$s_6$] {};
\node (t2) [transition] [below=\eodist of v2,label=right:$\;\tau$] {};
\node (v3) [place]  [below left=\eodist of t2,label=left:$s_7\;$] {};
\node (v4) [place]  [below right=\eodist of t2,label=right:$\;s_8$] {};

\draw  [->] (v2) to (t2);
\draw  [->] (t2) to (v3);
\draw  [->] (t2) to (v4);

\end{tikzpicture}
\caption{Some simple nets with silent moves}
\label{tau-fig2}
\end{figure}
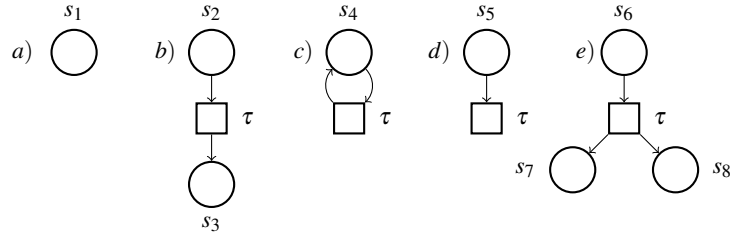

\begin{example}
Consider the nets in Figure \ref{tau-fig2}. Of course, $s_1 \approx_p s_2$, as well as $s_1 \approx_p s_4$.
However, $s_2 \not \approx_p s_5$, because $s_2$ cannot respond to the non-$\tau$-sequential move
$s_5 \deriv{\tau} \theta$. For the same reason, $s_2 \not \approx_p s_6$. Note that silent transitions that are not
$\tau$-sequential are not considered as unobservable.
\fine
\end{example}

\noindent
By Definition \ref{bpb-bis-def}, branching place bisimilarity can be defined as follows:

$\approx_p = \bigcup \{ R^\oplus \mid R \mbox{ is a branching place bisimulation}\}.$

\noindent
By monotonicity of the additive closure (Proposition \ref{add-prop1}(2)), if $R_1 \subseteq R_2$, then
$R_1^\oplus \subseteq R_2^\oplus$. Hence, we can restrict our attention to maximal branching place bisimulations only:

$\approx_p = \bigcup \{ R^\oplus \mid R \mbox{ is a {\em maximal} branching place bisimulation}\}.$

\noindent
However, it is not true that 

$\approx_p = (\bigcup \{ R \mid R \mbox{ is a {\em maximal} branching place bisimulation}\})^\oplus$,
because the union of branching place bisimulations may be not a branching place bisimulation.

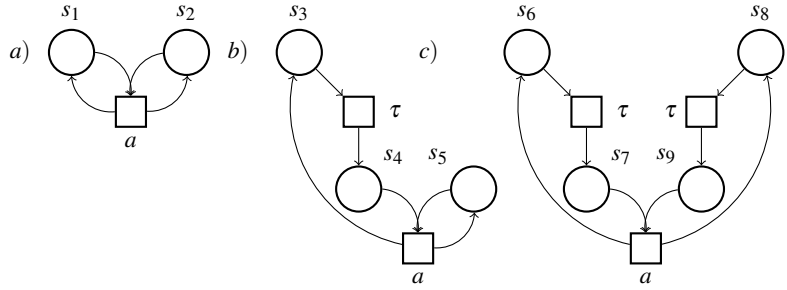
\begin{figure}[t]
\centering
\begin{tikzpicture}[
every place/.style={draw,thick,inner sep=0pt,minimum size=6mm},
every transition/.style={draw,thick,inner sep=0pt,minimum size=4mm},
bend angle=45,
pre/.style={<-,shorten <=1pt,>=stealth,semithick},
post/.style={->,shorten >=1pt,>=stealth,semithick}
]
\def\eofigdist{2.4cm}
\def\eodist{0.5}
\def\eodisty{0.9}

\node (a) [label=left:$a)\quad $]{};

\node (q1) [place] [label={above:$s_1$} ] {};
\node (q2) [place] [right=\eodisty of q1,label={above:$s_2$}] {};
\node (f1)[transition][below right=\eodist of q1,label=below:$a$]{};

\draw  [->, bend left] (q1) to (f1);
\draw  [->, bend left] (f1) to (q1);
\draw  [->, bend right] (q2) to (f1);
\draw  [->, bend right] (f1) to (q2);


\node (b) [right={2.5cm} of a,label=left:$b)\;\;$] {};

\node (p1) [place]  [right=\eofigdist of q1,label=above:$s_3$] {};
\node (s1) [transition] [below right=\eodist of p1,label=right:$\;\tau$] {};
\node (p2) [place] [below =\eodist of s1,label=above right:$s_4$]{};
\node (p3) [place] [right=\eodisty of p2,label=above left:$s_5$]{};
\node (s2) [transition] [below right=\eodist of p2,label=below:$a$] {};

\draw  [->] (p1) to (s1);
\draw  [->] (s1) to (p2);
\draw  [->, bend left] (p2) to (s2);
\draw  [->, bend left] (s2) to (p1);
\draw  [->, bend right] (p3) to (s2);
\draw  [->, bend right] (s2) to (p3);


\node (c) [right={2.3cm} of b,label=left:$c)\;\;$] {};

\node (v1) [place]  [right=\eofigdist of p1,label=above:$s_6$] {};
\node (t1) [transition] [below right=\eodist of v1,label=right:$\;\tau$] {};
\node (v2) [place] [below =\eodist of t1,label=above right:$s_7$]{};
\node (v3) [place] [right=\eodisty of v2,label=above left:$s_9$]{};
\node (t2) [transition] [below right=\eodist of v2,label=below:$a$] {};
\node (t3) [transition] [above=\eodist of v3,label=left:$\tau$] {};
\node (v4) [place] [above right=\eodist of t3,label=above:$s_8$]{};

\draw  [->] (v1) to (t1);
\draw  [->] (t1) to (v2);
\draw  [->, bend left] (v2) to (t2);
\draw  [->, bend left] (t2) to (v1);
\draw  [->, bend right] (v3) to (t2);
\draw  [->, bend right] (t2) to (v4);
\draw  [->] (v4) to (t3);
\draw  [->] (t3) to (v3);

\end{tikzpicture}
\caption{Some branching place bisimilar nets}
\label{tau-fig3}
\end{figure}

\begin{example}
Consider the nets in Figure \ref{tau-fig3}. It is easy to realize that $s_1 \oplus s_2 \approx_p s_3 \oplus s_5$,
because $R_1 = \{(s_1, s_3), (s_2, s_5), (s_1, s_4)\}$ is a branching place bisimulation.
In fact, to the move $t_1 = s_1 \oplus s_2 \deriv{a} s_1 \oplus s_2$, $s_3 \oplus s_5$ replies
with $s_3 \oplus s_5[\sigma \rangle s_4 \oplus s_5 [t_2\rangle s_3 \oplus s_5$, 
where $\sigma = t \, i(s_5)$ (with $t = (s_3, \tau, s_4)$ and $i(s_5) = (s_5, \tau, s_5)$) and 
$t_2 = (s_4 \oplus s_5, a, s_3 \oplus s_5)$, such that $(\pre{t_1}, s_4\oplus s_5) \in R_1^\oplus$,
$(\pre{t_1}, \pre{t_2}) \in R_1^\oplus$ and $(\post{t_1}, \post{t_2}) \in R_1^\oplus$.
Then, to the move $s_3 \oplus s_5[t\rangle s_4 \oplus s_5$, $s_1 \oplus s_2$ can reply by idling with
$s_1 \oplus s_2 [\sigma'\rangle s_1 \oplus s_2$, where $\sigma' = i(s_1)$, and
$(\pre{\sigma'}, \pre{t}) \in R_1^\oplus$, $(\post{\sigma'}, \pre{t}) \in R_1^\oplus$ and $(\post{\sigma'}, \post{t}) \in R_1^\oplus$.
 
Note that also the identity relation $\mathcal{I}_S$, where $S = \{s_1, s_2, s_3, s_4, s_5\}$ is a branching place bisimulation.
However, $R = R_1 \cup \mathcal{I}_S$ is not a branching place bisimulation, because, for instance, 
$(s_1 \oplus s_2, s_3 \oplus s_2) \in R^\oplus$, but these two markings are clearly not equivalent, as $s_1 \oplus s_2$ can do $a$,
while $s_3 \oplus s_2$ cannot.

Similarly, one can prove that $s_1 \oplus s_2 \approx_p s_6 \oplus s_8$ because 
$R_2 = \{(s_1, s_6), (s_2, s_8),$ $(s_1, s_7), (s_2, s_9)\}$ is a branching place bisimulation.
\fine
\end{example}

%
\section{Branching Place Bisimilarity is Decidable} \label{decid-br-place-sec}
%
 
In order to prove that $\approx_p$ is decidable, we first need a technical lemma which states that it is decidable to check 
if a place relation $R \subseteq S \times S$ is a branching place bisimulation.

\begin{lemma}\label{bpb-rel-dec-lem}
Given a P/T net $N = (S, A, T)$ and a place relation $R \subseteq S \times S$, it is decidable if $R$ 
is a branching place bisimulation.
\proof
We want to prove that $R$ is a branching place bisimulation if and only if the following two conditions are satisfied:
\begin{enumerate}
\item $\forall t_1 \in T$, $\forall m$ such that $(\pre{t_1}, m) \in R^\oplus$
	\begin{itemize}
	\item[$(a)$] either $t_1$ is $\tau$-sequential and there exists an acyclic $\tau$-sequential
	$\sigma$ such that $m = \pre{\sigma}$, $\Psi(\pre{t_1}, \sigma, R^\oplus)$,
	 $(\pre{t_1}, \post{\sigma}) \in R^\oplus$ and $(\post{t_1}, \post{\sigma}) \in R^\oplus$;
	\item[$(b)$] or there exist an acyclic $\tau$-sequential $\sigma$  and $t_2 \in T$, with 
	$\post{\sigma} = \pre{t_2}$, such that $m = \pre{\sigma}$, $l(t_1) = l(t_2)$, $\Psi(\pre{t_1}, \sigma, R^\oplus)$,
	$(\pre{t_1}, \pre{t_2}) \in R^\oplus$
	and $(\post{t_1}, \post{t_2}) \in R^\oplus$.
	\end{itemize}
\item $\forall t_2 \in T$, $\forall m$ such that $(m, \pre{t_2}) \in R^\oplus$
	\begin{itemize}
	\item[$(a)$] either $t_2$ is $\tau$-sequential and there exists an acyclic $\tau$-sequential $ \sigma$  
	such that $m = \pre{\sigma}$, $\Phi(\sigma, \pre{t_2}, R^\oplus)$,
	 $(\post{\sigma}, \pre{t_2}) \in R^\oplus$ and $(\post{\sigma}, \post{t_2}) \in R^\oplus$;
	\item[$(b)$] or there exist an acyclic $\tau$-sequential $\sigma$  and $t_1 \in T$, with 
	$\post{\sigma} = \pre{t_1}$, such that $m = \pre{\sigma}$, $l(t_1) = l(t_2)$, $\Phi(\sigma, \pre{t_2}, R^\oplus)$,
	$( \pre{t_1}, \pre{t_2}) \in R^\oplus$
	and $(\post{t_1}, \post{t_2}) \in R^\oplus$.
	\end{itemize}
\end{enumerate}

The implication from left to right is obvious: if $R$ is a branching place bisimulation, then for sure conditions 1 and 2 are satisfied,
because, as observed in Remark \ref{tr-seq-silent-rem}, if there exists a suitable $\tau$-sequential transition 
sequence $\sigma$, then there exists also a suitable {\em acyclic} $\tau$-sequential
$\sigma'$ such that $\pre{\sigma} = \pre{\sigma'}$ and $\post{\sigma} = \post{\sigma'}$.
For the converse implication, assume that conditions 1 and 2 are satisfied; then we have to prove that the branching
place bisimulation game for $R$ holds for all pairs $(m_1, m_2) \in R^\oplus$.

Let $ q = \{(s_1, s_1'), (s_2, s_2'), \ldots,$ $(s_k, s_k')\}$ be any multiset of associations
that can be used to prove that $(m_1, m_2) \in R^\oplus$. So this means that 
$m_1 = s_1 \oplus s_2 \oplus \ldots \oplus s_k$, $m_2 = s_1' \oplus s_2' \oplus \ldots \oplus s_k'$
and that $(s_i, s_i') \in R$ for $i = 1, \ldots, k$. 
If $m_1 [t_1 \rangle m_1'$, then $m_1' = m_1 \ominus \pre{t_1} \oplus \post{t_1}$.
Consider the multiset of associations $p = \{(\overline{s}_{1}, \overline{s}'_{1}),$ $\ldots, (\overline{s}_{h}, \overline{s}'_{h})\} \subseteq q$,
with $\overline{s}_{1} \oplus \ldots \oplus \overline{s}_{h}$ $= \pre{t_1}$.
Note that $(\pre{t_1}, \overline{s}'_{1} \oplus  \ldots \oplus \overline{s}'_{h}) \in R^\oplus$. 
Therefore, by condition 1, (by denoting by $m$ the multiset $\overline{s}'_{1} \oplus  \ldots \oplus \overline{s}'_{h}$)
	\begin{itemize}
	\item[$(a)$] either $t_1$ is $\tau$-sequential and there exists an acyclic $\tau$-sequential
	$\sigma$ such that $m = \pre{\sigma}$, $\Psi(\pre{t_1}, \sigma, R^\oplus)$,
	 $(\pre{t_1}, \post{\sigma}) \in R^\oplus$ and $(\post{t_1}, \post{\sigma}) \in R^\oplus$;
	\item[$(b)$] or there exist an acyclic $\tau$-sequential $\sigma$  and $t_2 \in T$, with 
	$\post{\sigma} = \pre{t_2}$, such that $m = \pre{\sigma}$, $l(t_1) = l(t_2)$, $\Psi(\pre{t_1}, \sigma, R^\oplus)$,
	$(\pre{t_1}, \pre{t_2}) \in R^\oplus$
	and $(\post{t_1}, \post{t_2}) \in R^\oplus$.
	\end{itemize}
In case $(a)$, since $\pre{\sigma} \subseteq m_2$, also $m_2 [\sigma \rangle m_2'$ is firable, where 
$m_2' = (m_2 \ominus \pre{\sigma}) \oplus \post{\sigma}$, so that $\Psi(\pre{t_1}, \sigma, R^\oplus)$,
$(\pre{t_1}, \post{\sigma}) \in R^\oplus$, $(\post{t_1}, \post{\sigma}) \in R^\oplus$
and, finally, $(m_1 \ominus \pre{t_1}, m_2 \ominus \pre{\sigma}) \in R^\oplus$, as required. 
Note that the last condition holds because, from the multiset $q$
of matching pairs for $m_1$ and $m_2$,
we have removed those in $p$.
In case $(b)$,  since $\pre{\sigma} \subseteq m_2$, also $m_2 [\sigma \rangle m [t_2\rangle m_2'$ is firable, 
where $m_2' = (m_2 \ominus \pre{\sigma}) \oplus \post{t_2}$, so that $l(t_1) = l(t_2)$, $\Psi(\pre{t_1}, \sigma, R^\oplus)$,
$(\pre{t_1}, \pre{t_2}) \in R^\oplus$, $(\post{t_1}, \post{t_2}) \in R^\oplus$
and, finally, $(m_1 \ominus \pre{t_1}, m_2 \ominus \pre{\sigma}) \in R^\oplus$, as required. 

If $m_2 [t_2 \rangle m_2'$, then we have to use an argument symmetric to the above, where condition 2 is used instead.
Hence, we have proved that conditions 1 and 2 are enough to prove that $R$ is a branching place bisimulation.

Finally, observe that the set $T$ is finite and, for each $t_1 \in T$,
the number of markings $m$ such that $(\pre{t_1}, m) \in R^\oplus$ and $(m, \pre{t_1}) \in R^\oplus$ is finite as well.
More precisely, this part of the procedure has worst-case time complexity 
$O(q \cdot n^{p})$,
where $q = |T|$, $n = |S|$ and  $p$ is the least number such that $| \pre{t}| \leq p$ for all $t \in T$, as the number 
of markings $m$ related via $R$ to $\pre{t_1}$ is  $n^{p}$ at most.

Moreover, for each pair $(t_1, m)$ satisfying the condition $(\pre{t_1}, m) \in R^\oplus$, 
we have to check conditions $(a)$ and $(b)$, each one checkable in a finite amount of time.
In fact, for case $(a)$, we have to check if there exists a place $s$ such that $(\pre{t_1}, s) \in R$ and $(\post{t_1}, s) \in R$,
which is reachable from $m$ by means of an acyclic 
$\tau$-1-sequential transition sequence $\sigma$; this condition is decidable because we 
have at most $n$ places to examine and for each 
candidate place $s$, we can check whether a suitable acyclic $\tau$-1-sequential $\sigma$ exists (i.e., satisfying
also the predicate 
$\Psi(\pre{t_1}, \sigma, R^\oplus)$).
Similarly, in case (b) we have to consider all the transitions $t_2$ such that 
$(\pre{t_1}, \pre{t_2}) \in R^\oplus$ and $(\post{t_1}, \post{t_2}) \in R^\oplus$ (and this can be checked with
worst-time complexity $O(q   \cdot  (p^2\sqrt{p}))$,
where $q = |T|$, $n = |S|$ and  $p$ is the least number such that $| \pre{t}| \leq p$ 
and $|\post{t}| \leq p$ for all $t \in T$)
and check whether at least one of these is reachable from 
$m$ by means of an acyclic $\tau$-sequential transition sequence $\sigma$ such that $\pre{\sigma} = m$, 
$\Psi(\pre{t_1}, \sigma, R^\oplus)$ and $\post{\sigma} = \pre{t_2}$
and, as observed in Remark \ref{tr-seq-silent-rem},
the existence of such a $\sigma$ is decidable.
Therefore, in a finite amount of time we can decide if a 
given place relation $R$ is actually a branching place bisimulation.
\fine
\end{lemma}

\begin{theorem}\label{bpl-bis-decid-th}{\bf (Branching place bisimilarity is decidable)}
Given a P/T net $N = (S, A, T)$, for each pair of markings $m_1$ and $m_2$, it is decidable whether $m_1 \approx_p m_2$.
\proof
If $|m_1| \neq |m_2|$, then $m_1 \not \approx_p m_2$ by Proposition \ref{fin-k-add}. 
Otherwise, we assume that $|m_1| = k = |m_2|$.
As $|S| = n$, the set of all the place relations over $S$ is of size $2^{n^2}$. 
Let us list all the place relations as follows: 
$R_1, R_2, \ldots, R_{2^{n^2}}$.
Hence, for $i = 1, \ldots, 2^{n^2}$, by Lemma \ref{bpb-rel-dec-lem} we can decide whether $R_i$ is a branching
place bisimulation and, in such a case,
we can check whether $(m_1, m_2) \in R_i^\oplus$ in $O(k^2 \sqrt{k})$ time. 
As soon as we found a branching place bisimulation $R_i$ such that $(m_1, m_2) \in R_i^\oplus$,
we stop concluding that $m_1 \approx_p m_2$. If none of the $R_i$ is a branching
place bisimulation such that $(m_1, m_2) \in R_i^\oplus$, then
we can conclude that $m_1 \not\approx_p m_2$. 
\fine
\end{theorem}

%
\section{A Small Case Study}\label{case-sec}
%

\begin{figure}[t]  
\centering

\begin{tikzpicture}[
every place/.style={draw,thick,inner sep=0pt,minimum size=6mm},
every transition/.style={draw,thick,inner sep=0pt,minimum size=4mm},
bend angle=42,
pre/.style={<-,shorten <=1pt,>=stealth,semithick},
post/.style={->,shorten >=1pt,>=stealth,semithick}
]
\def\eofigdist{5cm}
\def\eodist{0.5cm}
\def\eodisty{1.2cm}
\def\eodistz{2.2cm}
 
\node (p1) [place]  [label=above:$P_1$] {};
\node (p2) [place]  [right=\eofigdist of p1,label=above:$C$] {};
\node (t1) [transition] [below left=\eodisty of p1,label=left:$\tau$] {};
\node (p3) [place] [below =\eodist of t1,label=left:$P_2$] {};
\node (t2) [transition] [below right=\eodisty of p1,label=right:$a$] {};
\node (p4) [place] [below =\eodist of t2,label=right:$D_3$] {};
\node (t3) [transition] [below left=\eodist of p3,label=right:$a$] {};
\node (p5) [place] [below =\eodist of t3,label=left:$D_1$] {};
\node (t4) [transition] [below right=\eodist of p3,label=left:$b$] {};
\node (p6) [place] [below =\eodist of t4,label=right:$D_2$] {};
\node (t5) [transition] [below right=\eodist of p5,label=left:$del_a$] {};
\node (t6) [transition] [below right=\eodisty of p6,label=left:$del_b$] {};
\node (t7) [transition] [below =\eodisty of p4,label=left:$del_a$] {};
\node (p7) [place] [below =\eodistz of t7,label=left:$C_1$] {};
\node (t8) [transition] [below =\eodist of p7,label=left:$\tau$] {};
\node (p8) [place] [right =\eodisty of t8,label=below:$C_2$] {};
\node (t9) [transition] [right =\eodisty of p7,label=above:$cons$] {};
\node (t10) [transition] [right =\eodisty of p8,label=below:$cons$] {};
\node (p9) [place] [right =\eodisty of t9,label=right:$C_3$] {};
\node (t11) [transition] [above right =\eodisty of p9,label=left:$\tau$] {};

\draw  [->] (p1) to (t1);
\draw  [->] (t1) to (p3);
\draw  [->] (p3) to (t3);
\draw  [->] (p3) to (t4);
\draw  [->, bend left] (p1) to (t2);
\draw  [->, bend left] (t2) to (p1);
\draw  [->] (t2) to (p4);
\draw  [->] (t3) to (p5);
\draw  [->] (t4) to (p6);
\draw  [->, bend left] (t3) to (p1);
\draw  [->] (t4) to (p1);
\draw  [->] (p4) to (t7);
\draw  [->] (p2) to (t7);
\draw  [->] (p5) to (t5);
\draw  [->] (p6) to (t6);
\draw  [->, bend left] (p2) to (t5);
\draw  [->, bend left] (p2) to (t6);
\draw  [->] (t5) to (p7);
\draw  [->] (t6) to (p7);
\draw  [->] (t7) to (p7);
\draw  [->] (p7) to (t8);
\draw  [->] (p7) to (t9);
\draw  [->] (t9) to (p9);
\draw  [->] (t8) to (p8);
\draw  [->] (p8) to (t10);
\draw  [->] (t10) to (p9);
\draw  [->] (p9) to (t11);
\draw  [->] (t11) to (p2);

\end{tikzpicture}
\caption{An unbounded producer-consumer system}
\label{upc-a-place}
\end{figure}
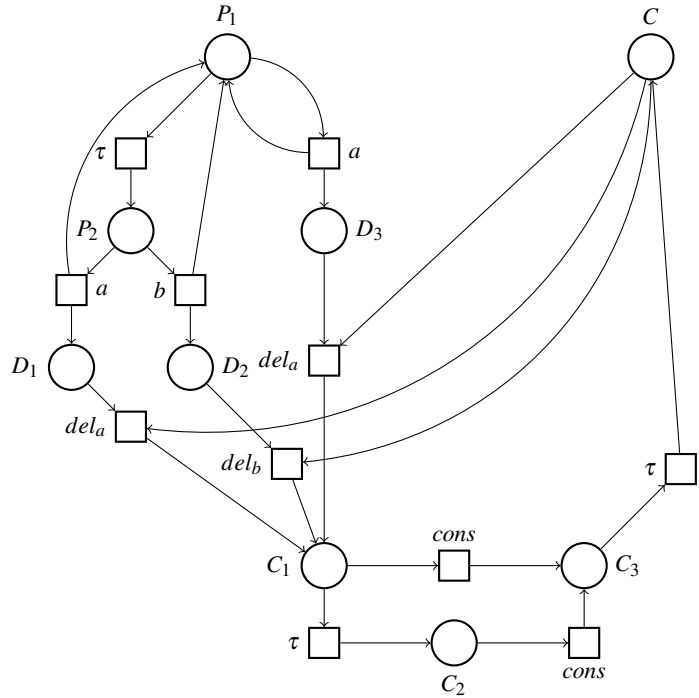

In Figure \ref{upc-a-place} a producer-consumer system is outlined. The producer $P_1$ can unboundedly produce item $a$, each time
depositing one token on place $D_3$, or it can perform some internal work (e.g., preparation of the production lines) and then choose to produce
item $a$ or item $b$, depositing one token on $D_1$ or $D_2$, respectively, and then start again from place $P_1$.
The consumer $C$ can synchronize with the deposit processes $D_1, D_2, D_3$ to perform the delivery of the selected item to $C_1$. This
sequential system has the ability to directly perform $cons$ reaching $C_3$ or it needs some preparatory internal work before 
performing $cons$ to reach the same place. Finally, $C_3$ can perform an internal transition reaching $C$. Note that the three silent transitions
are all $\tau$-sequential.

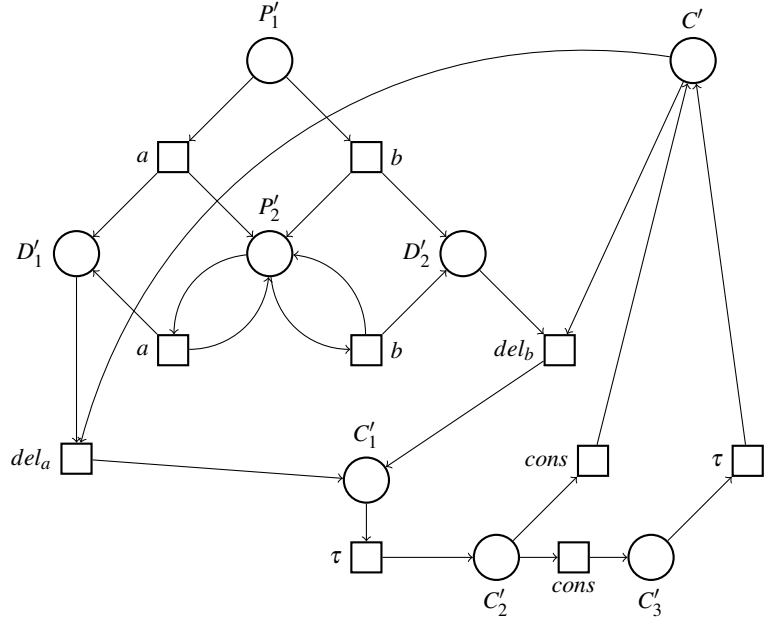
\begin{figure}[t]
\centering

\begin{tikzpicture}[
every place/.style={draw,thick,inner sep=0pt,minimum size=6mm},
every transition/.style={draw,thick,inner sep=0pt,minimum size=4mm},
bend angle=42,
pre/.style={<-,shorten <=1pt,>=stealth,semithick},
post/.style={->,shorten >=1pt,>=stealth,semithick}
]
\def\eofigdist{5cm}
\def\eodist{0.5cm}
\def\eodisty{1.2cm}
\def\eodistz{2.2cm}

\node (p1) [place]  [label=above:$P_1'$] {};
\node (p2) [place]  [right=\eofigdist of p1,label=above:$C'$] {};
\node (t1) [transition] [below left=\eodisty of p1,label=left:$a$] {};
\node (t2) [transition] [below right=\eodisty of p1,label=right:$b$] {};
\node (p3) [place] [below left=\eodisty of t1,label=left:$D_1'$] {};
\node (p4) [place] [below right=\eodisty of t1,label=above:$P_2'$] {};
\node (p5) [place] [below right=\eodisty of t2,label=left:$D_2'$] {};
\node (t3) [transition] [below right=\eodisty of p3,label=left:$a$] {};
\node (t4) [transition] [below right=\eodisty of p4,label=right:$b$] {};
\node (t5) [transition] [below =\eodistz of p3,label=left:$del_a$] {};
\node (t6) [transition] [below right=\eodisty of p5,label=left:$del_b$] {};
\node (p6) [place] [below=\eodisty of t4,label=above:$C_1'$] {};
\node (t7) [transition] [below =\eodist of p6,label=left:$\tau$] {};
\node (p7) [place] [right=\eodisty of t7,label=below:$C_2'$] {};
\node (t8) [transition] [above right =\eodisty of p7,label=left:$cons$] {};
\node (t9) [transition] [right =\eodist of p7,label=below:$cons$] {};
\node (p8) [place] [right=\eodist of t9,label=below:$C_3'$] {};
\node (t10) [transition] [above right =\eodisty of p8,label=left:$\tau$] {};

\draw  [->] (p1) to (t1);
\draw  [->] (p1) to (t2);
\draw  [->] (t1) to (p3);
\draw  [->] (t1) to (p4);
\draw  [->] (t2) to (p4);
\draw  [->] (t2) to (p5);
\draw  [->, bend right] (p4) to (t4);
\draw  [->, bend right] (t4) to (p4);
\draw  [->, bend right] (p4) to (t3);
\draw  [->, bend right] (t3) to (p4);
\draw  [->] (t3) to (p3);
\draw  [->] (t4) to (p5);
\draw  [->, bend right] (p2) to (t5);
\draw  [->] (p2) to (t6);
\draw  [->] (p3) to (t5);
\draw  [->] (p5) to (t6);
\draw  [->] (t5) to (p6);
\draw  [->] (t6) to (p6);
\draw  [->] (p6) to (t7);
\draw  [->] (t7) to (p7);
\draw  [->] (p7) to (t8);
\draw  [->] (t8) to (p2);
\draw  [->] (p7) to (t9);
\draw  [->] (t9) to (p8);
\draw  [->] (p8) to (t10);
\draw  [->] (t10) to (p2);

\end{tikzpicture}
\caption{Another unbounded producer-consumer system}
\label{upc-b-place}
\end{figure}

In Figure \ref{upc-b-place} another unbounded producer-consumer system is outlined. 
The producer $P_1'$ can choose to produce item $a$ or item $b$,
depositing one token on $D_1'$ or $D_2'$, respectively, and then become $P_2'$, which can unboundedly choose to produce $a$ or $b$.
The consumer $C'$ can synchronize with the deposit processes $D_1', D_2'$ to perform the delivery of the selected item to $C_1'$. This
sequential system first performs an internal transition and then it has the ability to perform $cons$ in two different ways: 
either directly reaching $C'$ or 
reaching $C_3'$, which performs an internal transition in order to reach $C'$. Note that the two silent transitions
are $\tau$-sequential.

It is not difficult to realize that the following place relation 

$
\begin{array}{rrcl}
\quad &R &  = &  \{(P_1, P_1'), (P_2, P_1'), (P_1, P_2'), (P_2, P_2'), (D_1, D_1'), (D_2, D_2'), (D_3, D_1'),\\
& && (C, C'), (C_1, C_1'), (C_2, C_2'), (C_3, C_3'), (C_1, C_2'), (C_3, C')\}
\end{array}
$

\noindent
is a branching place bisimulation, so that $P_1 \oplus C \approx_p P_1' \oplus C'$ as $(P_1 \oplus C, P_1' \oplus C') \in R^\oplus$. 
The fact that $R$ is a branching place bisimulation can be proved by exploiting Lemma \ref{bpb-rel-dec-lem}: it is enough 
to check that, for each transition $t_1$ of the first net and for each marking $m$ of the second net such 
that $(\pre{t_1}, m) \in R^\oplus$, the following hold:
\begin{itemize}
	\item[$(a)$] either $t_1$ is $\tau$-sequential and there exists an acyclic $\tau$-sequential
	$\sigma$ such that $m = \pre{\sigma}$, $\Psi(\pre{t_1}, \sigma, R^\oplus)$,
	 $(\pre{t_1}, \post{\sigma}) \in R^\oplus$ and $(\post{t_1}, \post{\sigma}) \in R^\oplus$;
	\item[$(b)$] or there exist an acyclic $\tau$-sequential $\sigma$  and $t_2 \in T$, with 
	$\post{\sigma} = \pre{t_2}$, such that $m = \pre{\sigma}$, $l(t_1) = l(t_2)$, $\Psi(\pre{t_1}, \sigma, R^\oplus)$,
	$(\pre{t_1}, \pre{t_2}) \in R^\oplus$
	and $(\post{t_1}, \post{t_2}) \in R^\oplus$.
\end{itemize}

And the symmetric condition for each transition $t_2$ of the second net and for each marking $m$ of the first 
net such that $(m, \pre{t_2}) \in R^\oplus$.

For instance, consider the $\tau$-sequential transition $(P_1, \tau, P_2)$. 
The only markings to consider are $P_1'$ and $P_2'$ and, by the either case $(a)$ above, 
it is enough to consider $\sigma = i(P_1')$ or $\sigma = i(P_2')$, respectively, to get the thesis. Similarly, for transition
$(C_1, cons, C_3)$ we have to consider only the markings $C_1'$ and $C_2'$; the former can respond by first performing the silent transition
to $C_2'$ and then $(C_2', cons, C_3')$, so that, by case $(b)$ above, we get the thesis by choosing $\sigma = (C_1', \tau, C_2')$;
in the latter case, we simply choose $\sigma = i(C_2')$. As a final example for this side of the proof, consider transition
$(D_1 \oplus C, del_a, C_1)$, so that the only marking to consider is $D_1' \oplus C'$, that can respond 
with $(D_1' \oplus C', del_a, C_1')$ to satisfy the required conditions.

Symmetrically, in case of transition $(P_1', b, P_2' \oplus D_2')$, the only markings to consider are $P_1$ and $P_2$. In the latter case,
$P_2$ can respond with transition $(P_2, b, P_1 \oplus D_2)$ and, by the or case $(b)$, we get the thesis
by choosing $\sigma = i(P_2)$. In the former case, $P_1$ can respond 
by first performing the internal $\tau$-sequential transition, reaching $P_2$, and then transition $(P_2, b, P_1 \oplus D_2)$; hence,
by the or case, we get the thesis by choosing $\sigma = (P_1, \tau, P_2)$.
Similarly, for transition $(C_2', cons, C')$ we have to consider markings $C_1$ and $C_2$. In the latter case, $C_2$ can respond
with $(C_2, cons, C_3)$ and the thesis is satisfied, by the or case, with $\sigma = i(C_2)$. In the former case, $C_1$ first performs
the silent transition to $C_2$ and then $(C_2, cons, C_3)$, and the thesis is satisfied by choosing $\sigma = (C_1, \tau, C_2)$.
As a final example for this side of the proof, consider transition
$(D_1' \oplus C', del_a, C_1')$, so that the two markings to consider are $D_1 \oplus C$ and $D_1 \oplus C_3$.
The former can simply respond by  $(D_1 \oplus C, del_a, C_1)$, while the latter first performs $\sigma = i(D_1) (C_3, \tau, C)$.

%
\section{A Coarser Variant: Branching D-place Bisimilarity}\label{br-d-place-sec}
%

We first recall from \cite{Gor21} a coarser variant of place bisimulation, called {\em d-place bisimulation}.
Then, we introduce {\em branching d-place} bisimulation. Finally, we prove that branching d-place bisimilarity $\approx_d$
is finer than branching fully-concurrent bisimilarity $\approx_{bfc}$. 

%
\subsection{D-place Bisimilarity}\label{d-place-ssec}
%

A coarser variant of place bisimulation,  introduced in \cite{Gor21} and called {\em d-place bisimulation}, may relate a 
place $s$ also to the empty marking $\theta$.
In order to provide the definition of d-place bisimulation, 
we need first to extend the domain of a place relation: 
the empty marking $\theta$ is considered as an additional place, so that a place 
relation is defined not on $S$, rather on $S \cup \{\theta\}$.
Hence, 
the symbols $r_1$ and $r_2$ that occur in the following definitions, 
can only denote either the empty marking $\theta$ or a 
single place $s$. 
Now we extend the idea of additive closure to these more general place relations, 
yielding {\em d-additive closure}.

\begin{definition}\label{hadd-eq}{\bf (D-additive closure)}
Given a P/T net $N = (S, A, T)$ and a {\em place relation} $R \subseteq (S\cup \{\theta\}) \times (S \cup \{\theta\})$, we define 
a {\em marking relation}
$R^\odot \, \subseteq \, {\mathcal M}(S) \times {\mathcal M}(S)$, called 
the {\em d-additive closure} of $R$,
as the least relation induced by the following axiom and rule.

$\begin{array}{lllllllllll}
 \bigfrac{}{(\theta, \theta) \in  R^\odot} & \; & \; 
 \bigfrac{(r_1, r_2) \in R \; \; (m_1, m_2) \in R^\odot }{(r_1 \oplus m_1, r_2 \oplus m_2) \in  R^\odot }  \\
\end{array}$
\\[-.2cm]
\fine
\end{definition}

Note that if two markings are related by $R^\odot$, then they 
may have different size; 
in fact, even if the axiom relates the empty marking to itself (so two markings with the same size),
as $R \subseteq (S\cup \{\theta\}) \times (S \cup \{\theta\})$, it may be the case that $(\theta, s) \in R$,
so that, assuming $(m_1', m_2') \in R^\odot$ with $|m_1'| = |m_2'|$, we get that the pair
$(m_1', s \oplus m_2')$ belongs to $R^\odot$, as $\theta$ is the identity for
the operator of multiset union. 
Hence, Proposition \ref{fin-k-add}, which is valid for place relations defined over $S$, is not valid for place relations 
defined over $S\cup \{\theta\}$. 
However, the properties in Propositions \ref{add-prop1}
and \ref{add-prop2} hold also for these more general place relations.
Note that checking whether $(m_1, m_2) \in R^\odot$ has complexity $O(k^2\sqrt{k})$, where $k$ is the size of the largest marking.

\begin{definition}\label{def-dplace-bis}{\bf (D-place bisimulation)}
Let $N = (S, A, T)$ be a P/T net. 
A {\em d-place bisimulation} is a relation
$R\subseteq (S\cup \{\theta\}) \times (S \cup \{\theta\})$ such that if $(m_1, m_2) \in R^\odot$
then
\begin{itemize}
\item $\forall t_1$ such that  $m_1[t_1\rangle m'_1$, $\exists t_2$ such that $m_2[t_2\rangle m'_2$ 
with $(\pre{t_1}, \pre{t_2}) \in R^\odot$, $l(t_1) = l(t_2)$,  $(\post{t_1}, \post{t_2}) \in R^\odot$ and, moreover, 
$(m_1', m_2') \in R^\odot$,
\item $\forall t_2$ such that  $m_2[t_2\rangle m'_2$, $\exists t_1$ such that $m_1[t_1\rangle m'_1$ 
with $(\pre{t_1}, \pre{t_2}) \in R^\odot$, $l(t_1) = l(t_2)$,  $(\post{t_1}, \post{t_2}) \in R^\odot$ and, moreover, 
$(m_1', m_2') \in R^\odot$.
\end{itemize}

Two markings $m_1$ and $m_2$ are  {\em d-place bisimilar}, denoted by
$m_1 \sim_{d} m_2$, if there exists a d-place bisimulation $R$ such that $(m_1, m_2) \in R^\odot$.
\fine
\end{definition}
 
D-place bisimilarity $\sim_d$ is a decidable equivalence relation \cite{Gor21}.
Moreover, in \cite{Gor21} it is proved that $\sim_d$ is
finer than fully-concurrent bisimilarity $\sim_{fc}$. This implication is strict,
as illustrated by the following example.

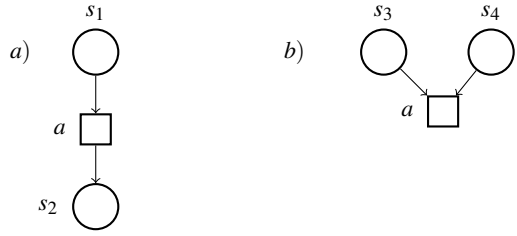
\begin{figure}[!t]
\centering
\begin{tikzpicture}[
every place/.style={draw,thick,inner sep=0pt,minimum size=6mm},
every transition/.style={draw,thick,inner sep=0pt,minimum size=4mm},
bend angle=30,
pre/.style={<-,shorten <=1pt,>=stealth,semithick},
post/.style={->,shorten >=1pt,>=stealth,semithick}
]
\def\eofigdist{3.2cm}
\def\eodist{0.5cm}
\def\eodisty{0.8cm}

\node (a) [label=left:$a)\qquad $]{};

\node (p1) [place]  [label=above:$s_1$] {};
\node (t1) [transition] [below =\eodist of p1,label=left:$a\;$] {};
\node (p2) [place] [below =\eodist of t1,label=left:$s_2\;$] {};

\draw  [->] (p1) to (t1);
\draw  [->] (t1) to (p2);

  \node (b) [right={3.1cm} of a,label=left:$b)\quad$] {};

\node (p3) [place]  [right=\eofigdist of p1, label=above:$s_3$] {};
\node (p4) [place]  [right=\eodisty of p3, label=above:$s_4$] {};
\node (t2) [transition] [below right =\eodist of p3,label=left:$a\;$] {};

\draw  [->] (p3) to (t2);
\draw  [->] (p4) to (t2);

\end{tikzpicture}
\caption{Two fc-bisimilar nets, but not d-place bisimilar}
\label{net-d2-place}
\end{figure}

\begin{example}\label{ex-dead}
Consider Figure \ref{net-d2-place}. Even if $s_1$ and $s_3 \oplus s_4$
are fc-bisimilar, we cannot find any d-place bisimulation relating these two markings. If we include the necessary 
pairs $(s_1, s_3)$ and $(\theta, s_4)$, then we would fail immediately, because the pair $(s_1, s_3)$ does not satisfy
the d-place bisimulation game, as $s_1$ can move, while $s_3$ cannot.
\fine
\end{example}

\begin{figure}[!t]
\centering
\begin{tikzpicture}[
every place/.style={draw,thick,inner sep=0pt,minimum size=6mm},
every transition/.style={draw,thick,inner sep=0pt,minimum size=4mm},
bend angle=30,
pre/.style={<-,shorten <=1pt,>=stealth,semithick},
post/.style={->,shorten >=1pt,>=stealth,semithick}
]
\def\eofigdist{3.2cm}
\def\eodist{0.5cm}
\def\eodisty{0.8cm}

\node (a) [label=left:$a)\qquad $]{};

\node (p1) [place]  [label=above:$s_1$] {};
\node (t1) [transition] [below =\eodist of p1,label=left:$a\;$] {};
\node (p2) [place] [below =\eodist of t1,label=left:$s_2\;$] {};
\node (t2) [transition] [below =\eodist of p2,label=left:$b\;$] {};
\node (p3) [place] [below  =\eodist of t2,label=left:$s_3\;$] {};

\draw  [->] (p1) to (t1);
\draw  [->] (t1) to (p2);
\draw  [->] (p2) to (t2);
\draw  [->] (t2) to (p3);

  \node (b) [right={3.1cm} of a,label=left:$b)\quad$] {};

\node (p4) [place]  [right=\eofigdist of p1, label=above:$s_4$] {};
\node (t3) [transition] [below  =\eodist of p4,label=left:$a\;$] {};
\node (p5) [place]  [below left =\eodist of t3, label=left:$s_5\;$] {};
\node (p6) [place]  [below right =\eodist of t3, label=right:$\;s_6$] {};
\node (t4) [transition] [below =\eodist of p6,label=left:$b\;$] {};

\draw  [->] (p4) to (t3);
\draw  [->] (t3) to (p5);
\draw  [->] (t3) to (p6);
\draw  [->] (p6) to (t4);

\end{tikzpicture}
\caption{Two d-place bisimilar nets}
\label{net-d-place}
\end{figure}
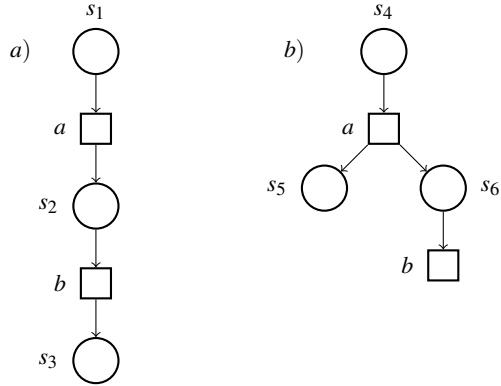

\begin{figure}[!t]
\centering
\begin{tikzpicture}[
every place/.style={draw,thick,inner sep=0pt,minimum size=6mm},
every transition/.style={draw,thick,inner sep=0pt,minimum size=4mm},
bend angle=30,
pre/.style={<-,shorten <=1pt,>=stealth,semithick},
post/.style={->,shorten >=1pt,>=stealth,semithick}
]
\def\eofigdist{2.7cm}
\def\eodist{0.4cm}
\def\eodisty{0.6cm}

\node (a) [label=left:$a)\quad $]{};

\node (p1) [place]  [label=above:$s_1$] {};

  \node (b) [right={3.1cm} of a,label=left:$b)\quad$] {};

\node (p2) [place]  [right=\eofigdist of p1, label=above:$s_2$] {};
\node (p3) [place]  [right=\eodisty of p2, label=above:$s_3$] {};
\node (p4) [place]  [right=\eodisty of p3, label=above:$s_4$] {};
\node (t2) [transition] [below =\eodist of p3,label=left:$a\;$] {};

\draw  [->] (p2) to (t2);
\draw  [->] (p3) to (t2);
\draw  [->] (p4) to (t2);

\end{tikzpicture}
\caption{Relation $\{(s_1, s_2), (\theta, s_3)\}$ is a d-place bisimulation}
\label{net-d3-place}
\end{figure}
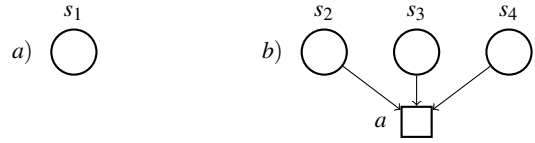

\begin{example}\label{ex-dplace}
Consider the net in Figure \ref{net-d-place}. It is easy to realize that $R = \{(s_1, s_4), (\theta, s_5),$ $
(s_2, s_6),$ $(s_3, \theta)\}$ is a d-place bisimulation. Hence, this example shows that d-place bisimilarity is strictly coarser
than place bisimilarity, and that it does not preserves the causal nets, 
because $s_1$ and $s_4$ generate different causal nets.
The places that are related to $\theta$ (i.e., $s_3$ and $s_5$) are deadlocks, i.e., they have empty post-set.
However, it may happen that a d-place bisimulation can also relate a place with non-empty post-set to $\theta$.
In fact, consider the net in Figure \ref{net-d3-place}. It is easy to observe that the relation 
$R = \{(s_1, s_2), (\theta, s_3)\}$ is a d-place bisimulation, as for all the pairs $(m_1, m_2) \in R^\odot$,
both markings are stuck, so that the d-place bisimulation game is vacuously satisfied.
\fine
\end{example}

\begin{remark}{\bf (Condition on the pre-sets)}\label{pre-rem}
As a consequence of the observation of the previous examples, it is easy to note that
if a d-place bisimulation $R$ relates a place $s$ with non-empty post-set to $\theta$, then it is not possible
to find two transitions $t_1$ and $t_2$ such that for the proof of $(\pre{t_1}, \pre{t_2}) \in R^\odot$ it is 
necessary to use the pair $(s, \theta)$ (cf. Example \ref{ex-dead}). In other words,
the condition $(\pre{t_1}, \pre{t_2}) \in R^\odot$ in Definition \ref{def-dplace-bis} is actually $(\pre{t_1}, \pre{t_2}) \in \overline{R}^\oplus$,
where $\overline{R} = \{(r_1, r_2) \in R \mid r_1 \in S \wedge r_2 \in S\}$.
\fine
\end{remark}

%
\subsection{Branching D-place Bisimulation}\label{brd-place-ssec}
%

Branching d-place bisimulation is defined as branching place bisimulation (using {\em $\tau$-sequential transition sequences}, 
i.e., sequences composed of $\tau$-sequential net transitions and also idling transitions), where the additive closure $\oplus$ is replaced
by the d-additive closure $\odot$, except when considering the presets of the matched transitions where $R^\odot$ 
is actually $\overline{R}^\oplus$
(cf. Remark \ref{pre-rem}). 

\begin{definition}\label{bdpb-bis-def}{\bf (Branching d-place bisimulation)}
Given a P/T net $N = (S, A, T)$, a {\em branching d-place bisimulation} is a relation
$R\subseteq (S\cup \{\theta\}) \times (S \cup \{\theta\})$ such that if $(m_1, m_2) \in R^\odot$
\begin{enumerate}
\item $\forall t_1$ such that $m_1[t_1\rangle m_1'$
    \begin{itemize}
    \item[$(i)$] either $t_1$ is $\tau$-sequential and
                      $\exists \sigma, m_2'$ such that $\sigma$ is $\tau$-sequential, 
                      $m_2[\sigma\rangle m_2'$,  and  $\Psi(\pre{t_1}, \sigma, \overline{R}^\oplus)$,
                                   $(\pre{t_1}, \post{\sigma})  \in \overline{R}^\oplus$, $(\post{t_1}, \post{\sigma}) \in \overline{R}^\oplus$ and 
                                   $(m_1 \ominus \pre{t_1}, m_2 \ominus \pre{\sigma}) \in R^\odot$;
     \item[$(ii)$] or there exist $\sigma, t_2, m, m_2'$ such that 
                     $\sigma$ is $\tau$-sequential, $m_2[\sigma\rangle m [t_2\rangle m_2'$, 
                     $l(t_1) = l(t_2)$, $\post{\sigma} = \pre{t_2}$, $\Psi(\pre{t_1}, \sigma, \overline{R}^\oplus)$,
                     $(\pre{t_1}, \pre{t_2}) \in \overline{R}^\oplus$
                     $(\post{t_1}, \post{t_2}) \in R^\odot$, and moreover, 
                     $(m_1 \ominus \pre{t_1}, m_2 \ominus \pre{\sigma}) \in R^\odot$; 
   \end{itemize}
\item and, symmetrically, $\forall t_2$ such that $m_2[t_2\rangle m_2'$
%
\end{enumerate}

Two markings $m_1$ and $m_2$ are branching d-place bisimilar, 
denoted by $m_1 \approx_{d} m_2$,
if there exists a branching d-place bisimulation $R$ such that $(m_1, m_2) \in R^\odot$.
\fine
\end{definition} 

It is easy to observe that, in case 1$(i)$ (either case), by additivity of $R^\odot$ (also w.r.t. $\overline{R}^\oplus$), from 
$(m_1 \ominus \pre{t_1}, m_2 \ominus \pre{\sigma}) \in R^\odot$ and 
$(\pre{t_1}, \post{\sigma})  \in \overline{R}^\oplus$, we get $(m_1, m_2') \in R^\odot$, as well as, from
$(\post{t_1}, \post{\sigma}) \in \overline{R}^\oplus$ we get  $(m_1', m_2') \in R^\odot$. In a similar manner, for case 1$(ii)$ (or case),
from $(m_1 \ominus \pre{t_1}, m_2 \ominus \pre{\sigma}) \in R^\odot$, $\post{\sigma} = \pre{t_2}$ and 
$(\pre{t_1}, \pre{t_2}) \in \overline{R}^\oplus$, we get $(m_1, m) \in R^\odot$, as well as, from $(\post{t_1}, \post{t_2}) \in R^\odot$,
we get  $(m_1', m_2') \in R^\odot$.

Note also that a $\tau$-sequential  transition performed by one of the two markings may be matched by the other one also by idling: 
this is due to the {\em either} case when $\sigma = i(s_2)$ for a suitable token $s_2$  such that $(\pre{t_1}, \pre{\sigma})  \in \overline{R}^\oplus$, 
                                   $(\pre{t_1}, \post{\sigma})  \in \overline{R}^\oplus$, $(\post{t_1}, \post{\sigma}) \in \overline{R}^\oplus$ and 
                                   $(m_1 \ominus \pre{t_1}, m_2 \ominus \pre{\sigma}) \in R^\odot$, where $\pre{\sigma} = \post{\sigma} = s_2$.

\begin{figure}[!t]
\centering
\begin{tikzpicture}[
every place/.style={draw,thick,inner sep=0pt,minimum size=6mm},
every transition/.style={draw,thick,inner sep=0pt,minimum size=4mm},
bend angle=30,
pre/.style={<-,shorten <=1pt,>=stealth,semithick},
post/.style={->,shorten >=1pt,>=stealth,semithick}
]
\def\eofigdist{3.2cm}
\def\eodist{0.4cm}
\def\eodisty{0.8cm}

\node (a) [label=left:$a)\qquad $]{};

\node (p1) [place]  [label=above:$s_1$] {};
\node (t1) [transition] [below =\eodist of p1,label=left:$a\;$] {};
\node (p2) [place] [below =\eodist of t1,label=right:$s_2\;$] {};
\node (t2) [transition] [below =\eodist of p2,label=left:$b\;$] {};
\node (p3) [place] [below  =\eodist of t2,label=left:$s_3\;$] {};
\node (t2') [transition] [left =\eodist of p2,label=left:$\tau\;$] {};

\draw  [->] (p1) to (t1);
\draw  [->] (t1) to (p2);
\draw  [->] (p2) to (t2);
\draw  [->] (t2) to (p3);
\draw  [->, bend right] (p2) to (t2');
\draw  [->, bend right] (t2') to (p2);

  \node (b) [right={3.1cm} of a,label=left:$b)\quad$] {};

\node (p4) [place]  [right=\eofigdist of p1, label=above:$s_4$] {};
\node (t3) [transition] [below  =\eodist of p4,label=left:$a\;$] {};
\node (p5) [place]  [right =\eodisty of p4, label=above:$s_5\;$] {};
\node (p5') [place]  [below left =\eodist of t3, label=left:$s_6\;$] {};
\node (p6') [place]  [below right =\eodist of t3, label=right:$\;s_7$] {};
\node (t4') [transition] [below =\eodist of p6',label=left:$\tau\;$] {};
\node (p7) [place]  [below =\eodist of t4', label=right:$\;s_8$] {};
\node (t4) [transition] [below =\eodist of p7,label=left:$b\;$] {};

\draw  [->] (p4) to (t3);
\draw  [->] (t3) to (p5');
\draw  [->] (t3) to (p6');
\draw  [->] (p6') to (t4');
\draw  [->] (t4') to (p7);
\draw  [->] (p6') to (t4');
\draw  [->] (p7) to (t4);

\end{tikzpicture}
\caption{Two branching d-place bisimilar nets}
\label{net-bd-place}
\end{figure}
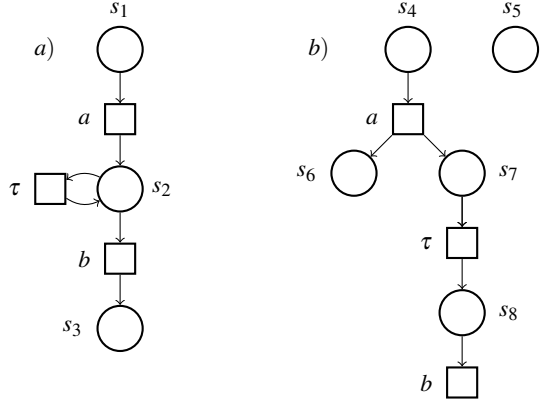

\begin{example}\label{ex-bdp-bis}
Consider the nets in Figure \ref{net-bd-place}. It is easy to realize that $s_1 \approx_d s_4 \oplus s_5$ because
$R = \{(s_1, s_4), (\theta, s_5),  (\theta, s_6), (s_2, s_7), (s_2, s_8), (s_3, \theta)\}$ is a branching d-place bisimulation
such that $(s_1, s_4 \oplus s_5) \in R^\odot$.
\fine
\end{example}

 Similarly to what done in Proposition \ref{prop-bpb-bis1} and Proposition \ref{prop-bpb-bis2}, we can also prove that the identity
 relation is a branching d-place bisimulation, that the inverse of a branching d-place bisimulation is a branching d-place bisimulation and that
 the relational composition of two branching d-place bisimulations is a branching d-place bisimulation.
 As a consequence, $\approx_d$ is also an equivalence relation. Moreover, similarly to what described in Remark \ref{rem-weak-stutt-place-bis},
 we can argue that $\approx_d$ enjoys the weak stuttering property, so that it fully respects the timing of choices.
 
  \noindent
By Definition \ref{bdpb-bis-def}, branching d-place bisimilarity can be defined as follows:

$\approx_d = \bigcup \{ R^\oplus \mid R \mbox{ is a branching d-place bisimulation}\}.$

\noindent
By monotonicity of the d-additive closure, if $R_1 \subseteq R_2$, then
$R_1^\odot \subseteq R_2^\odot$. Hence, we can restrict our attention to maximal branching d-place bisimulations only:

$\approx_d = \bigcup \{ R^\odot \mid R \mbox{ is a {\em maximal} branching d-place bisimulation}\}.$

\noindent
However, it is not true that 

$\approx_d = (\bigcup \{ R \mid R \mbox{ is a {\em maximal} d-place bisimulation}\})^\odot$,
because the union of branching d-place bisimulations may be not a branching d-place bisimulation.
Hence, its definition is not coinductive, so that we cannot adapt the well-known algorithms for computing
the largest bisimulation \cite{PT87,KS83}, as there is no one largest branching d-place bisimulation.
Nonetheless, we can adapt the decidability proof 
 in Section \ref{decid-br-place-sec}, to prove that also $\approx_d$ is decidable for finite P/T nets.
 The key point is that we can prove, similarly to what done in Lemma \ref{bpb-rel-dec-lem}, that $R$ is a 
 branching d-place bisimulation if and only if the following two finite conditions are satisfied:
\begin{enumerate}
\item $\forall t_1 \in T$, $\forall m$ such that $(\pre{t_1}, m) \in \overline{R}^\oplus$
	\begin{itemize}
	\item[$(a)$] either $t_1$ is $\tau$-sequential and there exists an acyclic $\tau$-sequential
	$\sigma$ such that $m = \pre{\sigma}$,  $\Psi(\pre{t_1}, \sigma, \overline{R}^\oplus)$,
	 $(\pre{t_1}, \post{\sigma}) \in \overline{R}^\oplus$ and $(\post{t_1}, \post{\sigma}) \in \overline{R}^\oplus$;
	\item[$(b)$] or there exist an acyclic $\tau$-sequential $\sigma$  and $t_2 \in T$, with 
	$\post{\sigma} = \pre{t_2}$, such that $m = \pre{\sigma}$, $l(t_1) = l(t_2)$,  $\Psi(\pre{t_1}, \sigma, \overline{R}^\oplus)$,
	$(\pre{t_1}, \pre{t_2}) \in \overline{R}^\oplus$
	and $(\post{t_1}, \post{t_2}) \in R^\odot$.
	\end{itemize}
\item $\forall t_2 \in T$, $\forall m$ such that $(m, \pre{t_2}) \in \overline{R}^\oplus$
	\begin{itemize}
	\item[$(a)$] either $t_2$ is $\tau$-sequential and there exists an acyclic $\tau$-sequential $ \sigma$  
	such that $m = \pre{\sigma}$,  $\Phi(\sigma, \pre{t_2}, \overline{R}^\oplus)$,
	 $(\post{\sigma}, \pre{t_2}) \in \overline{R}^\oplus$ and $(\post{\sigma}, \post{t_2}) \in \overline{R}^\oplus$;
	\item[$(b)$] or there exist an acyclic $\tau$-sequential $\sigma$  and $t_1 \in T$, with 
	$\post{\sigma} = \pre{t_1}$, such that $m = \pre{\sigma}$, $l(t_1) = l(t_2)$, $\Phi(\sigma, \pre{t_2}, \overline{R}^\oplus)$,
	$( \pre{t_1}, \pre{t_2}) \in \overline{R}^\oplus$
	and $(\post{t_1}, \post{t_2}) \in R^\odot$,
	\end{itemize}
\end{enumerate}

\noindent
that are decidable in exponential time.
Hence, by considering all the finitely many place relations for a finite P/T net, we can check whether each of them is a branching
d-place bisimulation and, in such a case, whether the considered markings are related by its d-additive closure.

 Of course, $\approx_d$ is coarser than $\approx_p$ because a branching place bisimulation is also a branching d-place bisimulation, but the reverse is not true; for instance, relation $R$ in Example \ref{ex-bdp-bis} is not a branching place bisimulation.

%
\subsection{Sensible Behavioral Equivalence}\label{bdpb>bfc-ssec}
%

In this section we argue that $\approx_d$ is a sensible (i.e., fully respecting causality and the branching structure) behavioral
equivalence, by proving that it is finer than branching fully-concurrent bisimilarity $\approx_{bfc}$.

\begin{theorem}\label{bdpbis>bfc-bis}{\bf (Branching d-place bisimilarity 
 is finer than branching fully concurrent bisimilarity)}
Let $N = (S, A, T)$ be a P/T net with silent moves.
If $m_1 \approx_{d} m_2$, then $m_1 \approx_{bfc} m_2$.

\proof
    If $m_1 \approx_{d} m_2$, then there exists a branching d-place bisimulation $R_1$ such that
$(m_1, m_2) \in R_1^\odot$.
Let us consider 
    \begin{equation*} \label{R2}
        \begin{split}
        R_2 \overset{def}{=} \lbrace (\pi_1, g, \pi_2) | & \pi_1 = (C_1, \rho_1) \text{ is a process of $N(m_{1})$,} \\
        & \pi_2 = (C_2, \rho_2) \text{ is a process of $N(m_{2})$},\\ 
        & \text{$g$ is an abstract event isomorphism between $C_1$ and $C_2$},\\ 
         & \text{and property } \Gamma(\pi_1,g,\pi_2) \text{ holds}
        \rbrace ,
        \end{split}
    \end{equation*}

\noindent
where property $\Gamma(\pi_1, g, \pi_2)$ states that there exists a multiset 

$q = \{(r_1, r_1'),$ $ (r_2, r_2'),$ $ \ldots,$ $(r_k, r_k')\}$

\noindent
of associations such that if $Max(C_1) = b_1 \oplus \ldots \oplus b_{k_1}$ and $Max(C_2) = b_1' 
\oplus \ldots \oplus b'_{k_2}$ (with $k_1, k_2 \leq k$),
then we have that 
\begin{enumerate}
\item $\rho_1(Max(C_1)) = r_1 \oplus \ldots \oplus r_k$ and $\rho_2(Max(C_2)) = r_1' \oplus \ldots \oplus r_k'$ 
(remember that some of the $r_i$ or $r_i'$ can be $\theta$),
\item for $i = 1, \ldots, k$, $(r_i, r_i') \in R_1$, so that $(\rho_1(Max(C_1)), \rho_2(Max(C_2))) \in R_1^\odot$,
\item and for $i = 1, \ldots, k$, if $r_i = \rho_1(b_j)$  for some $b_j \in Max(C_1) \cap \post{e_1}$, 
then
\begin{itemize}
\item[$(i)$] either $r_i' = \theta$, 
\item[$(ii)$] or $e_1$ (and each event preceding $e_1$) is unobservable and $r_i' = \rho_2(b'_{j'})$ for some $b'_{j'} \in Max(C_2)$ that is minimal 
(i.e., such that $b'_{j'} \in Min(C_2)$),
\item[$(iii)$]
or $r_i' = \rho_2(b'_{j'})$ for some $b'_{j'} \in Max(C_2) \cap \post{e_2}$ for some event $e_2$ such that 
\begin{itemize}
\item if $e_1$ is observable, then either $g(e_1) = e_2$ or $g(e_1) \leq_{\pi_2} e_2$
and all the events in the path from $g(e_1)$ (excluded) to $e_2$ (included) are $\tau$-sequential;
\item if $e_1$ is not observable, then for each observable $e_1'$ we have that $e_1' \leq_{\pi_1} e_1$ if and only if
$g(e_1') \leq_{\pi_2} e_2$.
\end{itemize}
\end{itemize}
And symmetrically, if $r_i' = \rho_2(b'_{j'})$ for some 
$b'_{j'} \in Max(C_2) \cap \post{e_2}$, then
\begin{itemize}
\item[$(i)$]
either $r_i = \theta$, 
\item[$(ii)$] or $e_2$ (and each event preceding $e_2$) is unobservable and 
$r_i = \rho_1(b_j)$ for some $b_j \in Max(C_1)$ that is minimal (i.e., such that $b_j \in Min(C_1)$),
\item[$(iii)$]
or $r_i = \rho_1(b_j)$ for some $b_j \in Max(C_1) \cap \post{e_1}$ for some event $e_1$ such that
\begin{itemize}
\item if $e_2$ is observable, then either $g(e_1) = e_2$ or there exists $e_1' \leq_{\pi_1} e_1$ such that $g(e_1') = e_2$
and all the events in the path from $e_1'$ (excluded) to $e_1$ (included) are $\tau$-sequential;
\item if $e_2$ is not observable, then for each observable $e_2'$ we have that $e_2' \leq_{\pi_2} e_2$ if and only if
$g^{-1}(e_2') \leq_{\pi_1} e_1$.
\end{itemize}
\end{itemize}
\end{enumerate}

Note that such a multiset $q$ has the property that for each $(r_i, r_i') \in q$, we have that either one of the two elements in the pair is
$\theta$, or both places are the image of suitable conditions with no observable predecessor events, or
both places are the image of conditions generated by (or causally dependent on) events related by 
the abstract event isomorphism $g$.

We want to prove that $R_2$ is a branching fully-concurrent bisimulation.
First of all, consider a triple of the form $(\pi_1^0, g^0, \pi_2^0)$, 
where $\pi_i^0 = (C_i^0, \rho_i^0)$, 
$C_i^0$ is the causal net without events and $\rho_1^0, \rho_2^0$ are
such that  $\rho_i^0(Min(C_i^0)) = \rho_i^0(Max(C_i^0)) = m_i$ for $i= 1, 2$
and $g^0$ is the empty function.
Then $(\pi_1^0, g^0, \pi_2^0)$ must belong to $R_2$,
because $(C_i^0, \rho_i^0)$ is a process of $N(m_i)$, for $i=1, 2$ and $\Gamma(\pi_1^0, g^0, \pi_2^0)$
trivially holds because,
by hypothesis, $(m_1, m_2) \in R_1^\odot$. 
Hence, if $R_2$ is a branching fully-concurrent bisimulation, then the triple 
$(\pi_1^0, g^0, \pi_2^0) \in R_2$ ensures that $m_1 \approx_{bfc} m_2$. 

Let us check that $R_2$ is a branching fc-bisimulation.
Assume $(\pi_1, g, \pi_2) \in R_2$, where $\pi_i = (C_i, \rho_i)$ for $i = 1, 2$, so that
$\Gamma(\pi_1, g, \pi_2)$ holds for some suitable multiset $q$ of associations.
In order to be a branching fc-bisimulation triple, it is necessary that
\begin{itemize}
\item[$i)$] 
$\forall t_1, e_1, \pi_1'$ such that $\pi_1 \deriv{e_1} \pi_1'$ with $\rho_1'(e_1) = t_1$, 
     \begin{itemize}
     \item {\em either} $l(e_1) = \tau$ and there exist $\sigma_2'$ (with $o(\sigma_2') = \epsilon$) and $\pi_2'$ 
     such that $\pi_2 \Deriv{\sigma_2'} \pi_2'$, $(\pi_1, g, \pi_2') \in R$ and $(\pi_1', g, \pi_2') \in R$;  
 \item {\em or}  $\exists \sigma'$ (with $o(\sigma') = \epsilon$), $e_2, \pi_2', \pi_2'', g'$ 
     such that         
      \begin{enumerate}
           \item $\pi_2 \Deriv{\sigma'} \pi_2' \deriv{e_2} \pi_2''$;
       \item if $l(e_1) = \tau$, then $l(e_2) = \tau$ and $g' = g$;
                otherwise, $l(e_1) = l(e_2)$ and  
                 $g' = g \cup \{(e_1, e_2)\}$; 
       \item and finally, $(\pi_1, g, \pi_2') \in R$ and  
              $(\pi_1', g', \pi_2'') \in R$;
    \end{enumerate}
  \end{itemize}
\item[$ii)$] symmetrically, if $\pi_2$ moves first.
\end{itemize}

Assume $\pi_1 = (C_1, \rho_1) \deriv{e_1} (C_1', \rho_1') = \pi_1'$ with $\rho_1'(e_1) = t_1$.
Now, let  $p = \{(\overline{r}_{1}, \overline{r}'_{1}),$ $\ldots,$ $ (\overline{r}_{h}, \overline{r}'_{h})\} \subseteq q$,
with $\overline{r}_{1}\oplus  \ldots \oplus \overline{r}_{h}$ $= \pre{t_1}$. 
Note that $(\pre{t_1}, \overline{r}'_{1} \oplus  \ldots \oplus \overline{r}'_{h}) \in R_1^\odot$.
Now we remove from $\overline{r}_{1}\oplus  \ldots \oplus \overline{r}_{h}$ those $\overline{r}_{i} = \theta$
to get $\overline{s}_{1}\oplus  \ldots \oplus \overline{s}_{h'}$ $= \pre{t_1}$, with $h' \leq h$.
Similarly, we filter out from $\overline{r}'_{1} \oplus  \ldots \oplus \overline{r}'_{h}$ only those related to places $\overline{s}_i$ in $\pre{t_1}$,
to get $\overline{m}_2 = \overline{s}'_{1} \oplus  \ldots \oplus \overline{s}'_{h'}$ such that 
$(\pre{t_1}, \overline{m}_{2}) \in \overline{R}_1^\oplus$.

By the characterization used in proving that a place relation is a branching d-place bisimulation in Section \ref{brd-place-ssec} 
(inspired to Lemma \ref{bpb-rel-dec-lem}),
since $R_1$ is a branching d-place bisimulation, from $(\pre{t_1},\overline{m}_2) \in \overline{R}_1^\oplus$ it follows that
\begin{itemize}
	\item[$(a)$] either $t_1$ is $\tau$-sequential and there exists an acyclic $\tau$-sequential
	$\sigma$ such that $\overline{m}_2 = \pre{\sigma}$, $\Psi(\pre{t_1}, \sigma, \overline{R}_1^\oplus)$,
	 $(\pre{t_1}, \post{\sigma}) \in \overline{R}_1^\oplus$ and $(\post{t_1}, \post{\sigma}) \in \overline{R}_1^\oplus$;
	\item[$(b)$] or there exist an acyclic $\tau$-sequential $\sigma$  and $t_2 \in T$, with 
	$\post{\sigma} = \pre{t_2}$, such that $\overline{m}_2 = \pre{\sigma}$, $l(t_1) = l(t_2)$, $\Psi(\pre{t_1}, \sigma, \overline{R}_1^\oplus)$,
	$(\pre{t_1}, \pre{t_2}) \in \overline{R}_1^\oplus$
	and $(\post{t_1}, \post{t_2}) \in R_1^\odot$.
 \end{itemize}

In the either-case $(a)$, since $(\pre{t_1},\overline{m}_2) \in \overline{R}_1^\oplus$ and $\overline{m}_2 = \pre{\sigma}$,
we can really extend $\pi_2$ by performing a suitable $\sigma'$ (with $o(\sigma') = \epsilon$) to a suitable process $\pi_2'$ 
 such that $\pi_2 \Deriv{\sigma'} \pi_2'$, $\rho_2'(\sigma') = \sigma$, $(\pi_1, g, \pi_2') \in R_2$ and $(\pi_1', g, \pi_2') \in R_2$,
 where the last two conditions hold because properties $\Gamma(\pi_1, g, \pi_2')$ and $\Gamma(\pi_1', g, \pi_2')$ trivially hold.
More precisely, $\Gamma(\pi_1, g, \pi_2')$ holds because 
from the multiset $q = \{(r_1, r_1'),$ $ (r_2, r_2'),$ $ \ldots,$ $(r_k, r_k')\}$ we remove the multiset 
$p = \{(\overline{s}, \overline{s}')\} \subseteq q$ (such that 
$ \pre{t_1} = \overline{s}$ and $\pre{\sigma} = 
 \overline{s}'$), and we
add the multiset $p' = \{(\overline{s}, \overline{s}'')\}$,
where $\post{\sigma} =  \overline{s}''$, so that the resulting multiset
of associations satisfies the three conditions required by property $\Gamma(\pi_1, g, \pi_2')$.
Similarly, $\Gamma(\pi_1', g, \pi_2')$ holds because from the multiset $q = \{(r_1, r_1'),$ $ (r_2, r_2'),$ $ \ldots,$ $(r_k, r_k')\}$ we remove the multiset 
$p = \{(\overline{s}, \overline{s}')\} \subseteq q$, and we
add the multiset $p'' = \{(\underline{s}, \overline{s}'')\}$,
where $\post{t_1} = \underline{s}$ and $\post{\sigma} =  \overline{s}''$, so that the resulting multiset
of associations satisfies the three conditions required by property $\Gamma(\pi_1', g, \pi_2')$.

In the or-case $(b)$, we can really extend $\pi_2$ by performing a suitable $\sigma'$ (with $o(\sigma') = \epsilon$) to a suitable 
process $\pi_2'$ such that $\pi_2 \Deriv{\sigma'} \pi_2'$, $\rho_2'(\sigma') = \sigma$ and $(\pi_1, g, \pi_2') \in R_2$; the last conditions
can be proved similarly as above; in particular, property $\Gamma(\pi_1, g, \pi_2')$ holds
because 
from the multiset $q = \{(r_1, r_1'),$ $ (r_2, r_2'),$ $ \ldots,$ $(r_k, r_k')\}$ we remove the multiset 
$p = \{(\overline{s}_{1}, \overline{s}'_{1}),$ $\ldots,$ $ (\overline{s}_{h'}, \overline{s}'_{h'})\} \subseteq q$
(such that $ \pre{t_1} = \overline{s}_{1}\oplus  \ldots \oplus \overline{s}_{h'}$ and $\pre{\sigma} = \overline{m}_2 = 
\overline{s}'_{1} \oplus  \ldots \oplus \overline{s}'_{h'}$) and we add
the multiset $p' = \{(\overline{s}_{1}, \overline{s}''_{1}), \ldots, (\overline{s}_{h'}, \overline{s}''_{h'})\}$,
where $\post{\sigma} =  \overline{s}''_{1} \oplus \ldots \oplus  \overline{s}''_{h'}$, so that the resulting multiset, say $q'$,
of associations satisfies the three conditions required by property $\Gamma(\pi_1, g, \pi_2')$, indeed.

Furthermore, as property $\Gamma(\pi_1, g, \pi_2')$ holds for the resulting multiset $q'$ 
and, moreover, $p' \subseteq q'$ is the multiset of associations ensuring that $(\pre{t_1}, \pre{t_2}) \in \overline{R}_1^\oplus$, 
it is possible to single out an event $e_2$ 
such that  $\pi_2' = (C_2', \rho_2') \deriv{e_2} (C_2'', \rho_2'') = \pi_2''$
(where $\rho_2''$ is such that $\rho_2''(e_2) = t_2$, with $l(t_1) = l(t_2)$) and such that
the set of observable events generating (or causing) the conditions of $\pre{e_1}$ (which are
mapped by $\rho_1$ to $\pre{t_1}$) are isomorphic, via $g$,
to the set of observable events generating (or causing) the conditions of $\pre{e_2}$ (which are
mapped by $\rho_2'$ to $\pre{t_2}$). 
Therefore, 
the new generated events $e_1$ and $e_2$ have isomorphic observable predecessors via $g$. 
So, by defining $g' = g \cup \{(e_1, e_2)\}$ (in case $l(t_1) \neq \tau$; otherwise, $g' = g$ and this case is trivial),
we can conclude that $g'$ is an abstract event isomorphism between $C_1'$ and $C_2''$,
so that $(\pi_1', g', \pi_2'') \in R_2$. This last condition holds because property $\Gamma(\pi_1', g', \pi_2'')$ holds.
In fact, from the multiset of associations $q'$ we remove the associations in $p'$ and 
add any multiset $p''$ of associations
that can be used to prove that $(\post{t_1}, \post{t_2}) \in R_1^\odot$. The resulting multiset $q''$ 
satisfies property $\Gamma(\pi_1', g', \pi_2'')$, as $q''$ can be used to prove that
 $(\rho_1'(Max(C_1')), \rho_2''(Max(C_2''))) \in R_1^\odot$
and 
for each $(r_i, r_i') \in q''$, we have that either one of the two elements in the pair is
$\theta$, or both places are the image of suitable conditions with no observable predecessor events, or
both places are the image of conditions generated by (or causally dependent on) events related by 
the abstract event isomorphism $g'$.

The case when $\pi_2 = (C_2, \rho_2)$ moves first is symmetrical and so omitted.
Therefore, $R_2$ is a branching fully-concurrent bisimulation and, since $(\pi_1^0, g^0, \pi_2^0) \in R_2$, we have that
$m_1 \approx_{bfc} m_2$.
\fine
\end{theorem}

However, the reverse implication of Theorem \ref{bdpbis>bfc-bis} does not hold in 
general: it may happen that  $m_1 \approx_{bfc} m_2$ but $m_1 \not \approx_{d} m_2$, as the following 
example shows.

\begin{figure}[t]
\centering
\begin{tikzpicture}[
every place/.style={draw,thick,inner sep=0pt,minimum size=6mm},
every transition/.style={draw,thick,inner sep=0pt,minimum size=4mm},
bend angle=45,
pre/.style={<-,shorten <=1pt,>=stealth,semithick},
post/.style={->,shorten >=1pt,>=stealth,semithick}
]
\def\eofigdist{2.3cm}
\def\eodist{0.5}
\def\eodisty{0.8}

\node (q1) [place] [label={above:$s_1$} ] {};
\node (t1) [transition] [right=\eodist of q1,label=above:$\tau$] {};
\node (q2) [place] [above right=\eodisty of t1,label=above:$s_2$] {};
\node (q3) [place] [below right=\eodisty of t1,label=above:$s_3$] {};
\node (t2) [transition] [right=\eodist of q2,label=above:$\tau$] {};
\node (t21) [transition] [above right=\eodisty of q2,label=above:$b$] {};
\node (t4) [transition] [above right=\eodisty of q3,label=above:$a$] {};
\node (t5) [transition] [below right=\eodisty of q3,label=above:$\tau$] {};
\node (q4) [place] [above right=\eodisty of t5,label=above:$s_4$] {};
\node (t6) [transition] [right=\eodist of q4,label=above:$\tau$] {};
\node (t7) [transition] [above right=\eodisty of q4,label=above:$b$] {};

\draw [->] (q1) to (t1);
\draw [->] (t1) to (q2);
\draw [->] (t1) to (q3);
\draw [->] (q2) to (t2);
\draw [->] (q2) to (t21);
\draw [->] (q3) to (t4);
\draw [->] (q3) to (t5);
\draw [->] (t5) to (q4);
\draw  [->, bend left] (t5) to (q3);
\draw [->] (q4) to (t6);
\draw [->] (q4) to (t7);

\end{tikzpicture}

\caption{A P/T net with $s_1 \approx_{bfc} s_3$ but $s_1 \not\approx_{d} s_3$}
\label{stutt1-fig}
\end{figure}
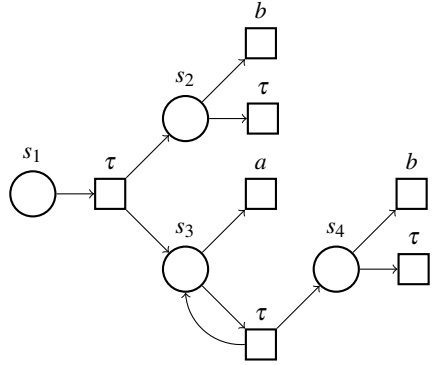

\begin{example}\label{counter-ex1}
Consider the net in Figure \ref{stutt1-fig}. It is not difficult to realize that $s_1 \approx_{bfc} s_3$. 
Informally, if $s_1 \deriv{\tau} s_2 \oplus s_3$,
$s_3$ can reply with $s_3 \deriv{\tau} s_3 \oplus s_4$ and $s_2 \oplus s_3 \approx_{bfc} s_3 \oplus s_4$, as required. 
Symmetrically, besides the move above, $s_3$ can also do $s_3 \deriv{a} \theta$, and $s_1$ can reply with $s_1 \Deriv{\tau} s_3 \deriv{a} \theta$ with $s_3 \approx_{bfc} s_3$ and $\theta \approx_{bfc} \theta$.
However, $s_1 \not \approx_{d} s_3$: if $s_3 \deriv{a} \theta$, then $s_1$ can only respond with $s_1 \deriv{\tau} s_2 \oplus s_3 \deriv{\tau} s_3 \deriv{a} \theta$, but the silent path $s_1 \deriv{\tau} s_2 \oplus s_3 \deriv{\tau} s_3$ is not composed of $\tau$-sequential transitions only (actually, none 
of the two is $\tau$-sequential).
\fine
\end{example}

\begin{figure}[t]
\centering

\begin{tikzpicture}

\node (e1) at (7,25) {$\sim_{p}$};
\node (e2) at  (4,24) {$\sim_{cn}$};
\node (e3) at (9,24) {$\approx_{p}$};
\node (e4) at (7,24) {$\sim_d$};
\node (e7) at(2,23) {$\sim_{fc}$};
\node (e8) at (9,23) {$\approx_{d}$};
\node (e9) at(2,22) {$\sim_{int}$};
\node (e10) at (7,22) {$\approx_{bfc}$};
\node (e12) at (7, 21) {$\approx_{bri}$};

\draw  (e1) to (e2);
\draw  (e1) to (e3);
\draw  (e1) to (e4);
\draw  (e2) to (e7);
\draw  (e3) to (e8);
\draw  (e4) to (e7);
\draw  (e4) to (e8);
\draw  (e7) to (e9);
\draw  (e7) to (e10);
\draw  (e8) to (e10);
\draw  (e10) to (e12);
\draw  (e9) to (e12);

\end{tikzpicture}

\hrulefill

{\tiny 
{\bf Legenda}\\
$
\begin{array}{lllllllll}
\sim_{p} & \mbox{place bisimilarity} &\quad  \sim_{cn} & \mbox{causal-net bisimilarity}\\
\sim_{d} & \mbox{d-place bisimilarity} &\quad  \sim_{fc} & \mbox{fully-concurrent bisimilarity} \\
\approx_{p} & \mbox{branching place bisimilarity} & \quad \sim_{int} & \mbox{interleaving bisimilarity}\\
\approx_{d} & \mbox{branching d-place bisimilarity} & \quad \approx_{bfc} & \mbox{branching fully-concurrent bisimilarity}\\
\approx_{bri} & \mbox{branching interleaving bisimilarity} & \quad\\

\end{array}
$
}

\hrulefill

\caption{The diagram with the 9 behavioral equivalences studied in this paper}
\label{diag2}
\end{figure}
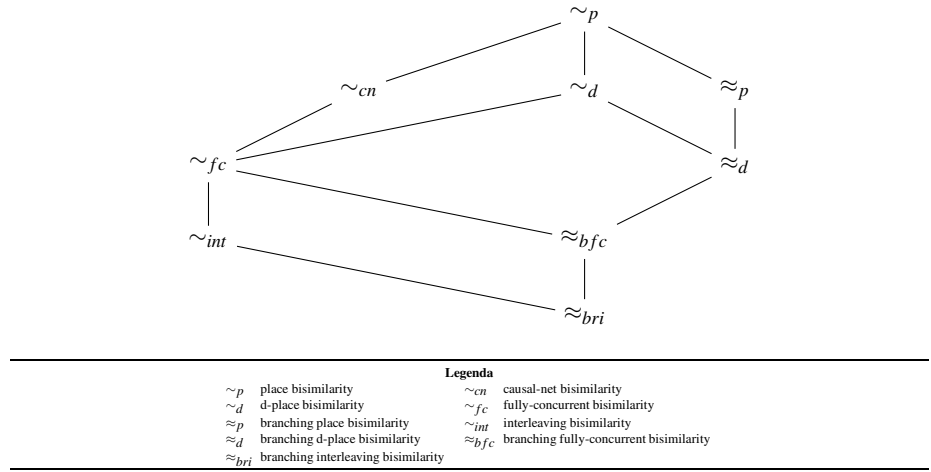

Figure \ref{diag2} shows the semantic inclusions among the 8 behavioral equivalences that we have considered in this paper, with the addition 
of {\em causal-net bisimilarity} $\sim_{cn}$ \cite{G15,Gor22} (which is equivalent to {\em structure-preserving bisimilarity} \cite{G15}) 
for completeness. 
The most discriminating of them is place bisimilarity $\sim_p$, while the coarsest one is branching interleaving bisimilarity $\approx_{bri}$. 
All the four place-based equivalences are decidable, while the others are undecidable ( with the exception of causal-net bisimilarity whose 
decidability is an open problem).


%
\section{Conclusion and Future Research}\label{conc-sec}
%

Place bisimilarity \cite{ABS91} is the only decidable \cite{Gor21}
behavioral equivalence for P/T nets which respects the expected causal behavior,
as it is slightly finer than {\em causal-net bisimilarity} \cite{G15,Gor22} (or, equivalently,
{\em structure preserving bisimilarity} \cite{G15}), 
in turn slightly finer than {\em fully-concurrent bisimilarity} \cite{BDKP91}. 
Thus, it is the only equivalence for which it is possible (at least, in principle) 
to verify algorithmically the (causality-preserving) correctness of an implementation by exhibiting a place 
bisimulation between its specification and implementation.

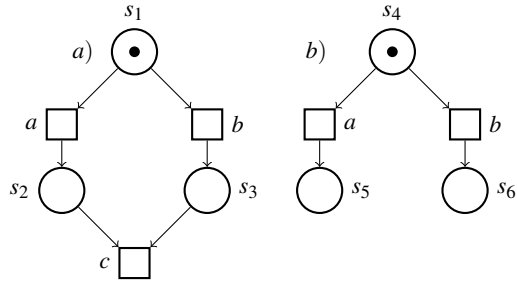
\begin{figure}[t]
\centering
\begin{tikzpicture}[
every place/.style={draw,thick,inner sep=0pt,minimum size=6mm},
every transition/.style={draw,thick,inner sep=0pt,minimum size=4mm},
bend angle=45,
pre/.style={<-,shorten <=1pt,>=stealth,semithick},
post/.style={->,shorten >=1pt,>=stealth,semithick}
]
\def\eofigdist{2.8cm}
\def\eodist{0.35}
\def\eodisty{0.75}

\node (a) [label=left:$a)\quad $]{};

\node (q1) [place,tokens=1] [label={above:$s_1$} ] {};
\node (f1)[transition][below left=\eodisty of q1,label=left:$a$]{};
\node (f2)[transition][below right=\eodisty of q1,label=right:$b$]{};
\node (q2) [place] [below=\eodist of f1,label={left:$s_2$}] {};
\node (q3) [place] [below=\eodist of f2,label={right:$s_3$}] {};
\node (f3)[transition][below right=\eodisty of q2,label=left:$c$]{};

\draw  [->] (q1) to (f1);
\draw  [->] (q1) to (f2);
\draw  [->] (f1) to (q2);
\draw  [->] (f2) to (q3);
\draw  [->] (q2) to (f3);
\draw  [->] (q3) to (f3);


\node (b) [right={2.7cm} of a,label=left:$b)\;\;$] {};

\node (p1) [place,tokens=1]  [right=\eofigdist of q1,label=above:$s_4$] {};
\node (s1) [transition] [below left=\eodisty of p1,label=right:$a$] {};
\node (s2) [transition] [below right=\eodisty of p1,label=right:$b$] {};
\node (p2) [place] [below =\eodist of s1,label= right:$s_5$]{};
\node (p3) [place] [below=\eodist of s2,label= right:$s_6$]{};

\draw  [->] (p1) to (s1);
\draw  [->] (p1) to (s2);
\draw  [->] (s1) to (p2);
\draw  [->] (s2) to (p3);

\end{tikzpicture}
\caption{Two non-place bisimilar nets}
\label{abs-net}
\end{figure}

It is sometimes argued that place bisimilarity is too discriminating. In particular, \cite{ABS91} and \cite{G15} argue
that a {\em sensible} equivalence should not distinguish markings whose behaviors are patently the same, such as
marked Petri nets that differ only in their unreachable parts. As an example, consider the net in Figure \ref{abs-net}, discussed in \cite{ABS91}.
Clearly, markings $s_1$ and $s_4$ are equivalent, also according to all the behavioral equivalences discussed in \cite{G15}, except
for place bisimilarity. As a matter of fact, a place bisimulation $R$ containing the pair $(s_1, s_4)$ would require also the pairs
$(s_2, s_5)$ and $(s_3, s_6)$, but then this place relation $R$ cannot be a place bisimulation because 
$(s_2, \oplus s_3, s_5 \oplus s_6) \in R^\oplus$, but $s_2 \oplus s_3$ can perform $c$, while this is not possible for $s_5 \oplus s_6$.
Nonetheless, we would like to argue in favor of place bisimilarity, despite this apparent paradoxical example.

As a matter of fact, our interpretation of place bisimilarity is that this equivalence is 
an attempt of giving semantics to {\em unmarked} nets, rather than to marked nets,
so that the focus shifts from the common (but usually undecidable) question {\em When are two markings equivalent?} to the more 
restrictive (but decidable) question {\em When are two places equivalent?}
A possible (preliminary, but not accurate enough) answer to the latter question may be: two places are equivalent if, 
whenever the same number of tokens are put on these two places,
the behavior of the marked nets is the same. If we reinterpret the example of Figure \ref{abs-net} in this perspective, we clearly see that
place $s_1$ and place $s_4$ cannot be considered as equivalent because, even if the markings $s_1$ and $s_4$
are equivalent, nonetheless the marking $2 \cdot s_1$ is not equivalent
to the marking $2 \cdot s_4$, as only the former can perform the trace $abc$.

A place bisimulation $R$ considers two places $s_1$ and $s_2$ as equivalent if $(s_1, s_2) \in R$, as,  by definition of place bisimulation, 
they must behave the same in any $R$-related context. Back to our example in Figure \ref{abs-net}, if $(s_1, s_4)$ would 
belong to $R$, then also $(2 \cdot s_1, 2 \cdot s_4)$ should belong to 
$R^\oplus$, but then we discover that the place bisimulation game does not hold for this pair of markings, so that $R$ cannot be a place bisimulation.

If we consider the duality between the process algebra FNM (a dialect of CCS, extended with multi-party interaction)
and P/T nets, proposed in \cite{Gor17}, we may find further arguments supporting this more restrictive interpretation of
net behavior. In fact, an {\em unmarked} P/T net $N$ can be described by an FNM system of equations, where each equation defines
a constant $C_i$ (whose body is a sequential process term $t_i$), representing place $s_i$. 
Going back to the nets in Figure \ref{abs-net}, according to this duality, the constant $C_1$ for place $s_1$
is not equivalent (in any reasonable sense) to the constant $C_4$ for place $s_4$ because these two constants 
describe all the potential behaviors of these two places, which are clearly different!
Then, the marked net $N(m_0)$ is described by a parallel term composed of as many instances of $C_i$ as the tokens that are present in $s_i$
for $m_0$, encapsulated by a suitably defined restriction operator $\restr{L}-$. Continuing the example,
it turns out that $\restr{L}C_1$ is equivalent to $\restr{L}C_4$ because the markings $s_1$ and $s_4$ are equivalent, 
but  $\restr{L}(C_1 \para C_1)$ is not equivalent to
$\restr{L}(C_4 \para C_4)$ because the markings $2 \cdot s_1$ is not equivalent
to the marking $2 \cdot s_4$, as discussed above.

Moreover, there are at least the following three important technical differences between 
place bisimilarity and other coarser, causality-respecting equivalences, such as fully-concurrent bisimilarity \cite{BDKP91}.
\begin{enumerate}
    \item A fully-concurrent bisimulation is a complex relation --
        composed of cumbersome triples of the form (process, bijection, process)  --
        that must contain infinitely many triples if the net system offers never-ending behavior. (Indeed, not even one single case study of a system with never-ending behavior
        has been developed for this equivalence.)
        On the contrary,
        a place bisimulation is always a very simple finite relation over the finite set of places. (And a simple case study is described in \cite{Gor21}.)
        
    \item A fully-concurrent bisimulation  proving that $m_1$ and $m_2$ are equivalent
            is a relation specifically designed for the initial markings $m_1$ and $m_2$. If we want to prove that,
            e.g., $n \cdot m_1$ and $n \cdot m_2$ are fully-concurrent bisimilar (which 
            may not hold!), we have to construct a new fully-concurrent bisimulation to this aim. 
            Instead, a place bisimulation $R$ 
            relates those places which are considered equivalent under all the possible 
            $R$-related contexts. 
            Hence, if $R$ justifies that  $m_1 \sim_{p} m_2 $ 
            as $(m_1, m_2) \in R^\oplus$, then
            for sure the same $R$ justifies that $n \cdot m_1$ and $n \cdot m_2$ are 
            place bisimilar, as also 
            $(n \cdot m_1, n \cdot m_2) \in R^\oplus$.
            
    \item Finally, while place bisimilarity is decidable \cite{Gor21}, fully-concurrent bisimilarity is undecidable on finite P/T nets \cite{Esp98}. 
\end{enumerate}

The newly defined {\em branching place bisimilarity} is the only extension 
of the place bisimilarity idea to P/T nets with silent moves
that has been proved decidable, even if the time complexity of its decision procedure we have proposed
is exponential in the size of the net. 
Thus, it is the only equivalence for P/T nets with silent transitions for which it is possible (at least, in principle) 
to verify algorithmically the correctness of an implementation by exhibiting a branching (d-)place bisimulation between its
specification and implementation, as we did for the small case study in Section \ref{case-sec}.

We have also proposed a slight weakening of branching place bisimilarity $\approx_p$, called {\em branching d-place bisimilarity} $\approx_d$, which may 
relate places to the empty marking $\theta$
and which is still decidable. Actually, we conjecture that branching d-place bisimilarity is the coarsest, 
sensible equivalence relation which is decidable on finite P/T nets with silent moves. 

Of course, these behavioral relations may be subject to the same criticisms raised to place bisimilarity
and also its restrictive assumption that only $\tau$-sequential transitions can be abstracted away can be criticized,
as its applicability to real case studies may appear rather limited.
In the following, we try to
defend our point of view.

First, on the subclass of BPP nets, branching place bisimilarity coincides with 
{\em branching team bisimilarity} \cite{Gor20c}, a very satisfactory equivalence which is actually coinductive and, for this reason, 
also very efficiently decidable in polynomial time. Moreover, on the subclass of {\em finite-state machines}
(i.e., nets whose transitions have singleton pre-set and singleton, or empty, post-set), branching team bisimilarity
has been axiomatized \cite{Gor19b} on the process algebra CFM \cite{Gor17}, which can represent all (and only) the 
finite-state machines, up to net isomorphism.

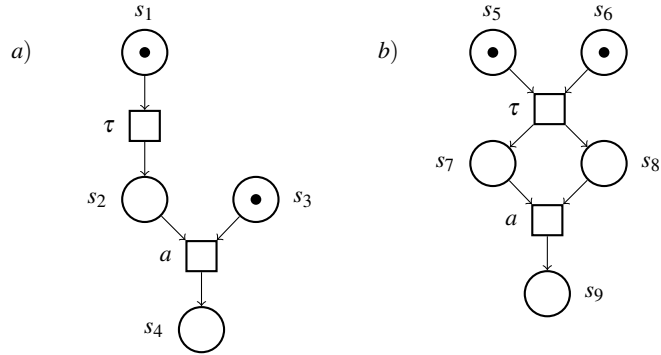
\begin{figure}[t]
\centering
\begin{tikzpicture}[
every place/.style={draw,thick,inner sep=0pt,minimum size=6mm},
every transition/.style={draw,thick,inner sep=0pt,minimum size=4mm},
bend angle=30,
pre/.style={<-,shorten <=1pt,>=stealth,semithick},
post/.style={->,shorten >=1pt,>=stealth,semithick}
]
\def\eofigdist{4cm}
\def\eodist{0.45cm}
\def\eodisty{0.85cm}

\node (a) [label=left:$a)\qquad \qquad$]{};

\node (p1) [place,tokens=1]  [label=above:$s_1$] {};
\node (t1) [transition] [below =\eodist of p1,label=left:$\tau\;$] {};
\node (p2) [place] [below =\eodist of t1,label=left:$s_2\;$] {};
\node (p3) [place,tokens=1] [right =\eodisty of p2,label=right:$\;s_3$] {};
\node (t2) [transition] [below right=\eodist of p2,label=left:$a\;$] {};
\node (p4) [place] [below =\eodist of t2,label=left:$s_4\;$] {};

\draw  [->] (p1) to (t1);
\draw  [->] (t1) to (p2);
\draw  [->] (p2) to (t2);
\draw  [->] (p3) to (t2);
\draw  [->] (t2) to (p4);

  \node (b) [right={3.7cm} of a,label=left:$b)\quad$] {};

\node (p5) [place,tokens=1]  [right=\eofigdist of p1, label=above:$s_5$] {};
\node (p6) [place,tokens=1]  [right=\eodisty of p5, label=above:$s_6$] {};
\node (t4) [transition] [below right=\eodist of p5,label=left:$\tau\;$] {};
\node (p7) [place]  [below =\eodisty of p5,label=left:$s_7\;$] {};
\node (p8) [place]  [below =\eodisty of p6,label=right:$\;s_8$] {};
\node (t5) [transition] [below left=\eodist of p8,label=left:$a\;$] {};
\node (p9) [place]  [below =\eodist of t5,label=right:$\;s_9$] {};

\draw  [->] (p5) to (t4);
\draw  [->] (p6) to (t4);
\draw  [->] (t4) to (p7);
\draw  [->] (t4) to (p8);
\draw  [->] (p7) to (t5);
\draw  [->] (p8) to (t5);
\draw  [->] (t5) to (p9);

\end{tikzpicture}
\caption{Two  branching fully-concurrent P/T nets}
\label{net-bfc-place}
\end{figure}

Second, branching (d-)place bisimilarity is a sensible behavioral equivalence relation, as it 
does respect the causal behavior of P/T nets. In fact, we have proved that {\em  branching fully-concurrent 
bisimilarity} \cite{Pin93,Gor20c} (which is undecidable) is strictly coarser than $\approx_d$, 
because it may equate nets whose silent transitions are not $\tau$-sequential (and also may relate markings of different size),
as illustrated in Example \ref{counter-ex1}.
As a further example,
consider the net in Figure \ref{net-bfc-place}. Of course, the markings $s_1 \oplus s_3$ and $s_5 \oplus s_6$
are branching fully-concurrent bisimilar: to the move $s_1 \oplus s_3 [t_1 \rangle s_2 \oplus s_3$, where
$t_1 = (s_1, \tau, s_2)$, $s_5 \oplus s_6$ can reply with $s_5 \oplus s_6[t_2\rangle s_7 \oplus s_8$,
where $t_2 = (s_5 \oplus s_6, \tau, s_7 \oplus s_8)$ and the reached markings are clearly equivalent.
However, $s_1 \oplus s_3 \not\approx_p s_5 \oplus s_6$ because $s_1 \oplus s_3$ cannot reply
to the move $s_5 \oplus s_6[t_2\rangle s_7 \oplus s_8$, as $t_2$ is not $\tau$-sequential (i.e., it can be seen as 
the result of a synchronization), while $t_1$ is $\tau$-sequential.

We already argued in the introduction that it is very much questionable whether a synchronization 
can be considered as unobservable, even if this idea is rooted 
in the theory of concurrency from the very beginning.
As a matter of fact, in CCS \cite{Mil89} and in the 
$\pi$-calculus \cite{MPW,SW}, the result of
a synchronization is a silent, $\tau$-labeled (hence unobservable) transition. 
However, the silent label $\tau$ is used in these 
process algebras for two different purposes:
\begin{itemize}
\item
First, to ensure that a synchronization is strictly binary: 
since the label $\tau$ cannot be used for synchronization, by labeling a synchronization transition by $\tau$
any further synchronization of the two partners with other parallel components is prevented (i.e.,
multi-party synchronization is disabled). 
\item
Second, to describe that the visible effect of the transition is null: a $\tau$-labeled transition can be considered 
unobservable and can be abstracted away, to some extent.
\end{itemize}

Nonetheless,  it is possible to modify slightly these process algebras by introducing
two different actions for these different purposes.  
In fact, the result of a binary synchronization can be some {\em observable} label, say $\lambda$ 
(or even $\lambda(a)$, if the name of the communication channel $a$ is
considered as visible), for which 
no co-label exists, so that further synchronization is impossible.
While the action $\tau$, that can be used as a prefix for the prefixing operator, is 
used to denote some local, internal (hence unobservable) computation.
In this way, a net semantics for these process algebras (in the style of, e.g., \cite{Gor17}) would generate
$\tau$-sequential P/T nets, that are amenable to be compared by means of branching (d-)place bisimilarity.

As a final comment, we want to discuss an apparently insurmountable limitation of our approach. 
In fact, the extension of the 
place bisimulation idea to nets with silent transitions that are not $\tau$-sequential seems very hard, or even impossible. 
Consider again the two P/T nets 
in Figure \ref{net-bfc-place}. If we want that $s_1 \oplus s_3$ be related to $s_5 \oplus s_6$, 
we need to include the pairs
$(s_1, s_5)$ and $(s_3, s_6)$. If the marking $s_5 \oplus s_6$ silently reaches
$s_7 \oplus s_8$, then $s_1 \oplus s_3$ can respond by idling 
(and in such a case we have to include the pairs $(s_1, s_7)$ and $(s_3, s_8)$) or by performing
the transition $s_1 \deriv{\tau} s_2$ (and in such a case we have to include the pairs $(s_2, s_7)$ and $(s_3, s_8)$).
In any case, the candidate place relation $R$ should be of the form 
$\{(s_1, s_5), (s_3, s_6), (s_3, s_8), \ldots\}$.
However, this place relation cannot be a place bisimulation of any sort because, on the one hand, 
$(s_1 \oplus s_3, s_5 \oplus s_8) \in R^\oplus$
but, on the other hand, $s_1 \oplus s_3$ can eventually perform $a$, while $s_5 \oplus s_8$ is stuck.

Nonetheless, this negative observation is coherent with our intuitive interpretation of (branching) place bisimilarity as 
a way to give semantics to
{\em unmarked} nets. In the light of the duality between
P/T nets and the FNM process algebra discussed above \cite{Gor17}, 
a place is interpreted as a sequential process type (and each token
in this place as an instance of a sequential process of that type); 
hence, a (branching) place bisimulation
essentially states which kinds of sequential processes (composing the distributed system represented by the Petri net)
are to be considered equivalent. In our example above, it makes no sense to consider place $s_1$ and place $s_5$ as equivalent,
because the corresponding FNM constants $C_1$ and $C_5$ have completely different behavior: 
$C_5$ can interact (with $C_6$), while $C_1$ can only perform 
some internal, local transition. 

Future work will be devoted to find more efficient algorithms for checking branching place bisimilarity. 
One idea could be to build directly the set of maximal branching place bisimulations, rather than to scan all the 
place relations to check whether
they are branching place bisimulations, as we did in the proof of Theorem \ref{bpl-bis-decid-th}.

\end{document}